\documentclass[aps,pra,showpacs,twoside,twocolumn,10pt]{revtex4-2}
\usepackage[colorlinks=true, citecolor=red, urlcolor=blue ]{hyperref}
\usepackage{epsfig,newlfont,amssymb,amsfonts,amsmath,bm,subfigure,palatino,mathtools,amsthm,braket,soul,enumitem,color,graphics,graphicx,times,physics,bbold}
\usepackage[normalem]{ulem}
\usepackage{xcolor}
\usepackage{physics}
\usepackage{dsfont}
\usepackage{mathrsfs}
\usepackage{verbatim}
\usepackage{amsthm,amssymb}
\usepackage{comment}
\usepackage{bigints}

\begin{document}

\title{Violation of the thermodynamic uncertainty relation in quantum collisional models}

\author{ Arghya Maity$^{1,2}$ and Ahana Ghoshal$^{3}$ }
\affiliation{$^{1}$  School of Physical and Mathematical Sciences, Nanyang Technological University, 21 Nanyang Link, Singapore 637371, Singapore \\
$^{2}$ Harish-Chandra Research Institute, A CI of Homi Bhabha National
Institute, 
Chhatnag Road, Jhunsi, Allahabad - 211019, India \\
$^{3}$ Naturwissenschaftlich-Technische Fakult\"{a}t, Universit\"{a}t Siegen, Walter-Flex-Stra\ss e 3, 57068 Siegen, Germany}

\begin{abstract} 
The thermodynamic uncertainty relation (TUR) is a fundamental principle in non-equilibrium thermodynamics that relates entropy production to fluctuations in a system, establishing a trade-off between the precision of an observable and the thermodynamic cost. Investigating TUR violations challenges classical thermodynamic limits, offering the potential for improved precision-entropy trade-offs, which is crucial for enhancing performance and optimization in quantum technologies.
In this work, we investigate the thermodynamic uncertainty relation within a quantum collisional model, which offers the advantage of discretizing interactions into successive collisions with auxiliaries, allowing for precise tracking of dynamics and the incorporation of memory effects and non-Markovian behavior.
We consider three types of dynamics in the collisional model: one is Markovian evolution, achieved by taking the continuous time limit and imposing the stability condition, while the other two are non-Markovian dynamics—one arising from increasing the collision time between the system and the auxiliaries, and the other from incorporating interactions between the auxiliaries. For the Markovian dynamics, we examine the classical and quantum TUR bounds in the non-equilibrium steady-state regime, and also the finite-time TUR bound.
We identify two distinct regimes of classical TUR violation: in some cases, the maximum violation occurs in the steady state, while in others, it is necessarily transient—appearing only at early times and vanishing with further evolution. 
For the two non-Markovian approaches, we find that both the degree and type of non-Markovianity crucially affect TUR violations. The second approach shows more pronounced violations during transient times, 
while the first approach has much 
stronger violations in the steady-state regime for a certain parameter window.
\end{abstract}

\maketitle

\section{Introduction}
\label{Sec.1}

The set of thermodynamic uncertainty relations (TURs)~\cite{Barato2015,Gingrich2016,Barato2019,Falasco2020,Hasegawa2021,Menczel2021,Falasco2022} builds upon and extends the principles of the Fluctuation-Dissipation Theorem (FDT)~\cite{Callen1951,Kubo1966}, providing a framework to explore the trade-off between fluctuations and dissipation in nonequilibrium systems. 
While the FDT traditionally connects equilibrium fluctuations to the linear response of the system to perturbations, TURs generalize this concept to non-equilibrium steady states and beyond. The standard form of TUR establishes that the precision of thermodynamic currents, such as heat or particle flow, is inversely related to the entropy production, thereby linking fluctuations in these currents to the irreversibility of the underlying processes. This means that achieving higher precision in measuring a current necessitates greater thermodynamic cost (of maintaining the non-equilibrium steady state) which is quantified by entropy production rate.
 Initially, TURs were formulated for classical time-homogeneous Markov processes~\cite{Carollo_PRL_2019,Goold_PRR_2019,Hasegawa_PRL_2020,Horowitz2020,Hasegawa_PRL_2021}, capturing essential aspects of nonequilibrium steady states. 
 Subsequent refinements and extensions have broadened their scope to include finite-time processes~\cite{Gingrich2016,Patrick_PRE_2017, Horowitz_PRE_2017, Otsubo2020}, periodically driven systems~\cite{Proesmans_EPL_2017,Barato_NJP_2018, Holubec_PRL_2018}, and quantum systems within the linear response regime~\cite{Macieszczak_PRL_2018}. Recently, the short-time current and short-time TUR have also been studied explicitly which focuses on the behavior of TUR in the limit of small observation times~\cite{Supriya_PRL_2020,Hasegawa_PRE_2020, Jeong_PRL_2020, Supriya_CP_2021, Otsubo2022}. TURs have also been adapted to account for correlations between multiple currents~\cite{moreira2024}, discrete or time-dependent driving forces~\cite{Barato2016, Proesmans_EPL_2017, Koyuk2020, Cangemi2021}, underdamped Langevin dynamics~\cite{Fischer2018, Lee2019, Vu2019}, systems involving measurement and feedback mechanisms~\cite{Potts2019}, and for more general finite-time nonstationary processes~\cite{Landi_PRL_2019}.
TURs establish fundamental limits for the efficiency and precision of biomolecular processes~\cite{Barato2015,Pietzonka_2016,Jack2020,Song2020} and thermal machines~\cite{Campisi2016,Holubec2017,Seifert_PRL_2018}, where thermal and quantum fluctuations play a significant role due to their small size. 
 Their applicability extends to Brownian clocks~\cite{Barato2016}, which are stochastic oscillators driven by non-equilibrium processes and to active matter systems, such as bacterial swarms or synthetic colloidal particles, where non-equilibrium driving forces produce steady-state currents~\cite{Seifert_2019_AR}.
Experimental efforts to realize and test the TURs are ongoing, driving forward the understanding of its practical implications~\cite{Hwang2018,Paneru2020,Pal2020,Friedman2020,Supriya_CP_2021}. 

TUR violations occur when the trade-off between the precision of thermodynamic currents and entropy production does not adhere to the predicted bounds, indicating that a system achieves higher precision at lower dissipation than expected. 
These violations can arise under certain conditions, such as the breaking of time-reversal symmetry 
~\cite{Taddei_PRB_2023, Saito_PRL_2018, Cangemi2021} and the influence of quantum coherence~\cite{Krzysztof_PRB_2018,Bijay_PRB_2018,Cangemi_PRB_2020,Mitchison_PRE_2021, Potts_PRE_2021, Keiji_PRL_2022,mohan2024}. 
While TURs in classical systems are derived under assumptions such as Markovian dynamics and the absence of quantum effects, quantum systems can ``violate" these bounds by utilizing quantum resources, enabling higher precision with lower dissipation. 
Quantum coherence has already been shown to play an important role in TUR violations~\cite{Krzysztof_PRB_2018,Bijay_PRB_2018,Cangemi_PRB_2020,Mitchison_PRE_2021, Potts_PRE_2021, Keiji_PRL_2022,mohan2024}. The validity of TURs in transport systems has been studied, and the conditions for its validity have been derived~\cite{Saito_PRL_2018,Saryal2019}. TUR violations in steady-state quantum thermoelectric junctions have also been investigated~\cite{Segal_PRE_2019}.

A quantum collisional model represents the interaction between a system and its environment as a series of discrete collisions with auxiliary systems. 
Each auxiliary system interacts with the main system for a finite duration before being replaced by a new auxiliary.
Unlike the traditional system-bath approach, which assumes continuous interactions with an environment, the collisional model offers greater flexibility by discretizing interactions, allowing for precise tracking of system dynamics~\cite{Landi2014,Ciccarello2022,Cusumano2022}. Additionally, it facilitates the incorporation of memory effects and non-Markovian behavior, making it an effective tool for studying open quantum systems and investigating decoherence processes~\cite{Ciccarello2013,Bernardes2014,Kretschmer_2016,Lorenzo2017,Man2019,Baris_Entropy_2022, Tanmay_NJP_2024}.
Collision models have proven to be a valuable tool for quantum transport, addressing limitations in traditional Lindblad master equations \cite{Levy_EPL_2014,Chisholm_NJP_2021, Tian_Entropy_2021}. They have been applied to various areas, including quantum synchronization \cite{Karpat_PRA_2019}, multipartite entanglement generation \cite{Paternostro_PRA_2019}, and on different quantum devices~\cite{Molitor_PRA_2020, Taranto_PRAP_2020, Abah_JPC_2020, Stella_PRL_2021}. 
These models have been instrumental in advancing our understanding of equilibrium and non-equilibrium dynamics \cite{Scarani_PRL_2002, Bruneau_JMP_2014, Grimmer_PRA_2016, Seah_PRE_2019, Cattaneo_PRL_2021}, and thermodynamics under strong coupling conditions \cite{Strasberg_PRL_2019}.
Collisional models have been implemented in various experimental platforms, including photonic systems~\cite{Bernardes_SR_2016}, trapped ions~\cite{Barreiro_Nature_2011}, quantum computers, and other experimental setups~\cite{Cuevas_SR_2019, IBM_NPJ_2020, Burger_Entropy_2022, Sabrina_PRXQ_2023, Erbanni_PRA_2023}.
A recent implementation showcased the capability of collisional models to explore both steady-state and periodic nonequilibrium steady-state dynamics in an $XXZ$ spin system~\cite{X_Science_2024}. Additionally, significant advancements have been made in using collisional models to study various aspects of quantum thermodynamics. For example, they have been applied to investigate non-equilibrium steady states~\cite{Dragi_PRL_2009, Seah_PRE_2019, Campbell_PLA_2020}, quantum battery charging~\cite{Stella_PRL_2021}, and thermometry~\cite{Landi_PRL_2019}.

In this paper, we investigate the thermodynamic uncertainty relation within the framework of a quantum collisional model. Understanding TUR violations is crucial for evaluating the performance and stability of quantum devices, while collisional models play a significant role in their design. Hence, by studying TUR violations within quantum collisional models, we uncover pathways to surpass traditional limits, paving the way for more efficient and advanced quantum technologies.
%
In the collisional model, we examine the TUR violation for three types of dynamics. The first type is Markovian evolution, which is obtained by taking the continuous-time limit and imposing the stability condition. The second and third types involve non-Markovian dynamics. The second arises from increasing the collision time between the system and the auxiliaries. The third type is generated by incorporating interactions between the auxiliaries themselves. For Markovian evolution, we examine three key aspects: the classical TUR bound in nonequilibrium steady states, the finite-time TUR bound for each collision, and the quantum TUR bound. We observe that the classical TUR bound is violated, with the maximum violation occurring either in the transient or steady-state regime, depending on parameters such as the magnetic field strength and the interaction strengths between the system and the auxiliaries. This gives rise to two distinct behaviors: in some regimes, the violation grows with evolution time and reaches its maximum in the steady state, while in others, it is necessarily transient, manifesting only at early times and vanishing as the system evolves.
The necessarily transient feature matters because it shows how TUR violation can be exploited within finite-time windows, which is especially useful for quantum devices that cannot rely on steady-state operation or where transient performance is more advantageous than their steady-state counterpart. We then analyze the quantum TUR bound using the same parameters that cause the classical violations. 
Our findings show that the model consistently adheres to the quantum TUR bound, underscoring its alignment with the principles of quantum thermodynamics. For the two non-Markovian approaches, we investigate the finite-time classical TUR, along with the degree of non-Markovianity using the well-known Breuer-Laine-Piilo measure~\cite{BLP_PRL_2009, BLP_PRA_2010}.
In non-Markovian dynamics with extended collision times, two distinct behaviors emerge, similar to the Markovian case: in some parameter regimes, the violation grows with increasing evolution time and persists into the steady state, while in others, it is necessarily transient.
In contrast, non-Markovian dynamics with interacting auxiliaries exhibit TUR violations that are more pronounced at short evolution times, leading to a broader necessarily transient regime. These contrasting behaviors highlight that both the degree and nature of non-Markovianity play a crucial role in shaping the extent and character of TUR violations.

The rest of the paper is organized as follows. In Sec.~\ref{Sec-2}, we introduce the physical setup used for analyzing the TUR bounds, along with the Markovian and non-Markovian formulations considered in this work. 
In Sec.~\ref{Sec.3}, we outline the method for calculating the TUR parameters for both the Markovian and the non-Markovian approaches, including the evaluation of the quantum TUR parameter. In Sec.~\ref{Sec.4}, we present our main results, showing the behavior of the TUR parameters for all Markovian and non-Markovian approaches, with a focus on the violation of the classical TUR bound. We conclude our findings in Sec.~\ref{Sec.5}.

\section{The model}
\label{Sec-2}

Consider a system $S$ modeled as a spin-$\frac{1}{2}$ particle that interacts sequentially with a set of auxiliary systems $\{A_1, A_2, \ldots, A_j\}$, where each auxiliary system is also modeled as a spin-$\frac{1}{2}$ particle, acting as an environment for the system $S$. The free Hamiltonians for the system and the auxiliaries are given by 
\begin{equation}
    \mathcal{H}_\alpha = \frac{1}{2} \hbar \omega_\alpha \sigma^z_\alpha,
\end{equation}
where $\alpha = S$ for the system and $\alpha = A_j$ for the $j^{\text{th}}$ auxiliary. Here, $\omega_\alpha$ represents the characteristic angular frequency associated with the $\alpha$-th system (either the system $S$ or the auxiliary $A_j$).
In addition, the spin system $S$ is driven by an external oscillating field given by
 \begin{equation}  
 \mathcal{V}(t) = \hbar \nu \big [ \sigma_S^x \cos(\omega t_d) + \sigma_S^y \sin(\omega t_d) ], 
 \end{equation}
where $\sigma^i_S \ (i=x,y,z)$ is the Pauli matrices acting on the spin space of the system, $\omega$ is the rotational frequency of the magnetic field, $\nu$ represents the field strength with units of frequency, and $t_d$ is the driving time. 
Thus, the total Hamiltonian of the system becomes 
\begin{equation}
\mathcal{H}_{\text{tot}}(t_d) =  \mathcal{H}_S + \mathcal{V}(t_d).
\end{equation}
Here, we assume that the auxiliary systems do not interact with one another. Initially, both the system and the auxiliaries are assumed to be in thermal equilibrium at their respective inverse temperatures $\beta_\alpha=1/k_B T_\alpha$. Therefore, the initial states of the system and the $j^{\text{th}}$ auxiliary are given by $\rho_{\alpha}^0 = \exp(-\beta_{\alpha}\mathcal{H}_{\alpha})/\tr[\exp(-\beta_{\alpha}\mathcal{H}_{\alpha})]$. Here $k_B$ is the Boltzmann constant. Here, we consider all environments to have a uniform temperature $T_A$ and a uniform characteristic angular frequency $\omega_A$, ensuring that nonequilibrium behavior arises solely from the time-dependent driving $\mathcal{V}(t)$ of the system. While nonequilibrium driving can result from various factors, in our case, it is exclusively induced by the time-dependent driving.
The system interacts with each auxiliary for a duration 
$\tau$, governed by the Hamiltonian
\begin{equation}
\label{eq:int}
V_{SA_j}=g_1(\sigma_S^{+}\sigma_{A_j}^{-}+\sigma_S^{-}\sigma_{A_j}^{+})+g_2\sigma_S^z \sigma_{A_j}^z,
\end{equation}
where $\sigma^{\pm}_\alpha=(\sigma^x_\alpha\pm i\sigma^y_\alpha)/2$. 
The first term in this interaction Hamiltonian facilitates the exchange of excitations between the system and the auxiliary, while the second term accounts for pure decoherence. The coupling constants $g_1$ and $g_2$ have units of $\hbar/\sqrt{\tau}$. Hence, the composite Hamiltonian of the system and an auxiliary system is expressed as
\begin{equation}
\mathcal{H}_{SA_j}=\mathcal{H}_{\text{tot}}(t_d)+\mathcal{H}_{A_j}+V_{SA_j}/\sqrt{\tau},
\end{equation}
where the interaction term is rescaled by a factor of $1/\sqrt{\tau}$ to account for the finite contribution of the system-environment interaction~\cite{Dragi_PRL_2009,Landi2014,Rodrigues_PRL_2019}. The composite system-auxiliary state evolves under this Hamiltonian $\mathcal{H}_{SA_j}$ during each collision.
To simplify the analysis and eliminate the time dependence of the system Hamiltonian we shift to the rotating frame using the unitary transformation $\mathcal{H}_{\text{tot}}^{r} = e^{\frac{i}{\hbar}\Theta t_d}\mathcal{H}_{\text{tot}}(t_d) e^{-\frac{i}{\hbar}\Theta^{\dagger} t_d}$ where $ \Theta = \frac{1}{2} \hbar \omega \sigma^z_S $. This results in the effective Hamiltonian $ \widetilde{\mathcal{H}} = \mathcal{H}_{\text{tot}}^{r} - \Theta $. 
Hence, after the $j^{\text{th}}$ interaction, the reduced state of the system in the rotating frame is represented by
\begin{equation}
     \rho_{S_j}^r 
    =  \tr_{A_{1,2,..,j}}\Big [ U_{SA_j}^{\prime}\ldots U_{SA_1}^{\prime}  \Big( \rho_{S}^0 \bigotimes_{j=1}^{j} \rho_{A_j}^0 \Big ) U_{SA_1}^{\prime\dagger}\ldots U_{SA_j}^{\prime\dagger}     \Big ],
\label{eq:reduced_state}
\end{equation}
where the unitary $U_{SA_j}^{\prime}=\exp[-\frac{i}{\hbar}\mathcal{H}_{SA_j}^{\prime}\tau]$ with $\mathcal{H}_{SA_j}^{\prime}=\widetilde{\mathcal{H}}+\mathcal{H}_{A_j}+V_{SA_j}/\sqrt{\tau}$.
This collisional model effectively captures both Markovian and non-Markovian dynamics by adjusting the interaction of the system with the auxiliaries. Markovian evolution arises from short, memoryless interactions, while non-Markovian effects emerge from extended interaction times or correlations between auxiliaries. We have considered all three cases to examine TUR violations, Markovian evolution and two distinct non-Markovian approaches, discussed in the next sections. 

\subsection{Continuous-time limit and the Markovian approach}
\label{Sec.2A}

Although the collisional model is inherently a discrete-in-time framework, a continuous-time master equation can be derived in the limit when the duration of interactions $\tau$ becomes sufficiently small, i.e., $\tau\rightarrow 0$. This continuous formulation of the collisional model aligns with the Gorini-Kossakowski-Sudarshan-Lindblad (GKSL) master equation~\cite{Sudarshan_JMP_1976, Lindblad_CMP_1976, Petruccione_book, Alicki_2007, Rivas_Huelga_book, Lidar_2020_lecture}. The Markovian GKSL master equation for this case reads
\begin{align}
     &\widetilde{\mathcal{L}}^{r}_j=\underset{\tau\rightarrow 0}{\lim}\frac{\rho_{S_j}^{M^r}-\rho_{S_{j-1}}^{M^r}}{\tau}  =  -\frac{i}{\hbar}[\widetilde{\mathcal{H}}^{\prime}, \rho^{M^r}_{S_{j-1}} ]  + \mathcal{D}^{M^r}_{j},\nonumber\\
     &\text{with}\;\;\widetilde{\mathcal{H}}^{\prime} =  \widetilde{\mathcal{H}}   +\tr_{A_j}\big(V_{SA_j} \rho_{A_j}^0\big)/\sqrt{\tau}\nonumber\\
     &\text{and}\;\; \mathcal{D}_j^{M^r}=-\frac{1}{2\hbar^2}\tr_{A_j}[V_{SA_j},[V_{SA_j}, \rho_{S_{j-1}}^{M^r}  \rho_{A_j}^0]]. \label{eq:after_shift}
\end{align}
Here, $\rho_{S_j}^{M^r}$  and $\rho_{S_{j-1}}^{M^r}$ denote the reduced states of the system after the $j^{\text{th}}$ and $(j-1)^{\text{th}}$ interactions, respectively, in the limit of small $\tau$. 
See App.~\ref{CM} for a detailed discussion on obtaining Eq.~(\ref{eq:after_shift}) from Eq.~(\ref{eq:reduced_state}). For the particular choice of $V_{SA_j}$, given in Eq.~(\ref{eq:int}), the term $\mathcal{D}_j^{M^r}$ comes out to be~\cite{Cusumano2022}
\begin{align}
\label{eq:D_Mar}
   \mathcal{D}_{j}^{M^r}   &=  \frac{g_{1}^{2}}{\hbar^2} \gamma_1 \left [  \sigma^{-}_S \rho_{S_{j-1}}^{M^r} \sigma^{+}_S -\frac{1}{2} \left \{ \sigma^{+}_S\sigma^{-}_S, \rho_{S_{j-1}}^{M^r} \right \} \right ]\nonumber\\
   &\quad \quad + \frac{g_{1}^{2}}{\hbar^2} \gamma_2 \left [  \sigma^{+}_S \rho_{S_{j-1}}^{M^r} \sigma^{-}_S -\frac{1}{2} \left \{ \sigma^{-}_S\sigma^{+}_S, \rho_{S_{j-1}}^{M^r} \right \} \right ] \nonumber\\
   &\quad \quad \quad \quad+ \frac{g_{2}^{2}}{\hbar^2} \left [ \sigma^{z}_S \rho_{S_{j-1}}^{M^r} \sigma^{z}_S - \rho_{S_{j-1}}^{M^r}  \right ],
\end{align}
where the dissipation rate coefficients are  $\gamma_1 = \text{tr}[\sigma^{-}_{A_j}\sigma^{+}_{A_j} \rho_{A_j}^0 ]$ and $\gamma_2 = \text{tr}[\sigma^{+}_{A_j}\sigma^{-}_{A_j} \rho_{A_j}^0 ]$. 


While the continuous-time limit of the collisional model yields the GKSL master equation for Markovian dynamics, real-world systems often deviate from this idealization. To address such complexities, we consider non-Markovian scenarios with memory effects or extended interactions. The following sections outline two approaches to model non-Markovianity for a more accurate description of system-environment interactions.
\subsection{Non-Markovian approach: I}
\label{Sec.2B}

The first type of non-Markovian approach we consider is when the interaction time between the system and the environment is not sufficiently small. In this scenario, the influence of the environment on the system cannot be considered as an instantaneous interaction, as assumed in Markovian models. Instead, the state of the system evolves over a finite duration of time for each interaction, during which the environment has a more prolonged effect on the system. As a result, the reduced dynamics of the system cannot be approximated by the GKSL form. Rather, the evolution of the system is described by Eq.~(\ref{eq:reduced_state}). Therefore, after the $j^{\text{th}}$ interaction, the reduced state of the system becomes $\rho_{S_j}^{NM_1^r} = \rho_{S_j}^{r}$.

\subsection{Non-Markovian approach: II}
\label{Sec.2C}
The second type of non-Markovian approach we consider arises from interactions between the auxiliary systems. In this scenario, after the system $S$ collides with auxiliary $A_1$, $A_1$ interacts with auxiliary $A_2$ before $S$ collides with $A_2$. This series of interactions continues, creating correlations between the auxiliaries that persist over time. Such interactions between auxiliaries introduce memory effects into the dynamics of the system, making the evolution non-Markovian~\cite{Ciccarello2022,Baris_Entropy_2022}. Consequently, the reduced dynamics of the system cannot be described by a simple Markovian master equation. 
 In the rotating frame, the reduced state of the system immediately before its collision with the $(j+1)^{\text{th}}$ auxiliary in the rotating frame is given by $\rho_{S_j}^{NM_2^r}=\tr_{A_jA_{j+1}}\big[\rho^{NM_2^r}_{S_jA_{j}A_{j+1}}\big]$, where
\begin{equation*}
    \label{eq:NM_II}
    \rho^{NM_2^r}_{S_jA_{j}A_{j+1}}=
    W_{{A_j}A_{j+1}}U_{SA_j}^{\prime}\big(\rho^{NM_2^r}_{S_{j-1}A_j}\rho_{A_j}^0\big)U_{SA_j}^{\prime\dagger}W_{{A_j}A_{j+1}}^{\dagger}.
\end{equation*}
Here, the operator $W_{{A_j}A_{j+1}}$ is a unitary operator that describes the interaction between the $j^{\text{th}}$ and $(j+1)^{\text{th}}$ auxiliaries.
In the framework, the unitary operator $W_{A_j A_{j+1}}$ representing the interaction between two consecutive auxiliaries, is modeled as a partial SWAP operator given by~\cite{Baris_Entropy_2022}
\begin{equation}
    W_{A_j A_{j+1}}(\epsilon)= \cos(\epsilon)\mathcal{I}_4+i\sin(\epsilon)\mathcal{S}.
    \end{equation}
Here, $\mathcal{I}_4$ denotes the four-dimensional identity matrix, and $\mathcal{S}$ is the SWAP operator defined in the energy eigenbasis of the two interacting auxiliaries.
The SWAP operator $\mathcal{S}$ is represented as
\begin{equation}
\mathcal{S}=\left(
\begin{array}{cccc}
 1 & 0  & 0  & 0\\
 0  & 0& 1 & 0  \\
 0  & 1 & 0 & 0  \\
 0 & 0  & 0  & 1
\end{array}
\right).
\end{equation} 
The parameter $\epsilon$ quantifies the strength of the interaction between the auxiliaries.

\section{Calculation of thermodynamic uncertainty relation}
\label{Sec.3}
To analyze the thermodynamic properties of a system, it is essential to examine key thermodynamic quantities. Two fundamental quantities relevant to the calculation of the TUR are the entropy flux ($J$) and the entropy production rate ($\sigma$). These quantities, after the $j^{\text{th}}$ collision, are defined as follows~\cite{Spohn_JMP_1978,Lebowitz,Petruccione_book,Groot}
\begin{align}
    &J_j^X=-\frac{1}{T_{A}}\tr[\mathcal{H}_S \widetilde{\mathcal{D}}_{j}^X],\nonumber\\ 
    &\sigma_j^X= \frac{dS_j^X}{dt}+J_j^X,
    \label{eq:J_sigma}
\end{align}
where $X\in \{M^r, NM_1^r, NM_2^r\}$ denotes the evolution process: $M^r$ for the Markovian case, and $NM_1^r$ or $NM_2^r$ for the two non-Markovian cases.
The term $S_j^X$ represents the von Neumann entropy of the system after the $j^{\text{th}}$ collision, defined as $S_j^X=-k_B \tr[\rho_{S_j}^{X^r} \ln(\rho_{S_j}^{X^r})]$. The dissipators, $\widetilde{\mathcal{D}}_j^X$, for all three approaches are explicitly given in Eqs.~(\ref{eq:D_Mar}),~(\ref{eq:D_NM_1}) in App.~\ref{appen:1}, and~(\ref{eq:D_NM_2}) in App.~\ref{appen:2}, respectively. 

The thermodynamic uncertainty relation provides a fundamental bound on current fluctuations around their mean, connecting these fluctuations to the entropy production rate in a nonequilibrium steady state (NESS)~\cite{Horowitz2020}. In classical, time-homogeneous Markovian systems, this TUR bound is expressed as~\cite{ Seifert_PRE_2016,Seifert_PRL_2018}
\begin{eqnarray}
 \mathcal{Q}_{\text{cl}}= \frac{\text{var}(\mathcal{J}_s)}{\langle \mathcal{J}_s \rangle^{2}}  \frac{\sigma}{k_B} \geq 2.
   \label{eq:STUR1}
\end{eqnarray}
where $\langle \mathcal{J}_s \rangle$ is the scaled average current of particles, charge, heat, or other quantities of interest, and
$\text{var}(\mathcal{J}_s)$ is its scaled variance, both obtained by dividing the average current and variance by $t$ in the limit $t \rightarrow \infty$. Additionally, $\sigma$ represents the entropy production rate in the NESS.  
Here $t$ represents the process duration. In the collisional model described above, $t = n \tau$, where $n$ represents the total number of collisions. Originally conceived for NESS, the TUR has also been shown to hold for finite-time processes~\cite{Patrick_PRE_2017, Horowitz_PRE_2017}, highlighting its versatility across diverse thermodynamic settings. In such cases, the finite-time TUR
takes the form
\begin{equation}
   \mathcal{Q}^{\text{FT}}_{\text{cl}}= \frac{\text{var}(\mathcal{J}_t)}{\langle \mathcal{J}_t \rangle^{2}}  \frac{\sigma}{k_B} t \geq 2.
   \label{eq:STUR_FT}
\end{equation}
Here all quantities with the subscript $t$ indicate that they are evaluated at time $t$. 
This form underscores the relevance of TUR beyond steady-state conditions, offering a robust framework for understanding current fluctuations in broader nonequilibrium processes.
Studies indicate that the classical thermodynamic uncertainty relations may break down under certain quantum conditions, sparking interest in finding their quantum counterparts. Recent investigations have focused on how quantum coherence contributes to these violations~\cite{Potts_PRE_2021, Keiji_PRL_2022, Segal_PRE_2021, Krzysztof_PRB_2018, Bijay_PRB_2018, Segal_PRE_2019}, revealing that classical TURs can be breached in specific driven systems, such as time-periodic Markov processes, or within the quantum domain. While the underlying mechanisms remain only partially understood, these findings suggest an intriguing opportunity: quantum effects might be harnessed to mitigate harmful fluctuations in current without sacrificing the performance of thermodynamic engines. Our focus is on studying these TUR violations within quantum collisional model systems under various Markovian and non-Markovian scenarios. These investigations are particularly significant for gaining deeper insights into their implications for quantum device construction.

\begin{figure*}[hbt]
        \includegraphics[width=0.33\textwidth]{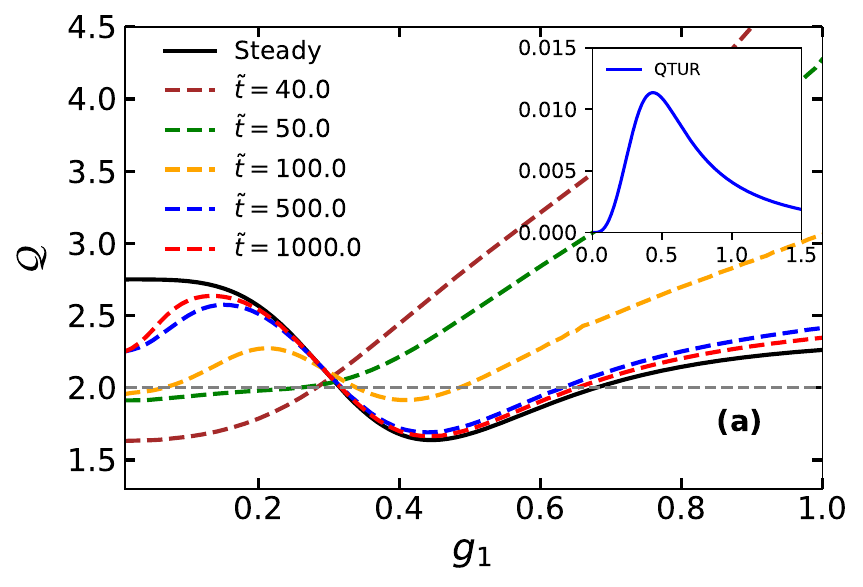}
         \includegraphics[width=0.34\textwidth]{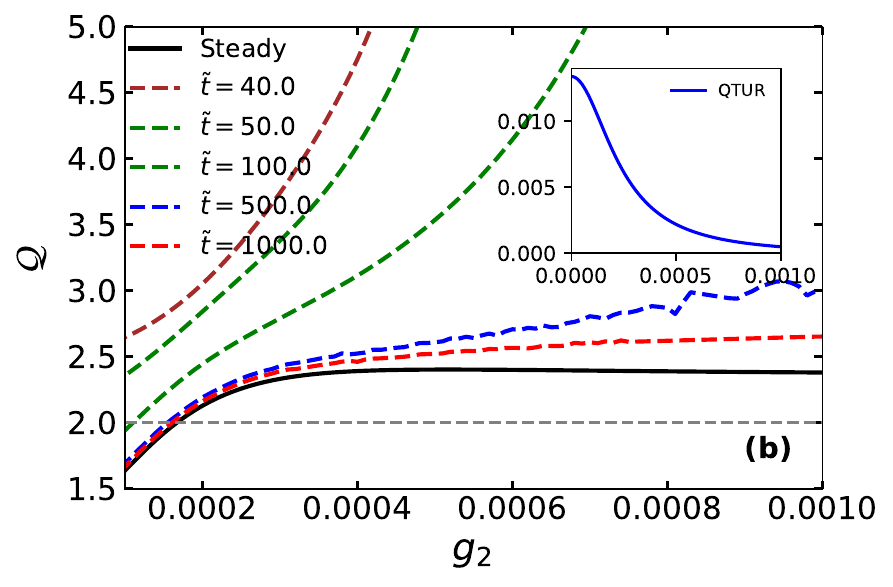}
         \includegraphics[width=0.32\textwidth]{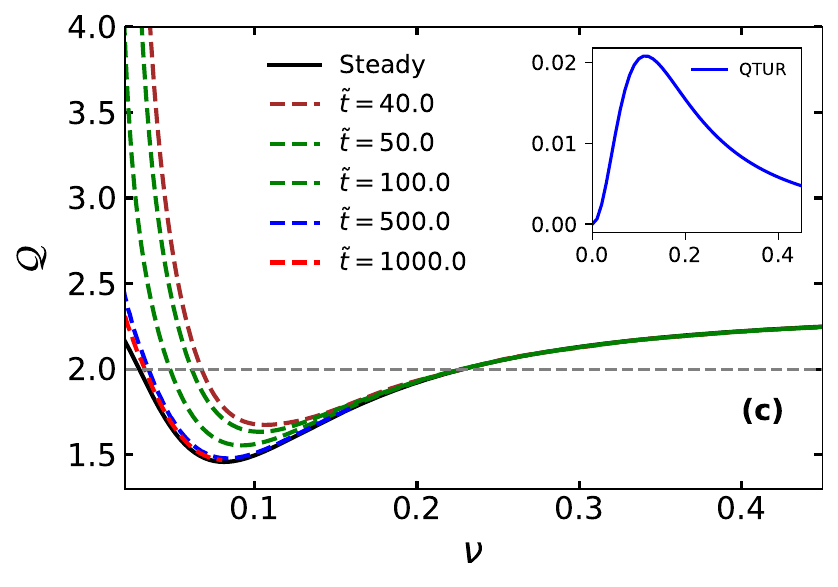}
    \caption{Classical and quantum TUR for collisional model governed by Markovian evolution. Here, we depict the classical TUR parameters for the steady-state regime $(\mathcal{Q}_{\text{cl}})$, the finite-time regime $(\mathcal{Q}_{\text{cl}}^{\text{FT}})$, and the quantum TUR parameter $(\mathcal{Q}_q)$ as functions of (a) $g_1$, with $g_2 = 0.0001\frac{\hbar}{\sqrt{t^{\prime}}}$ and $\nu = \frac{0.05}{t^{\prime}}$, (b) $g_2$, with $g_1 = 0.45\frac{\hbar}{\sqrt{t^{\prime}}}$ and the same $\nu$ as in (a); and (c) $\nu$, with the same $g_1$ as in (b) and the same $g_2$ as in (a). In each panel, the solid black lines represent $\mathcal{Q}_{\text{cl}}$, the dashed lines correspond to the $\mathcal{Q}_{\text{cl}}^{\text{FT}}$ at different finite times $\tilde{t}$, and the solid blue lines in the insets represent $\mathcal{Q}_q$. 
    Here we choose, $\omega=\omega_S= \frac{1.0}{t^{\prime}} $, $\omega_{A}=\frac{1.25}{t^{\prime}}$, $T_S=\frac{0.1\hbar}{k_Bt^{\prime}}$, $T_{A}=\frac{0.5\hbar}{k_Bt^{\prime}}$, and $\tau=1 \cross 10^{-5} t^{\prime}$. The quantities $g_1$ and $g_2$, plotted along the $x$ axes of panel (a) and (b), have units of $\hbar/\sqrt{t^{\prime}}$, and $\nu$ in panel (c) has the unit of $1/t^{\prime}$. The quantities plotted along the $y$ axes are dimensionless.}
    \label{fig:mainfig}
\end{figure*}

To analyze the TUR, we need to determine the various moments of the currents. Since the system exchanges energy with the auxiliaries, the relevant current is the heat current, whose moments can be obtained using Full Counting Statistics (FCS)~\cite{Esposito_RMP_2009,Bruderer_2014_NJP,Kacper_2023_PRR, Landi_PRX_Quantum_2024}. FCS is a theoretical framework used to analyze the statistical distribution of transferred particles, energy, or charge in a quantum or classical system over a given time interval. It provides a detailed description of transport properties beyond average currents by capturing fluctuations and higher-order moments of the transfer process.
The detailed procedure for calculating $\mathcal{Q}_{\text{cl}}$ and $\mathcal{Q}_{\text{cl}}^{\text{FT}}$ across all three Markovian and non-Markovian scenarios, along with a discussion on the invariance of the TUR under the considered rotation, is presented in App.~\ref{appen:4}. Note that, in the steady-state TUR bound (Eq.~(\ref{eq:STUR1})), the time parameter does not appear explicitly because the average current and the variance of the current are scaled by dividing them by $t$ in the limit $t \rightarrow \infty$.
For this scaling by $t$, the time parameter does not appear explicitly in the steady-state TUR bound (see App.~\ref{appen:4} for more details).



In quantum systems, coherence~\cite{Krzysztof_PRB_2018, Bijay_PRB_2018, Cangemi_PRB_2020, Mitchison_PRE_2021} and particle exchange correlations~\cite{Bijay_PRB_2018, Saito_PRL_2018, Segal_PRE_2019} significantly modify classical thermodynamic uncertainty bounds,  altering our understanding of heat flow, energy exchange, and entropy production~\cite{G_PRB_2006, Latune_PRA_2020, Scandi_PRR_2020}. The quantum thermodynamic uncertainty relation (QTUR), derived using the quantum Cramér-Rao bound~\cite{Hasegawa_PRL_2020}, refines the classical TUR by introducing quantum-specific parameters, as
\begin{equation}
\mathcal{Q}_{\text{cl}} \geq \mathcal{Q}_{q} = \frac{1}{\Upsilon + \Psi} \frac{\sigma}{k_B},
\label{EQ:QTUR}
\end{equation}
where $\Upsilon$ quantifies quantum dynamical activity, and $\Psi$ reflects the impact of coherent dynamics. The mathematical expressions of $\Upsilon$ and $\Psi$ are given in App.~\ref{Ap:QTUR}. While a quantum system may violate the classical thermodynamic uncertainty relation bound, it will always adhere to the quantum thermodynamic uncertainty relation.

In the next section, we aim to study the classical and quantum thermodynamic uncertainty relation bounds for the collisional model.
As we discussed before, studying the thermodynamic uncertainty relation within the collisional model framework is a promising endeavor due to its potential to advance both theoretical understanding and practical applications in quantum thermodynamics. 
The collisional model has been successfully implemented in various experimental platforms, including photonic systems, trapped ions, and quantum computers, showcasing its versatility in modeling open quantum systems. 
By leveraging the applications of collisional models, the study of TUR can lead to the development of more efficient quantum devices, such as heat engines, sensors, and error-correction mechanisms. Thus, exploring TUR within the collisional model framework not only deepens our understanding of quantum thermodynamic principles but also paves the way for practical advancements in quantum technologies.

\section{Violation of TUR for the Markovian and non-Markovian scenarios}
\label{Sec.4}

In this section, we investigate the behavior of the TUR parameters $\mathcal{Q}_{\text{cl}}$ and $\mathcal{Q}_q$ across different Markovian and non-Markovian scenarios discussed earlier. We begin by analyzing these parameters in the NESS of the system, where the reduced dynamics are governed by the Markovian GKSL master equation described in Eq.~(\ref{eq:after_shift}). Next, we explore the classical TUR bound in the finite-time regime for both the Markovian case and the two non-Markovian approaches.

\begin{figure*}[hbt]
 \centering
\includegraphics[width=0.34\textwidth]{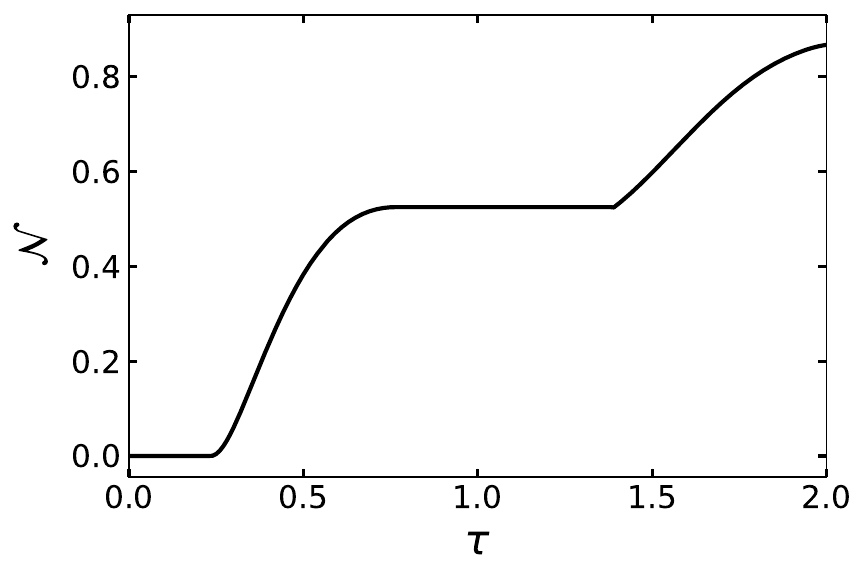}
\includegraphics[width=0.325\textwidth]{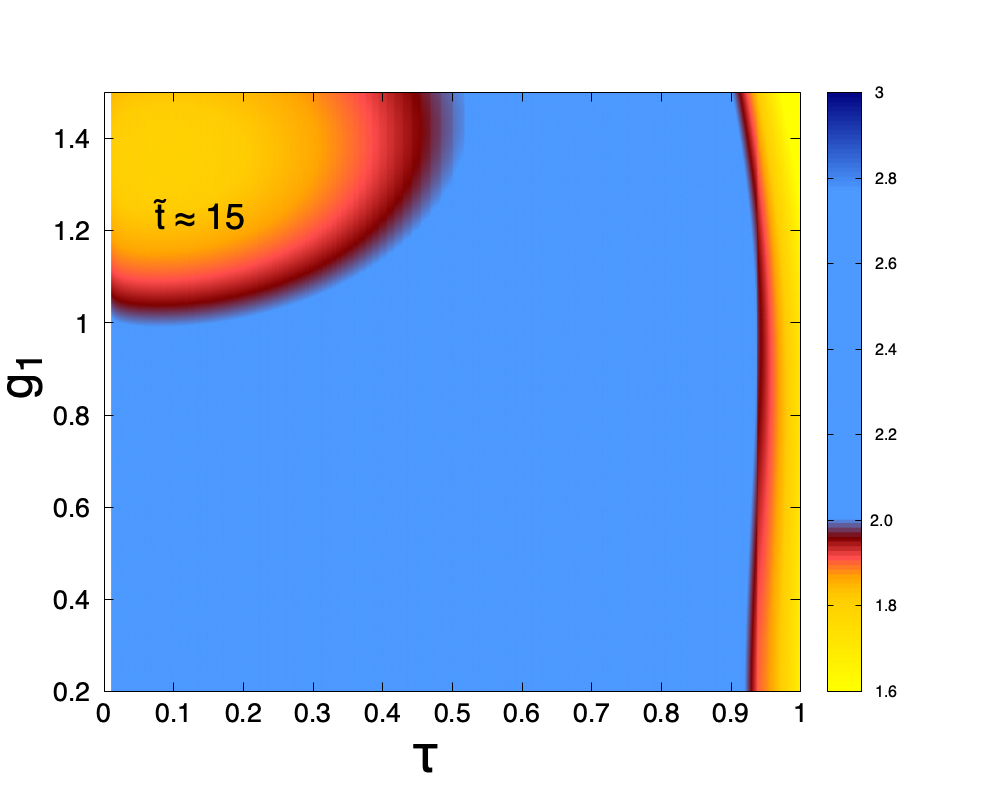}
\includegraphics[width=0.325\textwidth]{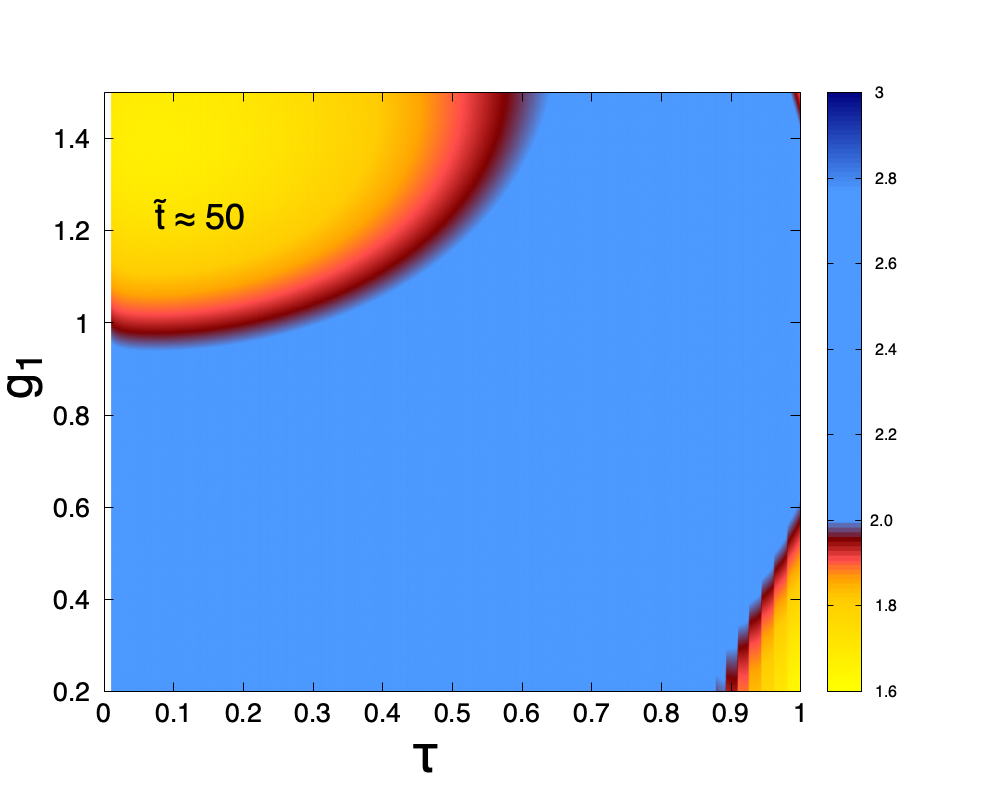}
\caption{
Non-Markovianity and the violation of the classical TUR for the collisional model in non-Markovian approach I. Here Non-Markovianity is introduced by increasing the collision time $\tau$. In the leftmost panel, we plot the non-Markovianity quantifier $\mathcal{N}$ as a function of $\tau$ for $g_1 = 1.5\frac{\hbar}{\sqrt{t^{\prime}}}$ and total evolution time $\tilde{t} \approx 50$. In the middle and rightmost panels, we plot $\mathcal{Q}_{\text{cl}}^{\text{FT}}$ along the color bar with respect to $\tau$ and $g_1$, for total evolution times $\tilde{t} \approx 15$ and $\tilde{t} \approx 50$, respectively. 
Here we take, $\nu=\frac{0.7}{t^{\prime}}$, $g_2=0.01\frac{\hbar}{\sqrt{t^{\prime}}}$, $\omega=\omega_S=\omega_{A}= \frac{4.0}{t^{\prime}} $, $T_S=\frac{1.5\hbar}{k_Bt^{\prime}}$, and $T_{A}=\frac{2.5\hbar}{k_Bt^{\prime}}$. The quantity $\tau$, plotted along the $x$ axis, has units of $t^{\prime}$, while the quantity $g_1$, plotted along the $y$ axis in the middle and the rightmost panels, has units of $\frac{\hbar}{\sqrt{t^{\prime}}}$. The quantities $\mathcal{N}$ and $\mathcal{Q}_{\text{cl}}^{\text{FT}}$ plotted here are dimensionless.}
\label{non-Marko_I}
\end{figure*}

\subsection{Violation of TUR for Markovian approach}
\label{result_Mark}

When the system evolves under the Markovian process described in Eq.~(\ref{eq:after_shift}), it eventually reaches a non-equilibrium steady state after a sufficiently long time. This steady state is obtained by solving the condition $\widetilde{\mathcal{L}}_j^r=0$. We analytically solve for the steady-state density matrix $\rho^{\text{ss}}$, with detailed expressions given in Eqs.~(\ref{steady_state}) and~(\ref{elements_ss}) of App.~\ref{appen:4}.

The behavior of $\mathcal{Q}_{\text{cl}}$ is illustrated in Fig.~\ref{fig:mainfig} by the black solid line, depicting its dependence on the system-auxiliary couplings $g_1$ and $g_2$ and the external field strength $\nu$.
Notably, $\mathcal{Q}_{\text{cl}}$ drops below the classical lower bound of $2.0$, reaching values as low as $1.637$, $1.236$, and $1.457$ by varying $g_1$, $g_2$, and $\nu$, respectively. These results demonstrate significant violations of the classical TUR, indicating nonclassical behavior due to quantum effects. This suggests that quantum systems can achieve a more favorable trade-off between the precision of thermodynamic quantities and the entropy production rate.

We now analyze the behavior of the TUR parameter during the finite-time evolution of the system. The dimensionless time is defined as $\tilde{t}=t/t^{\prime}$, where $t^{\prime}$ is a constant with the unit of time. In Fig.~\ref{fig:mainfig}, $\mathcal{Q}_{\text{cl}}^{\text{FT}}$ is shown for $\tilde{t}= 40.0$, $50.0$, $100.0$, $500.0$ and $1000.0$, corresponding to $40$, $50$, $100$, $500$, and $1000$ units of $10^5$ collisions, respectively.
The behavior of $\mathcal{Q}_{\text{cl}}^{\text{FT}}$ is strongly parameter dependent, with TUR violations manifesting either in the transient or the steady-state regime depending on the chosen parameter values. As shown in Fig.~\ref{fig:mainfig}(a), small $g_1$ values lead to necessarily transient violations prominent at early times (e.g., $\tilde{t} = 40$) but diminishing with further evolution. In contrast, for larger evolution times ($\tilde{t} \geq 100$), the violation emerges at higher values of $g_1$ ($\approx 0.3–0.6$) and grows with $\tilde{t}$ until it saturates at the steady state. 
Similar trends emerge when varying $g_2$ and $\nu$, with $g_1$ determining whether the maximum violation occurs at transient or at steady-state regime. Fig.~\ref{fig:mainfig}(b) and (c) illustrate steady-state-dominant violations, while App.~\ref{Ap:Ness_tran} shows necessarily transient behavior, where violations vanish at long times. 
This necessarily transient feature is important, as many quantum technologies operate over finite timescales where short-time operation can minimize noise and even outperform steady-state behavior~\cite{Brask_PRE_2015,Mitchison_NJP_2015,Sreetama_EPL_2019,Ghosh_PRA_2021,Maity2024}. 

We now examine the QTUR in the steady-state regime. Any quantum system governed by the GKSL master equation is expected to satisfy the QTUR bound, given in Eq.~(\ref{EQ:QTUR}). Our analysis of $\mathcal{Q}_q$ with respect to $g_1$, $g_2$, and $\nu$ (blue solid lines in the inset of Fig.~\ref{fig:mainfig}) shows that even in regions where the classical TUR bound is violated, the system adheres to the QTUR bound, i.e., $\mathcal{Q}_{q} \ll \mathcal{Q}_{\text{cl}}$. This highlights that the QTUR bound remains a fundamental, unviolated constraint for quantum systems.

\subsection{Non-Markovianity and the violation of classical TUR}
\label{Sec.4B}

In this section, we analyze the classical finite-time TUR parameter, $\mathcal{Q}_{\text{cl}}^{\text{FT}}$, for the two non-Markovian approaches. Non-Markovian dynamics, characterized by memory effects, can enhance the precision of observables relative to entropy production, potentially leading to violations of classical TUR bounds~\cite{Razzoli_QST_2024}. To establish a connection between the violation of the classical TUR and the degree of non-Markovianity, we use a well-known quantifier of non-Markovianity introduced by Breuer \textit{et al.}~\cite{BLP_PRL_2009, BLP_PRA_2010}. Considering that the collisional model represents a time-discrete dynamic process, the non-Markovianity quantifier is defined as~\cite{Luoma_2016,Kretschmer_2016}
\begin{equation}
    \mathcal{N} = \max_{\rho^{1}_0, \rho^{2}_0} \sum_{D_{j+1}-D_{j} > 0} D_{j+1}-D_j,
\end{equation}
where $j\in[0,n]$ with $n$ being the total number of collisions.
Here $D=\frac{1}{2}\tr\big(|\rho^1 - \rho^2|\big)$ represents the trace distance between any two states $\rho^1$ and $\rho^2$, and $D_j$ denotes the trace distance after the $j^{\text{th}}$ collision.
 The maximization is performed over all possible pairs of initial states $\rho^{1}_0$ and $\rho^{2}_0$. To optimize this measure, initially $\rho^1_0$ and $\rho^2_0$ are chosen to be orthogonal, as orthogonal states are known to maximize the trace distance~\cite{Piilo_2012}. 
 This measure quantifies the total amount of information that flows from the environment back to the system during the evolution. A positive value of $\mathcal{N}$ $(\mathcal{N} > 0 )$ indicates non-Markovian behavior, characterized by a temporary reversal in the monotonic decay of the trace distance, which is otherwise expected to decrease monotonically in Markovian dynamics.

 \begin{figure*}[hbt]
 \centering
\includegraphics[width=0.34\textwidth]{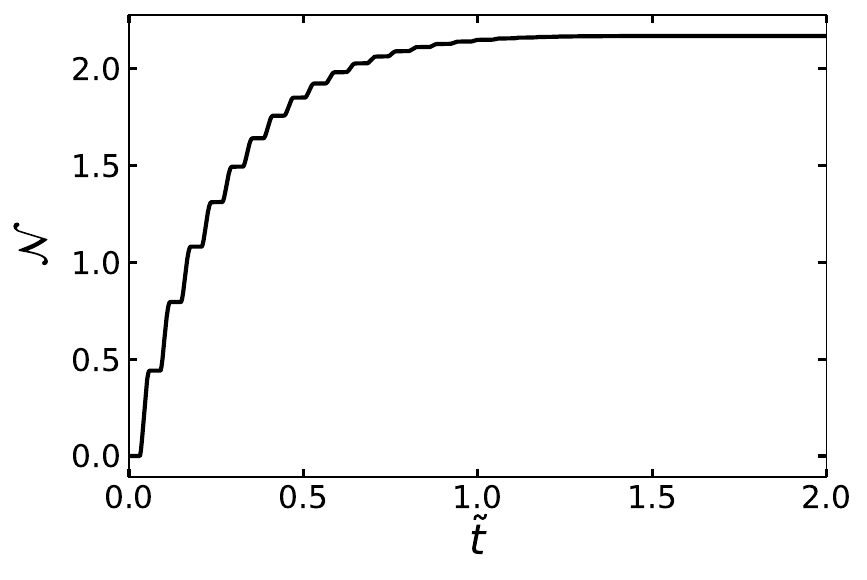}
 \includegraphics[width=0.325\textwidth]{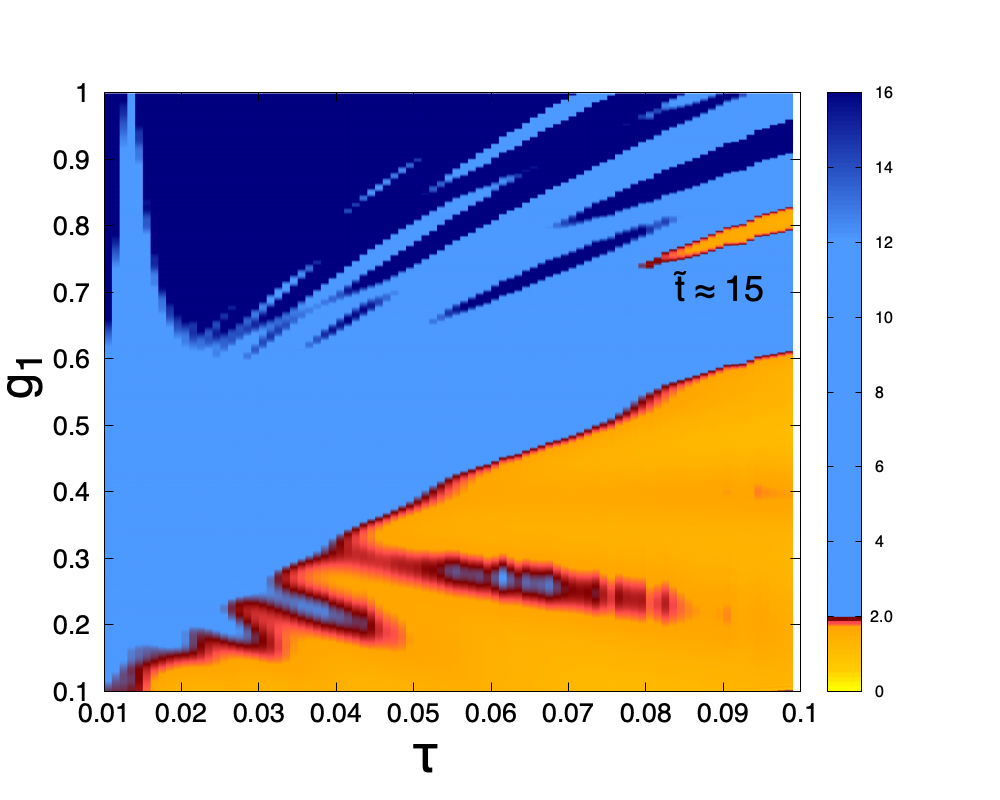}
 \includegraphics[width=0.325\textwidth]{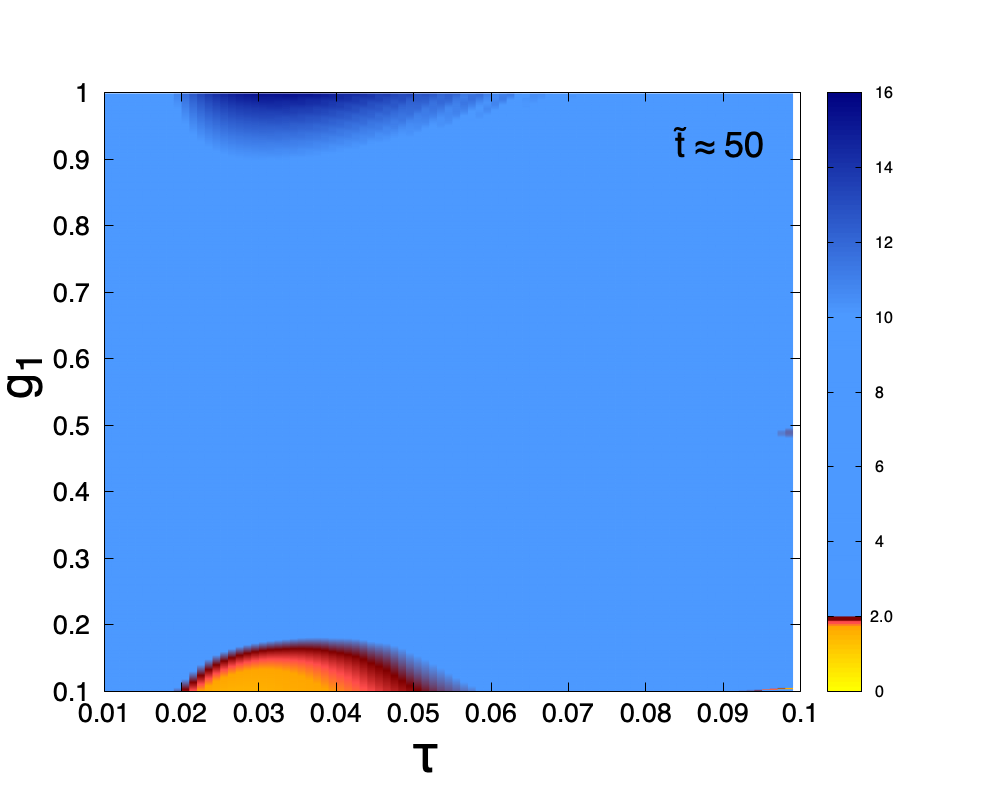}
 \caption{Non-Markovianity and the violation of classical TUR for the collisional model in non-Markovian approach II. In this approach, non-Markovianity arises from interactions between auxiliary systems after each collision with the main system. In the leftmost panel, we plot the non-Markovianity quantifier $\mathcal{N}$ with $t$ for $g_1 =1.5\frac{\hbar}{\sqrt{t^{\prime}}}$ and $\tau=1 \cross 10^{-3}t^{\prime}$. In the middle and the rightmost panel, we plot $\mathcal{Q}_{\text{cl}}^{\text{FT}}$ along the color bar with respect to $\tau$ and $g_1$ for $\tilde{t} \approx 15$ and $\tilde{t} \approx 50$, respectively. In both the panels, we set $\epsilon = 0.95 \frac{\pi}{2}$.  
All other parameters are the same as in Fig.~\ref{non-Marko_I}. On the $x$ axes, the quantities $\tilde{t}$ and $\tau$ are plotted; $\tilde{t}$ is dimensionless, while $\tau$ carries units of $t^{\prime}$, while the quantity $g_1$, plotted along the $y$ axis of the middle and the rightmost panels, has units of $\frac{\hbar}{\sqrt{t^{\prime}}}$. The quantities $\mathcal{N}$ and $\mathcal{Q}_{\text{cl}}^{\text{FT}}$ plotted here are dimensionless.
}
\label{fig:non-mark_II}
\end{figure*}

In the \textit{non-Markovian approach I}, the collision time $\tau$ between the system and auxiliaries is chosen to be significantly longer than in the Markovian case. 
In the leftmost panel of Fig.~\ref{non-Marko_I}, we illustrate the behavior of the non-Markovianity quantifier $\mathcal{N}$ as a function of the increasing collision time $\tau$. 
Our results indicate that by simply increasing the collision time, the system transitions from a Markovian regime to a non-Markovian one. It is important to note that this non-Markovianity is observable only within the time interval $\tau$, i.e., during the collision process itself. However, if the quantifier is evaluated after each complete collision, no non-Markovianity is detectable, as the memory effects in this case are localized within the collision time frame.
 
The middle and rightmost panel of Fig.~\ref{non-Marko_I} show the behavior of $\mathcal{Q}_{\text{cl}}^{\text{FT}}$ as a function of the coupling constant $g_1$ and the collision time $\tau$ for $\tilde{t}\approx 15$ and $\tilde{t} \approx 50$, respectively. In both cases, we observe TUR violations in two distinct regions: when $\tau$ is very small and when $\tau$ is large. For small $\tau$, TUR violations occur only at large values of $g_1$, whereas for large $\tau$, the violation spans the entire range of $g_1$ for small values of $\tilde{t}$. As the number of collisions increases (i.e., with increasing $\tilde{t}$), the TUR-violating region at small $\tau$ becomes more pronounced, while the violation region at large $\tau$ gradually decreases. This reveals a narrow parameter window in $\tau$ and $g_1$—specifically at higher values—where the TUR violation is necessarily transient.

In \textit{non-Markovian approach II}, memory effects emerge from interactions among the auxiliary systems. 
For this scenario, the quantifier of non-Markovianity ($\mathcal{N}$) is depicted as a function of the total evolution time ($\tilde{t}$) in the leftmost panel of Fig.~\ref{fig:non-mark_II}. 
Unlike non-Markovian approach I, where non-Markovianity is detectable only during the collision, in this case, the system demonstrates non-Markovian behavior throughout the entire evolution period. Furthermore, the degree of non-Markovianity is significantly higher in this case.
In the middle and the rightmost panel of Fig.~\ref{fig:non-mark_II}, $\mathcal{Q}_{\text{cl}}^{\text{FT}}$ is plotted with respect to $\tau$ and $g_1$ for this scenario for $\tilde{t}\approx 15$ and $\tilde{t} \approx 50$, respectively. 
In this case, the violation of TUR is more pronounced at shorter evolution times, i.e., for smaller values of $\tilde{t}$, while the region of violation progressively shrinks as the number of collisions increases.

The gradient of yellow in the middle and the rightmost panels of Fig.~\ref{fig:non-mark_II} indicates regions of TUR violation, highlighting enhanced performance and stability of quantum devices due to reduced power fluctuations. In contrast, the blue regions correspond to large values of $\mathcal{Q}_{\text{cl}}^{\text{FT}}$, signaling suboptimal device performance. Previous studies have shown that non-Markovianity can lead to a significant violation of the TUR~\cite{Razzoli_QST_2024} or influence TUR-related parameters, indirectly indicating violation~\cite{Campbell_PRA_2018, Pati_PRA_2017, Marcantoni_SR_2017, Argentieri_EPL_2014, Baris_Entropy_2022, Xu_PRE_2018, Strasberg_PRE_2019}. In our investigation, we find that the two non-Markovian approaches exhibit qualitatively distinct behaviors, emphasizing that not just the degree, but the type of non-Markovianity crucially determines the extent and nature of TUR violation. Non-Markovian approach II, which incorporates stronger memory effects, exhibits TUR violations across the entire range of collision times $\tau$, with particularly pronounced violations at short collision times. This reveals a broad parameter regime of necessarily transient behavior, relevant for collision model--based quantum devices intended to function within finite-time regimes, where steady-state operation may not be feasible or optimal.
In contrast, non-Markovian approach I exhibits TUR violations only within a limited range of $\tau$, with a relatively narrow necessarily transient regime, offering less practical advantage for quantum device optimization in the transient regime. However, in the steady-state regime, non-Markovian approach I yields more pronounced TUR violations, making it more suitable for optimizing steady-state thermal device performance.

An important point to note is that Markovian dynamics are generally associated with a positive entropy production rate, as established by Spohn's inequality~\cite{Spohn_JMP_1978}. However, in the case of non-Markovian dynamics, this positivity in the transient regime is no longer guaranteed, since the assumptions underlying Spohn's inequality may no longer be valid. Several studies have explored this phenomenon and demonstrated that non-Markovian effects can lead to negative entropy production rates in the transient regime~\cite{Argentieri_EPL_2014, Pati_PRA_2017, Marcantoni_SR_2017, Campbell_PRA_2018}. Consequently, a negative transient entropy production rate can serve as a witness of non-Markovianity~\cite{BLP_PRL_2009, Rivas_RPP_2014, Breuer_RMP_2016}.

In contrast, within the nonequilibrium steady-state regime, the entropy production rate remains positive for both Markovian and non-Markovian dynamics. Consequently, $\mathcal{Q}_{\text{cl}}^{\text{FT}}$ also remains strictly positive, as it explicitly depends on the steady-state entropy production rate (see Eq.~(\ref{eq:STUR_FT})). However, evaluating the finite-time thermodynamic uncertainty relation in non-Markovian systems is often nontrivial, particularly because reaching a steady state may be delayed or even practically unattainable within experimentally relevant timescales. In some cases, the system may require an extremely long time to reach a steady state, or such a state may not exist at all.

In our study, we observe that the two non-Markovian cases considered in this work do reach a steady state (see Fig.~\ref{fig:non-mark_II_check} in App.~\ref{Ap:EPR}); however, this is not a general feature of non-Markovian dynamics. In scenarios where the steady-state entropy production rate is inaccessible, it becomes meaningful to consider the time-integrated entropy production over a finite time interval. This quantity remains positive according to the second law of thermodynamics~\cite{Esposito_NJP_2010,Landi_RMP_2021}, and provides an alternative for characterizing $\mathcal{Q}_{\text{cl}}^{\text{FT}}$ in the context of finite-time, where a steady state may be delayed or may not exist. For such cases, the finite-time thermodynamic uncertainty relation is given by~\cite{Horowitz_PRE_2017}
\begin{equation}
   \mathcal{Q}^{\text{FT}}_{\text{cl}}= \frac{\text{var}(\mathcal{J}_t)}{\langle \mathcal{J}_t \rangle^{2}}  \frac{\Sigma_t}{k_B} \geq 2,
\end{equation}
where $\Sigma_t$ is the total entropy production over the time interval $[0,t]$, defined as
\begin{equation}
    \Sigma_t=\int_0^t\sigma_t dt,
\end{equation}
where $\sigma_t$ is the instantaneous entropy production rate at time $t$.

\section{Conclusion}
\label{Sec.5}
We investigated the thermodynamic uncertainty relation within quantum collisional models, recognizing their critical role in experimental platforms such as photonic systems, trapped ions, quantum computers, and in quantum devices such as heat engines and batteries.
TUR violations reveal insights into system performance and stability, offering pathways to enhance efficiency and optimize quantum devices beyond classical limits.
In this work, we analyzed the violation of the TUR across three distinct types of dynamics in collisional models. The first type corresponds to Markovian evolution. 
For this dynamics, we examined three key aspects: the classical and quantum TUR bound in the NESS regime, and the finite-time TUR for each collision.
We find that the classical TUR bound can be violated, with the maximum violation occurring either during the transient or in the steady-state regime, depending on parameters such as magnetic field strength and system-auxiliary interactions. This leads to two distinct behaviors: in certain regimes, the violation increases over time and saturates in the steady state, while in others, it is necessarily transient—appearing only at early times and fading with continued evolution.
Despite classical TUR violations, our analysis confirms that systems governed by the GKSL master equation adhere to the quantum TUR bound, emphasizing its role as a strict constraint for quantum systems. 
We examined two types of non-Markovian dynamics in the collisional model. The first arises from extended collision times, which induce non-Markovianity detectable only during collisions.
The second type of non-Markovian dynamics, driven by interactions between auxiliaries, exhibits memory effects throughout the evolution of the system. 
The two types of non-Markovian dynamics exhibit distinct TUR behavior. For the first non-Markovian approach, two distinct dynamical aspect arise depending on the parameter regime. The violation of TUR may either increase over time and sustain in the steady state or remain strictly transient. However, in the second approach of non-Markovianity, TUR violations are strongest at short times and gradually diminish, giving rise to a more extensive necessarily transient regime. 
Previous studies in this direction have demonstrated that non-Markovianity can cause significant TUR violations. Our results highlight that these violations are influenced not only by the degree but also by the nature of non-Markovianity. Also, the necessarily transient TUR violations are particularly relevant for scenarios where only short-time operation is possible or preferred, allowing TUR violations to be leveraged within finite-time windows rather than relying on steady-state dynamics.
Therefore, studying TUR in various aspects of collisional models is crucial for advancing quantum technologies, offering insights into the unique advantages and efficiencies of quantum systems for designing next-generation quantum devices.

\section{Acknowledgments}
We acknowledge computations performed using Armadillo~\cite{Sanderson,Sanderson1},
NLOPT~\cite{NLOPT}
   and QIClib~\cite{QIClib}
   on the cluster computing facility of the Harish-Chandra Research Institute, India. 
   A. G. acknowledges support from the Alexander von Humboldt Foundation.

\appendix
\begin{widetext}

\section{The Markovian formulation of the collisional model}
\label{CM}
The reduced state of the system in the rotating frame, after the $j^{\text{th}}$ interaction in the collisional model, described in Sec.~\ref{Sec-2} of the  main text, is given by
\begin{equation}
     \rho_{S_j}^r 
    =  \tr_{A_{1,2,..,j}}\Big [ U_{SA_j}^{\prime}...U_{SA_1}^{\prime}  \Big( \rho_{S}^0 \bigotimes_{j=1}^{j} \rho_{A_j}^0 \Big ) U_{SA_1}^{\prime\dagger}...U_{SA_j}^{\prime\dagger}     \Big ],
    \label{eq:reduced_state_app}
\end{equation}
where the unitary $U_{SA_j}^{\prime}=\exp[-\frac{i}{\hbar}\mathcal{H}_{SA_j}^{\prime}\tau]$ with $\mathcal{H}_{SA_j}^{\prime}=\widetilde{\mathcal{H}}+\mathcal{H}_{A_j}+V_{SA_j}/\sqrt{\tau}$. Although this collisional model is inherently a discrete-in-time framework, a continuous-time master equation can be derived in the limit when the duration of interactions $\tau$ becomes sufficiently small, i.e., $\tau\rightarrow 0$. In this limit Eq.~(\ref{eq:reduced_state_app}) reduces to
\begin{align}
    \rho_{S_j}^{M^r} =   \rho_{S_{j-1}}^{M^r} -\frac{i}{\hbar} \tau [ \widetilde{\mathcal{H}}, \rho_{S_{j-1}}^{M^r} ]- \frac{i}{\hbar} \sqrt{\tau} \big[\tr_{A_j}\big(V_{SA_j}\rho_{A_j}^0\big) , \rho_{S_{j-1}}^{M^r}   \big]- \frac{1}{2\hbar^2} \tau \tr_{A_j}[V_{SA_j},[V_{SA_j}, \rho_{S_{j-1}}^{M^r}  \rho_{A_j}^0]],
\end{align}
keeping terms up to linear order in $\tau$. Here, $\rho_{S_j}^{M^r}$  and $\rho_{S_{j-1}}^{M^r}$ denote the reduced states of the system after the $j^{\text{th}}$ and $(j-1)^{\text{th}}$ interactions, respectively, in the limit of small $\tau$. We now define $\widetilde{\mathcal{L}}_j^r=\underset{\tau\rightarrow 0}{\lim}\frac{\rho_{S_j}^{M^r}-\rho_{S_{j-1}}^{M^r}}{\tau}$, which leads to
\begin{align}
    \widetilde{\mathcal{L}}_j^r = -\frac{i}{\hbar} [\widetilde{\mathcal{H}}, \rho_{S_{j-1}}^{M^r}] -\frac{i}{\hbar\sqrt{\tau}} [\tr_{A_j} \big(V_{SA_j} \rho_{A_j}^0\big), \rho_{S_{j-1}}^{M^r}]+ \mathcal{D}_j^{M^r}
     \label{eq:Lindblad}
\end{align}
where $\mathcal{D}_j^{M^r}=-\frac{1}{2\hbar^2}\tr_{A_j}[V_{SA_j},[V_{SA_j}, \rho_{S_{j-1}}^{M^r}  \rho_{A_j}^0]]$.
This continuous formulation of the collisional model aligns with the GKSL master equation~\cite{Sudarshan_JMP_1976, Lindblad_CMP_1976, Petruccione_book, Alicki_2007, Rivas_Huelga_book, Lidar_2020_lecture} for Markovian dynamics. , but includes an additional term. The second term in this equation, which is absent in the standard GKSL master equation, vanishes when all the assumptions for Markovian dynamics are fully satisfied.
Most of the assumptions underlying Markovian dynamics are inherently incorporated in the construction of the collisional model. For instance, the Born approximation is inherently included in the construction of the collisional model, as it assumes that after each collision, the system and the next auxiliary system are initialized in a product state. The Markov assumption, which ensures that the evolution of the state of the system is memoryless, is also incorporated by design, since the auxiliary systems do not interact with one another, preventing any backflow of information. Furthermore, the secular approximation, typically used to ensure complete positivity of the dynamics in traditional open quantum system approaches, is not required here because the collisional model naturally guarantees completely positive dynamics at each collision~\cite{Cusumano2022}. The only remaining assumption is the stability condition.
For the microscopic approach of the derivation, if $\mathcal{E}$ is the environment and $S$ is the system, then the stability condition is expressed as
\begin{equation}
    \tr_{\mathcal{E}}\big[H_{S\mathcal{E}}\rho_{S\mathcal{E}}(0)\big]=0,
    \tag{A4}
\end{equation}
where $H_{S\mathcal{E}}$ is the interaction Hamiltonian between the system and the environment, and $\rho_{S\mathcal{E}}(0)$ is the initial state of the composite system-environment setup. Similarly, in the collisional model approach the stability condition in the rotating frame becomes
\begin{equation}
    \label{eq:stability}\tr_{A_j}\big[V_{SA_j}\big(\rho_{S_{j-1}}^{M^r}\otimes \rho_{A_j}^0\big)\big]=0 \quad \forall j.
    \tag{A5}
\end{equation}
Under this condition, the second term in Eq.~(\ref{eq:Lindblad}) vanishes, simplifying the equation to the general form of a GKSL master equation. In most cases, the condition in Eq.~(\ref{eq:stability}) is naturally satisfied. However, even if it is not fulfilled, it is always possible to redefine the interaction Hamiltonian as
\begin{equation}
V^{\prime}_{SA_j}=V_{SA_j}-\tr_{A_j}\big(V_{SA_j} \rho_{A_j}^0\big)\otimes \mathbf{I}_{A_j},
\tag{A6}
\label{eq:int_H_shift}
\end{equation}
where $\tr_{A_j}\big[V^{\prime}_{SA_j}\big(\rho_{S_{j-1}}^{M^r}\otimes \rho_{A_j}^0\big)\big]=0$ $\forall j$, and simultaneously, the system Hamiltonian can be adjusted to account for a shifted energy origin~\cite{Rivas_Huelga_book}
\begin{equation}
\widetilde{\mathcal{H}}^{\prime}=\widetilde{\mathcal{H}}+\mathcal{H}_{\text{shift}}
,
\tag{A7}
\label{eq:int_sysH_shift}
\end{equation}
with $\mathcal{H}_{\text{shift}}=\frac{1}{\sqrt{\tau}}\tr_{A_j}\big(V_{SA_j} \rho_{A_j}^0\big)$. Here, $\mathbf{I}_{A_j}$ is the identity matrix on the Hilbert space of $A_j$. With these redefinitions, Eq.~(\ref{eq:Lindblad}) simplifies to
\begin{align}
    \widetilde{\mathcal{L}}_j^r = -\frac{i}{\hbar} [
    \widetilde{\mathcal{H}}, \rho_{S_{j-1}}^{M^r}]-\frac{i}{\hbar} [ \mathcal{H}_{\text{shift}}, \rho_{S_{j-1}}^{M^r}]
     + \mathcal{D}_j^{M^r}=[\widetilde{\mathcal{H}}^{\prime}, \rho_{S_{j-1}}^{M^r}]
     + \mathcal{D}_j^{M^r},
     \tag{A8}
     \label{eq:Lindblad1}
\end{align}
This equation now precisely corresponds to the GKSL master equation for Markovian dynamics. The term, $-\frac{i}{\hbar}
    [\widetilde{\mathcal{H}}, \rho_{S_{j-1}}^{M^r}]$, represents the unitary evolution of the system, while the other two terms capture the dissipative effects resulting from interactions with the environment. Hence, the total dissipative effect from the environments is captured by
    \begin{equation}
        \widetilde{\mathcal{D}}_j^{M^r}=-\frac{i}{\hbar} [ \mathcal{H}_{\text{shift}}, \rho_{S_{j-1}}^{M^r}]
     + \mathcal{D}_j^{M^r}.
     \tag{A9}
    \end{equation}

\section{Dissipator for non-Markovian approach: I}
\label{appen:1}

For the non-Markovian evolution of the system, when the collision times are not sufficiently small as in the Markovian case, the system evolves according to the dynamics given in Eq.~(\ref{eq:reduced_state}). Thus, the state of the composite system-auxiliary system after the $j^{\text{th}}$ collision is given by
\begin{equation}
    \rho_{SA_j}^{NM_1^r}=\rho_{SA_j}^r
      = \Big [ U_{SA_j}^{\prime}  \Big( \rho_{S_{j-1}}^{NM_1^r} \otimes \rho_{A_j}^0 \Big ) U_{SA_j}^{\prime\dagger}     \Big ]
\end{equation}
By expanding the unitaries in Taylor's series, we obtain
\begin{align}
  \rho_{SA_j}^{NM_1^r} &=\big[\mathbf{I}_{SA_j}-\frac{i}{\hbar}\tau \mathcal{H}_{SA_j}^{\prime}-\frac{1}{2\hbar^2}\tau^2 \mathcal{H}_{SA_j}^{\prime 2}+\mathcal{O}(\tau^3)\big]
  \rho_{S_{j-1}}^{NM_1^r} \rho_{A_j}^0 \big[\mathbf{I}_{SA_j}+\frac{i}{\hbar}\tau \mathcal{H}_{SA_j}^{\prime}-\frac{1}{2\hbar^2}\tau^2 \mathcal{H}_{SA_j}^{\prime 2}+\mathcal{O}(\tau^3)\big] \nonumber\\
  &=\rho_{S_{j-1}}^{NM_1^r} \rho_{A_j}^0-\frac{i}{\hbar}\tau [\widetilde{\mathcal{H}}+\mathcal{H}_{A_j},\rho_{S_{j-1}}^{NM_1^r} \rho_{A_j}^0]-\frac{i}{\hbar}\sqrt{\tau} [V_{SA_j},\rho_{S_{j-1}}^{NM_1^r} \rho_{A_j}^0]-\frac{\tau}{2\hbar^2}[V_{SA_j},[V_{SA_j},\rho_{S_{j-1}}^{NM_1^r} \rho_{A_j}^0]]+ \mathcal{O}(\tau^2) 
\end{align}
Here, $\mathbf{I}_{SA_j}$ denotes the identity operator acting on the composite Hilbert space of the system and the $j^{\text{th}}$ auxiliary. Thus, the reduced state of the system is given by
\begin{equation}
    \rho_{S_j}^{NM_1^r}=\rho_{S_{j-1}}^{NM_1^r}-\frac{i}{\hbar}\tau[\widetilde{\mathcal{H}}^{\prime},\rho_{S_{j-1}}^{NM_1^r}]+\tau \mathcal{D}_j^{NM_1^r} 
\end{equation}

Here, 
$\mathcal{D}_j^{NM_1^r}$ is given by
\begin{equation}
\label{eq:D_NM_1}
\mathcal{D}_j^{NM_1^r}=-\frac{1}{2\hbar^2}\tr_{A_j}[V_{SA_j},[V_{SA_j},\rho_{S_{j-1}}^{NM_1^r} \rho_{A_j}^0]]+\mathcal{O}(\tau)=\frac{\rho_{S_j}^{NM_1^r}-\rho_{S_{j-1}}^{NM_1^r}}{\tau}+\frac{i}{\hbar}[\widetilde{\mathcal{H}}^{\prime},\rho_{S_{j-1}}^{NM_1^r}],
\end{equation}
and the total dissipator is
\begin{equation}
        \widetilde{\mathcal{D}}_j^{NM
        ^r_1}=-\frac{i}{\hbar} [ \mathcal{H}_{\text{shift}}, \rho_{S_{j-1}}^{NM_1^r}]
     + \mathcal{D}_j^{NM_1^r}=\frac{\rho_{S_j}^{NM_1^r}-\rho_{S_{j-1}}^{NM_1^r}}{\tau}+\frac{i}{\hbar}[\widetilde{\mathcal{H}},\rho_{S_{j-1}}^{NM_1^r}],
    \end{equation}
which represents the dissipation due to the interaction with the $j^{\text{th}}$ environment. Unlike the Markovian case, where the dissipator $D_j^{M^r}$ is independent of $\tau$, $D_j^{NM_1^r}$ incorporates higher-order contributions of $\tau$. This dependence arises because, in the non-Markovian scenario, $\tau$ is not assumed to be negligibly small, and its higher-order terms significantly influence the dynamics of the system.
\section{Dissipator for non-Markovian approach: II}
\label{appen:2}

For the case where the auxiliary systems interact with each other, the state of the composite system-environment setup just before the $(j+1)^{\text{th}}$ collision is given by

\begin{align}
    &\rho^{NM_2^r}_{S_jA_{j}A_{j+1}}=W_{{A_j}A_{j+1}}U_{SA_j}^{\prime}\big(\rho^{NM_2^r}_{S_{j-1}A_j}\rho_{A_{j+1}}^0\big)U_{SA_j}^{\prime\dagger}W_{{A_j}A_{j+1}}^{\dagger}\nonumber\\
    &=W_{{A_j}A_{j+1}}\Big\{\big[\mathbf{I}_{SA_j}-\frac{i}{\hbar}\tau \mathcal{H}_{SA_j}^{\prime}-\frac{1}{2\hbar^2}\tau^2 \mathcal{H}_{SA_j}^{\prime 2}+\mathcal{O}(\tau^3)\big]
  \big(\rho^{NM_2^r}_{S_{j-1}A_j}\rho_{A_{j+1}}^0\big)\big[\mathbf{I}_{SA_j}+\frac{i}{\hbar}\tau \mathcal{H}_{SA_j}^{\prime}-\frac{1}{2\hbar^2}\tau^2 \mathcal{H}_{SA_j}^{\prime 2}+\mathcal{O}(\tau^3)\big]\Big\} W_{{A_j}A_{j+1}}^{\dagger}\nonumber\\
  &=W_{{A_j}A_{j+1}}\Big[\rho^{NM_2^r}_{S_{j-1}A_j} \rho_{A_{j+1}}^0-\frac{i}{\hbar}\tau [\widetilde{\mathcal{H}}+\mathcal{H}_{A_j},\rho_{S_{j-1}A_j}^{NM_2^r} \rho_{A_{j+1}}^0]\nonumber\\
  &~~~~~~~~~~~~ -\frac{i}{\hbar}\sqrt{\tau} [V_{SA_j},\rho^{NM_2^r}_{S_{j-1}A_j} \rho_{A_{j+1}}^0]-\frac{\tau}{2\hbar^2}[V_{SA_j},[V_{SA_j},\rho^{NM_2^r}_{S_{j-1}A_j} \rho_{A_{j+1}}^0]]+\mathcal{O}(\tau^2)\Big]W_{{A_j}A_{j+1}}^{\dagger} 
  \end{align}
The reduced state of the system after tracing out the $j^{\text{th}}$ and $(j+1)^{\text{th}}$ auxiliaries is given by
\begin{equation}
    \rho^{NM_2^r}_{S_j}=\tr_{A_jA_{j+1}}\Big[W_{{A_j}A_{j+1}}\big(\rho^{NM_2^r}_{S_{j-1}A_j} \rho_{A_{j+1}}^0\big)W_{{A_j}A_{j+1}}^{\dagger}-\frac{i}{\hbar}\tau W_{{A_j}A_{j+1}} [\widetilde{\mathcal{H}}^{\prime}+\mathcal{H}_{A_j},\rho^{NM_2^r}_{S_{j-1}A_j} \rho_{A_{j+1}}^0]W_{{A_j}A_{j+1}}^{\dagger}\Big]+\tau\mathcal{D}_j^{NM_2^r}.
\end{equation}
Here, 
$\mathcal{D}_j^{NM_2^r}$ is defined as 
\begin{align}
  \mathcal{D}_j^{NM_2^r}&= \tr_{A_jA_{j+1}}\Big[W_{{A_j}A_{j+1}} \Big\{-\frac{1}{2\hbar^2}[V_{SA_j},[V_{SA_j},\rho^{NM_2^r}_{S_{j-1}A_j} \rho_{A_{j+1}}^0]]+\mathcal{O}(\tau)\Big\} W_{{A_j}A_{j+1}}^{\dagger}\Big],
  \label{eq:D_NM_2}
\end{align}
and the total dissipator
\begin{equation}
\widetilde{\mathcal{D}}_j^{NM^r_2}=\frac{\rho^{NM_2^r}_{S_j}-\rho^{NM_2^r}_{S_{j-1}}}{\tau}+\frac{i}{\hbar}\tr_{A_jA_{j+1}}\Big[W_{{A_j}A_{j+1}}[\widetilde{\mathcal{H}}+\mathcal{H}_{A_j},\rho^{NM_2^r}_{S_{j-1}A_{j}}\rho_{A_{j+1}}^0]W_{{A_j}A_{j+1}}^{\dagger}\Big].
\end{equation}

\section{Calculation of thermodynamic uncertainty relation}
\label{appen:4}
The thermodynamic uncertainty relation provides a fundamental bound on current fluctuations around their mean, connecting these fluctuations to the entropy production rate in a NESS~\cite{Horowitz2020}. In classical, time-homogeneous Markovian systems, this TUR bound is expressed as~\cite{ Seifert_PRE_2016, Seifert_PRL_2018}
\begin{equation}
  \mathcal{Q}_{\text{cl}}= \frac{\text{var}(\mathcal{J}_s)}{\langle \mathcal{J}_s \rangle^{2}}  \frac{\sigma}{k_B} \geq 2.
   \label{eq:STUR}
\end{equation}
where $\langle \mathcal{J}_s \rangle$ is the scaled average current of particles, charge, heat, or other quantities of interest, and
$\text{var}(\mathcal{J}_s)$ is its scaled variance, both obtained by dividing the average current and variance by $t$ in the limit $t \rightarrow \infty$. Additionally, $\sigma$ represents the entropy production rate in the NESS.
Here $t$ represents the process duration. In the collisional model described in the main text, $t = n \tau$, where $n$ represents the total number of collisions. Originally conceived for NESS, the TUR has also been shown to hold for finite-time processes~\cite{Patrick_PRE_2017, Horowitz_PRE_2017}, highlighting its versatility across diverse thermodynamic settings. In such cases, the finite-time TUR for the collisional model takes the form
\begin{equation}
   \mathcal{Q}^{\text{FT}}_{\text{cl}}= \frac{\text{var}(\mathcal{J}_t)}{\langle \mathcal{J}_t \rangle^{2}}  \frac{\sigma}{k_B} t \geq 2.
\end{equation}
Here all quantities with the subscript $t$ indicate that they are evaluated at time $t$. 
We first focus on the classical TUR parameter for the nonequilibrium steady-state case of the collisional model, using the Markovian approach. 
To analyze the TUR, we need to determine the various moments of the currents. Since the system exchanges energy with the auxiliaries, the relevant current is the heat current, whose moments can be obtained using Full Counting Statistics~\cite{Esposito_RMP_2009,Bruderer_2014_NJP,Kacper_2023_PRR, Landi_PRX_Quantum_2024}. FCS is a theoretical framework used to analyze the statistical distribution of transferred particles, energy, or charge in a quantum or classical system over a given time interval. It provides a detailed description of transport properties beyond average currents by capturing fluctuations and higher-order moments of the transfer process. The incorporation of counting fields into the master equation enables FCS to systematically track and quantify the transfer of particles, energy, or charge within a system. 
Solving this modified master equation then gives the generating function for particle transfer statistics. Let $\chi$ denote the counting field associated with the corresponding auxiliary. For simplicity, we assume that the same counting field $\chi$ applies to all auxiliaries, reflecting a uniform approach to tracking the relevant interactions. 
The modified Liouvillian becomes,
\begin{align}
     & \widetilde{\mathcal{L}}^{r}_j (\chi)   =   -\frac{i}{\hbar}[\widetilde{\mathcal{H}}^{\prime}, \rho^{M^r}_{S_{j-1}} ] 
      + \frac{g_{1}^{2}}{\hbar^2} \gamma_1 \left [ \exp \{-i\omega_{A} \chi \} \sigma^{-}_S \rho^{M^r}_{S_{j-1}} \sigma^{+}_S -\frac{1}{2} \left \{ \sigma^{+}_S\sigma^{-}_S, \rho^{M^r}_{S_{j-1}} \right \} \right ] \nonumber \\
    & \phantom{ami poth}+ \frac{g_{1}^{2}}{\hbar^2} \gamma_2 \left [ \exp \{i\omega_{A} \chi \} \sigma^{+}_S \rho^{M^r}_{S_{j-1}} \sigma^{-}_S -\frac{1}{2} \left \{ \sigma^{-}_S\sigma^{+}_S, \rho^{M^r}_{S_{j-1}} \right \} \right ] + \frac{g_{2}^{2}}{\hbar^2} \left [ \sigma^{z}_S \rho^{r}_{S_{j-1}} \sigma^{z}_S - \rho^{M^r}_{S_{j-1}}  \right ].
\label{eq:add_cout_field}
\end{align}
Note that the counting field appears exclusively in the jump terms, precisely identifying the contributions from each jump channel and weighting them by $\omega_{A} $. This structure allows us to compute the cumulant generating function ($\mathcal{C}(\chi,t)$), which encapsulates the statistical properties of the current. The cumulant generating function is formally defined as
\begin{equation}
   \mathcal{C}(\chi,t) = \ln  \mathcal{M}(\chi,t),
\end{equation}
where
\begin{align}
    \mathcal{M}(\chi,t) = \tr[\rho^{M^r}_{S_{j}}(\chi)] & = \tr[e^{ \widetilde{\mathcal{L}}^{r}_j(\chi) t}\rho_{S}^0].
\end{align}
From $\mathcal{C}(\chi,t)$, the cumulants of the current can be defined as~\cite{Landi_PRX_Quantum_2024}
\begin{equation}
\label{eq:cumulant}
    \langle\langle N(t)^l\rangle \rangle =(-i \partial_\chi)^l \mathcal{C}(\chi,t)\Big|_{\chi=0}.
\end{equation}
The first two cumulants correspond to the mean and variance of the heat current, respectively: $\langle \mathcal{J}_t\rangle=\langle\langle N(t)^1\rangle \rangle$ and $\text{var}(\mathcal{J}_t)=\langle\langle N(t)^2\rangle \rangle$. In the long-time limit, which corresponds to the steady-state analysis, these cumulants can be obtained from the scaled cumulant generating function~\cite{Kacper_2023_PRR} as  
\begin{align}
    \mathcal{C}(\chi) = \frac{\partial^l}{\partial(i\chi)^l }\left (\lim_{t \to \infty} \frac{\mathcal{C}(\chi,t)} {t} \right ) \Big |_{\chi=0} = \frac{\partial^l}{\partial(i\chi)^l} \lambda (\chi ) \big |_{\chi=0},
\end{align}
where the $\lambda(\chi)$ is the eigenvalue of $ \widetilde{\mathcal{L}}^{r}_j (\chi)$ with the largest real part. In certain cases, directly computing $\lambda(\chi)$ can be challenging. To address this,  we use the method described in Ref.~\cite{Bruderer_2014_NJP} to calculate the scaled average heat current, $\langle \mathcal{J}_s\rangle$,
and the scaled variance of the heat current, $\text{var}(\mathcal{J}_s)$.
This approach leverages the characteristic polynomial of the Liouvillian operator for computation. The detailed procedure is as follows.

The Liouvillian super-operator corresponding to Eq.~(\ref{eq:add_cout_field}) is expressed as
\begin{align}
\widetilde{\mathcal{L}}^{r}_j (\chi)  = & -\frac{i}{\hbar} \big ( \mathcal{I}_2 \otimes \widetilde{\mathcal{H}}^{\prime} - (\widetilde{\mathcal{H}}^{\prime})^{T} \otimes \mathcal{I}_2 \big )
 + \frac{g_{1}^{2}}{\hbar^2} \gamma_1 \left [ \exp (-i\omega_{A} \chi ) (\sigma^{-}_S)^{*} \otimes \sigma^{-}_S -\frac{1}{2} \mathcal{I}_2\otimes \sigma^{+}_S\sigma^{-}_S  -\frac{1}{2} \left( \sigma^{+}_S\sigma^{-}_S\right )^{T} \otimes \mathcal{I}_2 \right ]
\nonumber \\
& + \frac{g_{1}^{2}}{\hbar^2} \gamma_2 \left [ \exp (-i\omega_{A} \chi ) (\sigma^{+}_S)^{*} \otimes \sigma^{+}_S -\frac{1}{2} \mathcal{I}_2\otimes \sigma^{-}_S\sigma^{+}_S  -\frac{1}{2} \left( \sigma^{-}_S\sigma^{+}_S\right )^{T} \otimes \mathcal{I}_2 \right ]
+ \frac{g_{2}^{2}}{\hbar^2} \left [ \sigma^{z}_S \otimes \sigma^{z}_S - \mathcal{I}_2 \otimes \mathcal{I}_2   \right ],
\end{align}
where $\mathcal{I}_2$ is the identity matrix of dimension $2$. Thus, the matrix form of the  Liouvillian super-operator is given by
\begin{equation}
   \widetilde{\mathcal{L}}^{r}_j (\chi)=\left(
\begin{array}{cccc}
 \frac{g_{1}^2}{\hbar^2} \left(\frac{1}{1+e^{\frac{\hbar\omega_{A}}{k_B T_{A}}}}-1\right) & -i \nu  & i \nu  & \frac{g_{1}^2}{\hbar^2}\frac{e^{i \chi \omega_{A}} }{1+e^{\frac{\hbar \omega_{A}}{k_B T_{A}}}} \\
 -i \nu  & \frac{-\Gamma +i \Delta }{\zeta } & 0 & i \nu  \\
 i \nu  & 0 & \frac{-\Gamma -i \Delta }{\zeta } & -i \nu  \\
 \frac{g_{1}^2}{\hbar^2}\frac{e^{-i \chi \omega_{A}} }{1+e^{-\frac{\hbar \omega_{A} }{k_B T_{A}}}} & i \nu  & -i \nu  & -\frac{g_{1}^2}{\hbar^2}\frac{1}{1 +e^{\frac{\hbar \omega_{A}}{k_B T_{A}}}} 
\end{array} 
\right),
\end{equation}
where
\begin{align}
 &\Gamma =  \sqrt{\tau} \left(0.5 g_1^2+2 g_2^2\right) \left[1+\exp \left(\frac{\hbar \omega_{A}}{k_B T_A}\right)\right],  \quad
    \Delta = \hbar^2  \sqrt{\tau} \left[1+\exp \left(\frac{\hbar \omega_{A}}{k_B T_A}\right)\right]  (\omega_S-\omega )+ 2 \hbar g_2 \left[1-\exp \left(\frac{\hbar \omega_{A}}{k_B T_A}\right)\right], \nonumber \\
    &\text{and} \quad \zeta = \hbar^2 \sqrt{\tau} \left[1+\exp \left(\frac{\hbar \omega_{A}}{k_B T_A}\right)\right]
    \nonumber.
\end{align}
Let us consider the characteristic polynomial of $\widetilde{\mathcal{L}}^{r}_j (\chi)$
\begin{equation}
    \sum_{k} a_{k} \lambda^{k}(\chi) = 0,
\end{equation}
where each coefficient $a_{k}$ depends on the variable $\chi$. The derivatives of $a_{k}$ with respect to $\chi$ are defined as
\begin{equation}
    a_{k}^{\prime} = -i \frac{\partial}{\partial \chi} a_{k} \Big |_{\chi=0} \quad \text{and} \quad a_{k}^{\prime \prime} = \left (- i \frac{\partial}{\partial \chi} \right )^{2} a_{k} \Big |_{\chi=0}.
\end{equation}
The mean and variance of the heat current can be expressed in terms of the coefficients $a_k$ as
\begin{equation}
     \langle \mathcal{J}_s \rangle =
     - \frac{a_{0}^{\prime}}{a_{1}} \quad \text{and} \quad \text{var}(\mathcal{J}_s)=
     -\frac{1}{a_1}\left [ a_{0}^{\prime \prime} +2  \langle \mathcal{J} \rangle \left ( a_{1}^{\prime} + a_{2} \langle \mathcal{J} \rangle   \right )    \right].
\end{equation}
It is important to note that the expressions for the mean and variance given above are valid for any system governed by GKSL master equation that has a unique steady state.
For the collisional model discussed in the main text, we have
\begin{equation*}
  a_{0}^{\prime}=  \frac{2\Gamma  g_{1}^{2} \nu^{2} \omega_A  \left( e^{\frac{ \hbar \omega_{A}  }{k_{B} T_{A}}}-1\right)}{\hbar^{2}\zeta  \left[1 +e^{\frac{\hbar\omega_{A} }{k_{B} T_{A}}}\right]} , \quad 
    a_{0}^{\prime \prime} =-   \frac{2\Gamma  g_{1}^{2} \nu^{2} \omega_A^2}{\hbar^{2}\zeta }
\end{equation*}
\begin{align*}
  a_{1}\big |_{\chi=0} =  \frac{4 \Gamma  \zeta  \nu ^2 \hbar ^2+g_{1}^2 \left(\Gamma ^2+\Delta ^2\right)}{\zeta ^2\hbar^2}, \quad   a_{1}^{\prime} =  \frac{2g_{1}^2 \nu ^2 \omega_{A} \left( e^{\frac{\hbar \omega_{A} }{k_{B} T_{A}}}-1\right)}{\hbar ^2 \left[1 + e^{\frac{\hbar\omega_{A}  }{k_{B} T_{A}}}\right]}, \;\; \text{and} \;\;  a_{2}\big |_{\chi=0} = \frac{\hbar^2(\Gamma ^2+\Delta ^2+4 \zeta ^2 \nu ^2)+2 \Gamma  \zeta  g_{1}^2}{\zeta ^2 \hbar^2}.
\end{align*}

For further details on these calculations, see Refs.~\cite{Potts_PRE_2021, Kacper_2023_PRR, Landi_PRX_Quantum_2024}. Hence, the quantity
\begin{equation}
   \frac{\text{var}(\mathcal{J}_s)}{\langle \mathcal{J}_s \rangle^{2}}
    = \frac{2 \zeta }{\Gamma} +\frac{\left(1 +e^{\frac{\hbar\omega_{A}}{k_{B} T_{A}}}\right)^2 \Big[2 \Gamma  \zeta  \nu ^2 \hbar ^2+g_{1}^2 \left(0.5\Gamma ^2+0.5\Delta ^2\right)\Big]}{\Gamma  \zeta  g_{1}^2 \nu ^2 \left(1 - e^{\frac{\hbar\omega_{A}}{k_{B} T_{A}}}\right)^2}
    - \frac{2(\Gamma ^2+\Delta ^2+4 \zeta ^2 \nu ^2) \hbar ^6+4 \Gamma  \zeta  g_{1}^2\hbar^4}{4 \Gamma  \zeta  \nu ^2 \hbar ^2+g_{1}^2 \left(\Gamma ^2 + \Delta ^2\right)}.
\end{equation}

The non-equilibrium steady state for this model can be determined by solving $\widetilde{\mathcal{L}}_j^r=0$. Let the steady-state density matrix be represented as
\begin{equation}
\label{steady_state}
   \rho^{\text{ss}}=\left(
\begin{array}{cc}
 \rho_{gg}^{\text{ss}} & \rho_{ge}^{\text{ss}}  \\
 \rho_{eg}^{\text{ss}} & \rho_{ee}^{\text{ss}}
\end{array} 
\right).
\end{equation}
Solving  $\widetilde{\mathcal{L}}_j^r=0$ yields the following steady-state elements
\begin{align}
\label{elements_ss}
    &\rho_{gg}^{\text{ss}} = \frac{g_1^2(\Delta^2 + \Gamma^2) + \Big(1 + 
       e^{\frac{\hbar\omega_A}{k_BT_A}}\Big)2\Gamma\zeta\hbar^2\nu^2}{\Big[
g_1^2(\Delta^2 + \Gamma^2) + 
      4\Gamma\zeta\hbar^2\nu^2\Big]\Big(1 + 
       e^{\frac{\hbar\omega_A}{k_BT_A}}\Big)}, \quad \quad \rho_{ee}^{\text{ss}}=1-\rho_{gg}^{\text{ss}},\nonumber\\
       &\rho_{eg}^{\text{ss}}=\frac{\zeta\nu(\rho_{gg}^{\text{ss}}-\rho_{ee}^{\text{ss}})}{\Delta+i\Gamma}, \phantom{valo lagchhe na dhur baba, ki je}\;\rho_{ge}^{\text{ss}}=\frac{\zeta\nu(\rho_{gg}^{\text{ss}}-\rho_{ee}^{\text{ss}})}{\Delta-i\Gamma}.
\end{align}
Using this steady-state solution, the entropy flux and entropy production rate, as described in Eq.~(\ref{eq:J_sigma}) of the main text, can be calculated by substituting the $j^{\text{th}}$ and $(j-1)^{\text{th}}$ states with the steady-state expressions. Then, the quantity $\mathcal{Q}_{\text{cl}}$ for this scenario can be determined using Eq.~\eqref{eq:STUR}.
To evaluate the TUR in the finite-time regime, where the system has not yet reached its steady state, the mean and variance of the current must be computed using Eq.~(\ref{eq:cumulant}). The finite-time TUR parameter, $\mathcal{Q}_{\text{cl}}^{\text{FT}}$, is then calculated using Eq.~(\ref{eq:STUR_FT}).

For the non-Markovian dynamics considered in this work, the state of the system does not evolve according to a GKSL master equation. Instead, it is governed by unitary dynamics. In these cases, the counting fields are incorporated into the system-environment composite unitary operator. Hence, for the non-Markovian approach I, the time evolution of the state of the system after the $j^{\text{th}}$ collision is given by~\cite{mohan2024}
\begin{align}
     \rho_{S_j}^{NM_1^r}(\chi) = \tr_{A_{1,2,..,j}}\Big [ U_{SA_j}^{\prime}(\chi)...U_{SA_1}^{\prime}(\chi)  \Big( \rho_{S}^0 \bigotimes_{j=1}^{j} \rho_{A_j}^0 \Big )  
\overline{U}_{SA_1}^{\prime}(-\chi)...\overline{U}_{SA_j}^{\prime}(-\chi)     \Big ],
\end{align}
where
\begin{align}
   & U_{SA_j}^{\prime}(\chi)=e^{-\frac{i}{2\hbar}\chi H_{A_j}} U_{SA_j}^{\prime} e^{\frac{i}{2\hbar}\chi H_{A_j}},\nonumber\\
&\overline{U}_{SA_j}^{\prime}(-\chi)=e^{\frac{i}{2\hbar}\chi H_{A_j}} U_{SA_j}^{\prime \dagger} e^{-\frac{i}{2\hbar}\chi H_{A_j}}.
\end{align}
Similarly, for the non-Markovian approach II, the evolution of the system state in the presence of counting fields is described by
\begin{equation}
\rho_{S_j}^{NM_2^r}(\chi)
=\tr_{A_jA_{j+1}}\Big[W_{{A_j}A_{j+1}}U_{SA_j}^{\prime}(\chi)\big(\rho^{NM_2^r}_{S_{j-1}A_j}(\chi)\rho_{A_{j+1}}^0\big)\overline{U}_{SA_j}^{\prime}(-\chi)W_{{A_j}A_{j+1}}^{\dagger}\Big].
\end{equation}
For both non-Markovian approaches, the mean and variance of the heat current are computed using Eq.~(\ref{eq:cumulant}). Subsequently, the finite-time TUR parameter, 
$\mathcal{Q}_{\text{cl}}^{\text{FT}}$, is evaluated using Eq.~(\ref{eq:STUR_FT}). 

It is important to note that the cumulant generating function depends on the trace of the density matrix of the system, which remains invariant under any rotation. On the other hand, since $[\mathcal{H}_S,\Theta]=0$, $J_j^X$ (defined in Eq.~(\ref{eq:J_sigma}) of the main text) remains invariant under the considered rotation. The von Neumann entropy of the system is also invariant under this rotation, as it depends only on the eigenvalues of the system. Consequently, the entropy production rate, $\sigma_j^X$, (see Eq.~(\ref{eq:J_sigma}) of the main text) is likewise invariant under the rotation. Hence, the TUR parameters for both the steady state and the finite-time regime,  $\mathcal{Q}_{\text{cl}}$ and $\mathcal{Q}_{\text{cl}}^{\text{FT}}$, respectively, are unaffected by the rotation. Thus, performing these calculations in the rotating frame does not introduce any discrepancies. For clarity, in the subsequent discussion on entropy flux and entropy production rate, we omit the superscript $r$ used for the rotating frame, as these quantities are invariant under the chosen rotation.

\section{Quantum Thermodynamic Uncertainty Relation}
\label{Ap:QTUR}
The quantum thermodynamic uncertainty relation for the steady-state dynamics of a system governed by the GKSL master equation is given by
\begin{equation}
   \mathcal{Q}_{\text{cl}} = \frac{\text{var}(\mathcal{J}_s)}{\langle \mathcal{J}_s \rangle ^{2}} \frac{\sigma}{k_B}\geq \frac{1}{ \Upsilon + \Psi }\frac{\sigma}{k_B} = \mathcal{Q}_{q}.
    \label{eq:QTUR_bound}
\end{equation}
where $\mathcal{Q}_{q}$ represents the quantum correction to the classical thermodynamic uncertainty relation bound. If the dynamical equation of the reduced system is expressed as
\begin{align}
    \dot{\rho} =\mathcal{L}= -\frac{i}{\hbar} [H, \rho] +\sum_{\delta}\Big[\mathcal{A}_{\delta}\rho \mathcal{A}_{\delta}^{\dagger}-\frac{1}{2}\{\mathcal{A}_{\delta}^{\dagger}\mathcal{A}_{\delta},\rho\}\Big],
\end{align}
where $H$ is the Hamiltonian of the system, $\mathcal{A}_{\delta}$ and $\mathcal{A}_{\delta}^{\dagger}$ are the jump operators and their adjoint operators, respectively, the parameter $\Upsilon$ 
is defined as
\begin{equation}
    \Upsilon = \sum_{\delta} \tr \Big ( \mathcal{A}_{\delta}^{\dagger} \mathcal{A}_{\delta} \rho_{st} \Big ),
\end{equation}
where $\rho_{st}$ denotes the steady-state density matrix of the system. This parameter quantifies the quantum dynamical activity, which can be interpreted as the average rate of transitions occurring in the steady state. The parameter $\Psi$, representing the contribution from coherent dynamics, is given by  
\begin{equation}
    \Psi = -4 \Big ( \langle \langle \mathcal{I} | \mathcal{L}_{L} \mathcal{L}^{\ddagger} \mathcal{L}_{R} | \rho_{st} \rangle \rangle +  \langle \langle \mathcal{I} | \mathcal{L}_{R} \mathcal{L}^{\ddagger} \mathcal{L}_{L} | \rho_{st} \rangle \rangle \Big ),
\end{equation}
where $|\rho_{st}\rangle \rangle$ represents the vectorized form of the steady-state density matrix $\rho_{st}$, and $|\mathcal{I}\rangle \rangle$ denotes the vectorized identity operator. In this expression, $\mathcal{L}_L$ and $\mathcal{L}_R$ represent the left and right contributions to the Liouvillian superoperator, respectively, and $\mathcal{L}^{\ddagger}$ is the Drazin inverse of the vectorized Liouvillian superoperator.
The right and left contributions to the Liouvillian are defined as 
\begin{align}
    &\mathcal{L}_{R} = -\frac{i}{\hbar} \mathcal{I} \otimes H + \frac{1}{2} \sum_{\delta} \Big ( \mathcal{A}_{\delta} ^{*}  \otimes \mathcal{A}_{\delta} - \mathcal{I} \otimes \mathcal{A}_{\delta}^{\dagger} \mathcal{A}_{\delta} \Big ),\nonumber\\
     &\mathcal{L}_{L} = \frac{i}{\hbar} H^{T} \otimes \mathcal{I} + \frac{1}{2} \sum_{\delta} \Big ( \mathcal{A}_{\delta} ^{*}  \otimes \mathcal{A}_{\delta} - ( \mathcal{A}_{\delta}^{\dagger} \mathcal{A}_{\delta})^{T} \otimes \mathcal{I} \Big ).
 \end{align}
The total Liouvillian is given by $ \mathcal{L} = \mathcal{L}_{R} + \mathcal{L}_{L}$. 

\section{Necessarily transient TUR violation for the Markovian approach}
\label{Ap:Ness_tran}

\begin{figure*}[hbt]
 \centering
\includegraphics[width=0.85\textwidth]{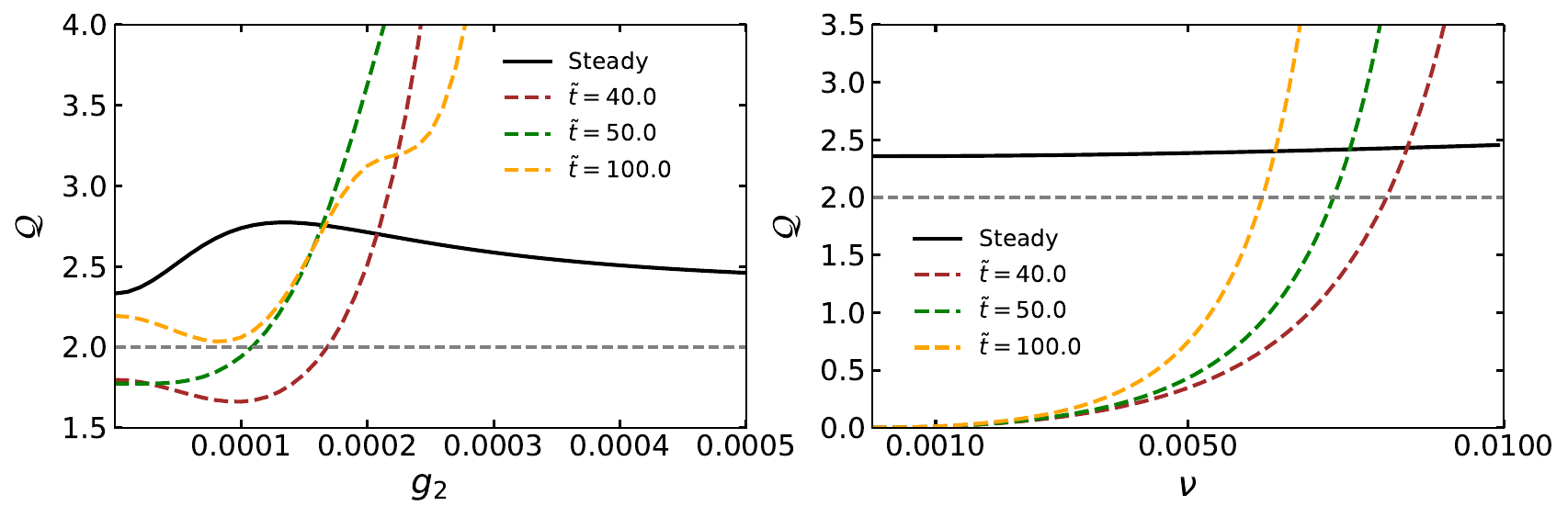}
\caption{
Necessarily transient TUR violation for Markovian evolution. Here, we depict the classical TUR parameters for the steady-state regime $(\mathcal{Q}_{\text{cl}})$ and the finite-time regime $(\mathcal{Q}_{\text{cl}}^{\text{FT}})$ as functions of  $g_2$ in the left panel and as function of $\nu$ in the right panel. For both the panels we set, $g_1 = 0.1\frac{\hbar}{\sqrt{t^{\prime}}}$ and all other parameters are the same as in Figs.~\ref{fig:mainfig}(b) and~(c), respectively. In each panel, the solid black lines represent $\mathcal{Q}_{\text{cl}}$ and the dashed lines correspond to the $\mathcal{Q}_{\text{cl}}^{\text{FT}}$ at different finite times $\tilde{t}$. The quantity $g_2$, plotted along the $x$ axes of the left panel, has units of $\hbar/\sqrt{t^{\prime}}$, and $\nu$ in the right panel has unit of $1/t^{\prime}$. The quantities plotted along the $y$ axes are dimensionless.}
\label{Transient}
\end{figure*}

In Fig.~\ref{fig:mainfig} of the main text, we have already shown  that the behavior of $\mathcal{Q}_{\text{cl}}^{\text{FT}}$ is strongly dependent on the parameter regime. Violations of TUR can occur in both the transient and steady-state regimes; however, whether the maximum violation occurs during the transient or in the steady-state regime depends on the specific parameter values chosen for the observation. In particular, when varying $g_2$ and $\nu$, the regime in which the maximum TUR violation occurs—transient or steady state—is determined by the corresponding value of $g_1$. In Fig.~\ref{Transient}, we present cases where the maximum TUR violation occurs at an early transient time and gradually vanishes as time progresses. This type of behavior, where TUR violations are present only at short times and disappear in the long-time limit, is referred to as necessarily transient. In such cases, the violation cannot be observed or harnessed in the steady state, thereby limiting it to a finite-time operational window. This feature is particularly important for designing collision model--based quantum devices intended to function within finite-time regimes, where steady-state operation may not be feasible or optimal.

\section{Transient regime analysis of entropy production rate in non-Markovian approaches}
\label{Ap:EPR}

\begin{figure*}[hbt]
 \centering
\includegraphics[width=0.95\textwidth]{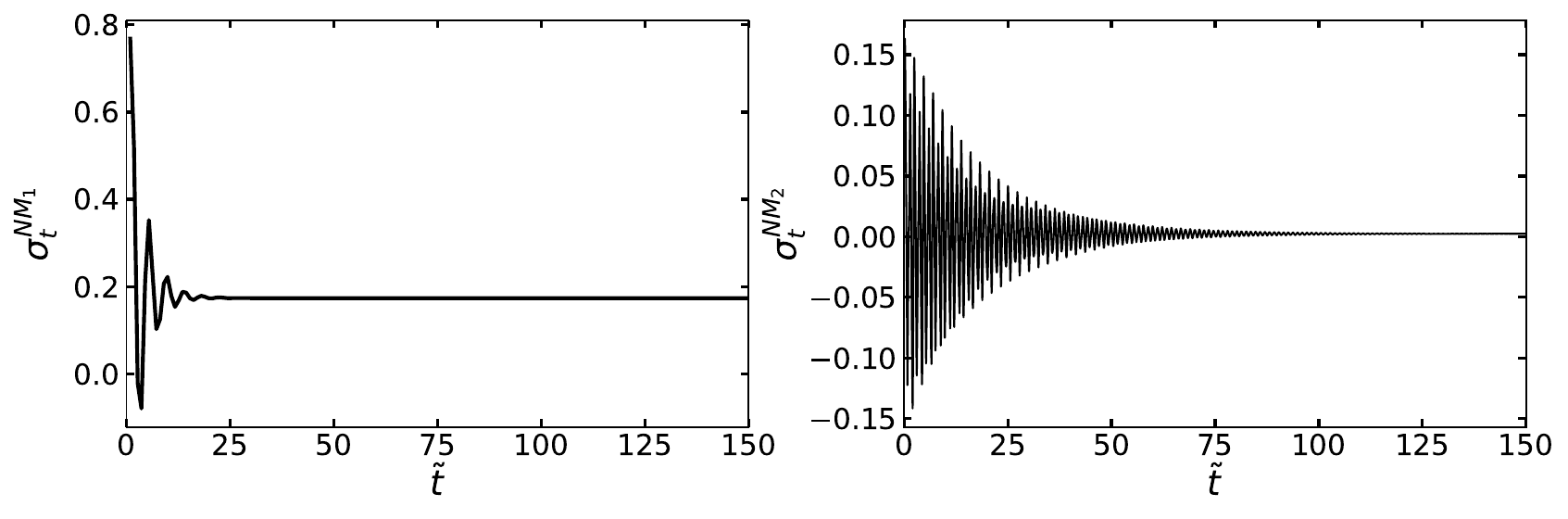}
 \caption{Transient evolution of entropy production rate for non-Markovian approaches. The left panel depicts the time evolution of the entropy production rate for non-Markovian approach I, denoted as $\sigma_t^{NM_1}$, at $\tau=0.9t^{\prime}$ and $g_1=1.5\frac{\hbar}{\sqrt{t^{\prime}}}$, with all other parameters remaining the same as in Fig.~\ref{non-Marko_I} of the main text. The right panel presents the corresponding plot for non-Markovian approach II, $\sigma_t^{NM_2}$, for $\tau=0.05t^{\prime}$, $g_1=1.5\frac{\hbar}{\sqrt{t^{\prime}}}$, and all other parameters set as in Fig. \ref{fig:non-mark_II} of the main text. The quantity plotted along the $x$ axis is a dimensionless quantity and the quantity plotted along the $y$ axis has units of $\frac{k_B}{t^{\prime}}$.
}
\label{fig:non-mark_II_check}
\end{figure*}

Note that while the entropy production rate can become negative during transient dynamics for non-Markovian evolution, the steady-state entropy production rate in a nonequilibrium setting must remain strictly positive. To verify this in the non-Markovian scenarios considered in the main text, we analyze the transient behavior of the entropy production rate, as shown in Fig.~\ref{fig:non-mark_II_check}. In both cases, the entropy production rate becomes negative during the transient regime but eventually saturates to a positive value at the steady state.


\end{widetext}

\bibliography{tur_cm}

\begin{thebibliography}{135}%
\makeatletter
\providecommand \@ifxundefined [1]{%
 \@ifx{#1\undefined}
}%
\providecommand \@ifnum [1]{%
 \ifnum #1\expandafter \@firstoftwo
 \else \expandafter \@secondoftwo
 \fi
}%
\providecommand \@ifx [1]{%
 \ifx #1\expandafter \@firstoftwo
 \else \expandafter \@secondoftwo
 \fi
}%
\providecommand \natexlab [1]{#1}%
\providecommand \enquote  [1]{``#1''}%
\providecommand \bibnamefont  [1]{#1}%
\providecommand \bibfnamefont [1]{#1}%
\providecommand \citenamefont [1]{#1}%
\providecommand \href@noop [0]{\@secondoftwo}%
\providecommand \href [0]{\begingroup \@sanitize@url \@href}%
\providecommand \@href[1]{\@@startlink{#1}\@@href}%
\providecommand \@@href[1]{\endgroup#1\@@endlink}%
\providecommand \@sanitize@url [0]{\catcode `\\12\catcode `\$12\catcode `\&12\catcode `\#12\catcode `\^12\catcode `\_12\catcode `\%12\relax}%
\providecommand \@@startlink[1]{}%
\providecommand \@@endlink[0]{}%
\providecommand \url  [0]{\begingroup\@sanitize@url \@url }%
\providecommand \@url [1]{\endgroup\@href {#1}{\urlprefix }}%
\providecommand \urlprefix  [0]{URL }%
\providecommand \Eprint [0]{\href }%
\providecommand \doibase [0]{https://doi.org/}%
\providecommand \selectlanguage [0]{\@gobble}%
\providecommand \bibinfo  [0]{\@secondoftwo}%
\providecommand \bibfield  [0]{\@secondoftwo}%
\providecommand \translation [1]{[#1]}%
\providecommand \BibitemOpen [0]{}%
\providecommand \bibitemStop [0]{}%
\providecommand \bibitemNoStop [0]{.\EOS\space}%
\providecommand \EOS [0]{\spacefactor3000\relax}%
\providecommand \BibitemShut  [1]{\csname bibitem#1\endcsname}%
\let\auto@bib@innerbib\@empty
\bibitem [{\citenamefont {Barato}\ and\ \citenamefont {Seifert}(2015)}]{Barato2015}%
  \BibitemOpen
  \bibfield  {author} {\bibinfo {author} {\bibfnamefont {A.~C.}\ \bibnamefont {Barato}}\ and\ \bibinfo {author} {\bibfnamefont {U.}~\bibnamefont {Seifert}},\ }\bibfield  {title} {\bibinfo {title} {Thermodynamic uncertainty relation for biomolecular processes},\ }\href {https://doi.org/10.1103/PhysRevLett.114.158101} {\bibfield  {journal} {\bibinfo  {journal} {Phys. Rev. Lett.}\ }\textbf {\bibinfo {volume} {114}},\ \bibinfo {pages} {158101} (\bibinfo {year} {2015})}\BibitemShut {NoStop}%
\bibitem [{\citenamefont {Gingrich}\ \emph {et~al.}(2016)\citenamefont {Gingrich}, \citenamefont {Horowitz}, \citenamefont {Perunov},\ and\ \citenamefont {England}}]{Gingrich2016}%
  \BibitemOpen
  \bibfield  {author} {\bibinfo {author} {\bibfnamefont {T.~R.}\ \bibnamefont {Gingrich}}, \bibinfo {author} {\bibfnamefont {J.~M.}\ \bibnamefont {Horowitz}}, \bibinfo {author} {\bibfnamefont {N.}~\bibnamefont {Perunov}},\ and\ \bibinfo {author} {\bibfnamefont {J.~L.}\ \bibnamefont {England}},\ }\bibfield  {title} {\bibinfo {title} {Dissipation bounds all steady-state current fluctuations},\ }\href {https://doi.org/10.1103/PhysRevLett.116.120601} {\bibfield  {journal} {\bibinfo  {journal} {Phys. Rev. Lett.}\ }\textbf {\bibinfo {volume} {116}},\ \bibinfo {pages} {120601} (\bibinfo {year} {2016})}\BibitemShut {NoStop}%
\bibitem [{\citenamefont {Barato}\ \emph {et~al.}(2019)\citenamefont {Barato}, \citenamefont {Chetrite}, \citenamefont {Faggionato},\ and\ \citenamefont {Gabrielli}}]{Barato2019}%
  \BibitemOpen
  \bibfield  {author} {\bibinfo {author} {\bibfnamefont {A.~C.}\ \bibnamefont {Barato}}, \bibinfo {author} {\bibfnamefont {R.}~\bibnamefont {Chetrite}}, \bibinfo {author} {\bibfnamefont {A.}~\bibnamefont {Faggionato}},\ and\ \bibinfo {author} {\bibfnamefont {D.}~\bibnamefont {Gabrielli}},\ }\bibfield  {title} {\bibinfo {title} {A unifying picture of generalized thermodynamic uncertainty relations},\ }\href {https://doi.org/10.1088/1742-5468/ab3457} {\bibfield  {journal} {\bibinfo  {journal} {Journal of Statistical Mechanics: Theory and Experiment}\ }\textbf {\bibinfo {volume} {2019}},\ \bibinfo {pages} {084017} (\bibinfo {year} {2019})}\BibitemShut {NoStop}%
\bibitem [{\citenamefont {Falasco}\ \emph {et~al.}(2020)\citenamefont {Falasco}, \citenamefont {Esposito},\ and\ \citenamefont {Delvenne}}]{Falasco2020}%
  \BibitemOpen
  \bibfield  {author} {\bibinfo {author} {\bibfnamefont {G.}~\bibnamefont {Falasco}}, \bibinfo {author} {\bibfnamefont {M.}~\bibnamefont {Esposito}},\ and\ \bibinfo {author} {\bibfnamefont {J.-C.}\ \bibnamefont {Delvenne}},\ }\bibfield  {title} {\bibinfo {title} {Unifying thermodynamic uncertainty relations},\ }\href {https://doi.org/10.1088/1367-2630/ab8679} {\bibfield  {journal} {\bibinfo  {journal} {New Journal of Physics}\ }\textbf {\bibinfo {volume} {22}},\ \bibinfo {pages} {053046} (\bibinfo {year} {2020})}\BibitemShut {NoStop}%
\bibitem [{\citenamefont {Hasegawa}(2021{\natexlab{a}})}]{Hasegawa2021}%
  \BibitemOpen
  \bibfield  {author} {\bibinfo {author} {\bibfnamefont {Y.}~\bibnamefont {Hasegawa}},\ }\bibfield  {title} {\bibinfo {title} {Thermodynamic uncertainty relation for general open quantum systems},\ }\href {https://doi.org/10.1103/PhysRevLett.126.010602} {\bibfield  {journal} {\bibinfo  {journal} {Phys. Rev. Lett.}\ }\textbf {\bibinfo {volume} {126}},\ \bibinfo {pages} {010602} (\bibinfo {year} {2021}{\natexlab{a}})}\BibitemShut {NoStop}%
\bibitem [{\citenamefont {Menczel}\ \emph {et~al.}(2021)\citenamefont {Menczel}, \citenamefont {Loisa}, \citenamefont {Brandner},\ and\ \citenamefont {Flindt}}]{Menczel2021}%
  \BibitemOpen
  \bibfield  {author} {\bibinfo {author} {\bibfnamefont {P.}~\bibnamefont {Menczel}}, \bibinfo {author} {\bibfnamefont {E.}~\bibnamefont {Loisa}}, \bibinfo {author} {\bibfnamefont {K.}~\bibnamefont {Brandner}},\ and\ \bibinfo {author} {\bibfnamefont {C.}~\bibnamefont {Flindt}},\ }\bibfield  {title} {\bibinfo {title} {Thermodynamic uncertainty relations for coherently driven open quantum systems},\ }\href {https://doi.org/10.1088/1751-8121/ac0c8f} {\bibfield  {journal} {\bibinfo  {journal} {Journal of Physics A: Mathematical and Theoretical}\ }\textbf {\bibinfo {volume} {54}},\ \bibinfo {pages} {314002} (\bibinfo {year} {2021})}\BibitemShut {NoStop}%
\bibitem [{\citenamefont {Falasco}\ \emph {et~al.}(2022)\citenamefont {Falasco}, \citenamefont {Esposito},\ and\ \citenamefont {Delvenne}}]{Falasco2022}%
  \BibitemOpen
  \bibfield  {author} {\bibinfo {author} {\bibfnamefont {G.}~\bibnamefont {Falasco}}, \bibinfo {author} {\bibfnamefont {M.}~\bibnamefont {Esposito}},\ and\ \bibinfo {author} {\bibfnamefont {J.-C.}\ \bibnamefont {Delvenne}},\ }\bibfield  {title} {\bibinfo {title} {Beyond thermodynamic uncertainty relations: nonlinear response, error-dissipation trade-offs, and speed limits},\ }\href {https://doi.org/10.1088/1751-8121/ac52e2} {\bibfield  {journal} {\bibinfo  {journal} {Journal of Physics A: Mathematical and Theoretical}\ }\textbf {\bibinfo {volume} {55}},\ \bibinfo {pages} {124002} (\bibinfo {year} {2022})}\BibitemShut {NoStop}%
\bibitem [{\citenamefont {Callen}\ and\ \citenamefont {Welton}(1951)}]{Callen1951}%
  \BibitemOpen
  \bibfield  {author} {\bibinfo {author} {\bibfnamefont {H.~B.}\ \bibnamefont {Callen}}\ and\ \bibinfo {author} {\bibfnamefont {T.~A.}\ \bibnamefont {Welton}},\ }\bibfield  {title} {\bibinfo {title} {Irreversibility and generalized noise},\ }\href {https://doi.org/10.1103/PhysRev.83.34} {\bibfield  {journal} {\bibinfo  {journal} {Phys. Rev.}\ }\textbf {\bibinfo {volume} {83}},\ \bibinfo {pages} {34} (\bibinfo {year} {1951})}\BibitemShut {NoStop}%
\bibitem [{\citenamefont {Kubo}(1966)}]{Kubo1966}%
  \BibitemOpen
  \bibfield  {author} {\bibinfo {author} {\bibfnamefont {R.}~\bibnamefont {Kubo}},\ }\bibfield  {title} {\bibinfo {title} {The fluctuation-dissipation theorem},\ }\href {https://doi.org/10.1088/0034-4885/29/1/306} {\bibfield  {journal} {\bibinfo  {journal} {Reports on Progress in Physics}\ }\textbf {\bibinfo {volume} {29}},\ \bibinfo {pages} {255} (\bibinfo {year} {1966})}\BibitemShut {NoStop}%
\bibitem [{\citenamefont {Carollo}\ \emph {et~al.}(2019)\citenamefont {Carollo}, \citenamefont {Jack},\ and\ \citenamefont {Garrahan}}]{Carollo_PRL_2019}%
  \BibitemOpen
  \bibfield  {author} {\bibinfo {author} {\bibfnamefont {F.}~\bibnamefont {Carollo}}, \bibinfo {author} {\bibfnamefont {R.~L.}\ \bibnamefont {Jack}},\ and\ \bibinfo {author} {\bibfnamefont {J.~P.}\ \bibnamefont {Garrahan}},\ }\bibfield  {title} {\bibinfo {title} {Unraveling the large deviation statistics of markovian open quantum systems},\ }\href {https://doi.org/10.1103/PhysRevLett.122.130605} {\bibfield  {journal} {\bibinfo  {journal} {Phys. Rev. Lett.}\ }\textbf {\bibinfo {volume} {122}},\ \bibinfo {pages} {130605} (\bibinfo {year} {2019})}\BibitemShut {NoStop}%
\bibitem [{\citenamefont {Guarnieri}\ \emph {et~al.}(2019)\citenamefont {Guarnieri}, \citenamefont {Landi}, \citenamefont {Clark},\ and\ \citenamefont {Goold}}]{Goold_PRR_2019}%
  \BibitemOpen
  \bibfield  {author} {\bibinfo {author} {\bibfnamefont {G.}~\bibnamefont {Guarnieri}}, \bibinfo {author} {\bibfnamefont {G.~T.}\ \bibnamefont {Landi}}, \bibinfo {author} {\bibfnamefont {S.~R.}\ \bibnamefont {Clark}},\ and\ \bibinfo {author} {\bibfnamefont {J.}~\bibnamefont {Goold}},\ }\bibfield  {title} {\bibinfo {title} {Thermodynamics of precision in quantum nonequilibrium steady states},\ }\href {https://doi.org/10.1103/PhysRevResearch.1.033021} {\bibfield  {journal} {\bibinfo  {journal} {Phys. Rev. Res.}\ }\textbf {\bibinfo {volume} {1}},\ \bibinfo {pages} {033021} (\bibinfo {year} {2019})}\BibitemShut {NoStop}%
\bibitem [{\citenamefont {Hasegawa}(2020)}]{Hasegawa_PRL_2020}%
  \BibitemOpen
  \bibfield  {author} {\bibinfo {author} {\bibfnamefont {Y.}~\bibnamefont {Hasegawa}},\ }\bibfield  {title} {\bibinfo {title} {Quantum thermodynamic uncertainty relation for continuous measurement},\ }\href {https://doi.org/10.1103/PhysRevLett.125.050601} {\bibfield  {journal} {\bibinfo  {journal} {Phys. Rev. Lett.}\ }\textbf {\bibinfo {volume} {125}},\ \bibinfo {pages} {050601} (\bibinfo {year} {2020})}\BibitemShut {NoStop}%
\bibitem [{\citenamefont {Horowitz}\ and\ \citenamefont {Gingrich}(2020)}]{Horowitz2020}%
  \BibitemOpen
  \bibfield  {author} {\bibinfo {author} {\bibfnamefont {J.~M.}\ \bibnamefont {Horowitz}}\ and\ \bibinfo {author} {\bibfnamefont {T.~R.}\ \bibnamefont {Gingrich}},\ }\bibfield  {title} {\bibinfo {title} {Thermodynamic uncertainty relations constrain non-equilibrium fluctuations},\ }\href {https://doi.org/10.1038/s41567-019-0702-6} {\bibfield  {journal} {\bibinfo  {journal} {Nature Physics}\ }\textbf {\bibinfo {volume} {16}},\ \bibinfo {pages} {15} (\bibinfo {year} {2020})}\BibitemShut {NoStop}%
\bibitem [{\citenamefont {Hasegawa}(2021{\natexlab{b}})}]{Hasegawa_PRL_2021}%
  \BibitemOpen
  \bibfield  {author} {\bibinfo {author} {\bibfnamefont {Y.}~\bibnamefont {Hasegawa}},\ }\bibfield  {title} {\bibinfo {title} {Irreversibility, {L}oschmidt echo, and thermodynamic uncertainty relation},\ }\href {https://doi.org/10.1103/PhysRevLett.127.240602} {\bibfield  {journal} {\bibinfo  {journal} {Phys. Rev. Lett.}\ }\textbf {\bibinfo {volume} {127}},\ \bibinfo {pages} {240602} (\bibinfo {year} {2021}{\natexlab{b}})}\BibitemShut {NoStop}%
\bibitem [{\citenamefont {Pietzonka}\ \emph {et~al.}(2017)\citenamefont {Pietzonka}, \citenamefont {Ritort},\ and\ \citenamefont {Seifert}}]{Patrick_PRE_2017}%
  \BibitemOpen
  \bibfield  {author} {\bibinfo {author} {\bibfnamefont {P.}~\bibnamefont {Pietzonka}}, \bibinfo {author} {\bibfnamefont {F.}~\bibnamefont {Ritort}},\ and\ \bibinfo {author} {\bibfnamefont {U.}~\bibnamefont {Seifert}},\ }\bibfield  {title} {\bibinfo {title} {Finite-time generalization of the thermodynamic uncertainty relation},\ }\href {https://doi.org/10.1103/PhysRevE.96.012101} {\bibfield  {journal} {\bibinfo  {journal} {Phys. Rev. E}\ }\textbf {\bibinfo {volume} {96}},\ \bibinfo {pages} {012101} (\bibinfo {year} {2017})}\BibitemShut {NoStop}%
\bibitem [{\citenamefont {Horowitz}\ and\ \citenamefont {Gingrich}(2017)}]{Horowitz_PRE_2017}%
  \BibitemOpen
  \bibfield  {author} {\bibinfo {author} {\bibfnamefont {J.~M.}\ \bibnamefont {Horowitz}}\ and\ \bibinfo {author} {\bibfnamefont {T.~R.}\ \bibnamefont {Gingrich}},\ }\bibfield  {title} {\bibinfo {title} {Proof of the finite-time thermodynamic uncertainty relation for steady-state currents},\ }\href {https://doi.org/10.1103/PhysRevE.96.020103} {\bibfield  {journal} {\bibinfo  {journal} {Phys. Rev. E}\ }\textbf {\bibinfo {volume} {96}},\ \bibinfo {pages} {020103} (\bibinfo {year} {2017})}\BibitemShut {NoStop}%
\bibitem [{\citenamefont {Otsubo}\ \emph {et~al.}(2020)\citenamefont {Otsubo}, \citenamefont {Ito}, \citenamefont {Dechant},\ and\ \citenamefont {Sagawa}}]{Otsubo2020}%
  \BibitemOpen
  \bibfield  {author} {\bibinfo {author} {\bibfnamefont {S.}~\bibnamefont {Otsubo}}, \bibinfo {author} {\bibfnamefont {S.}~\bibnamefont {Ito}}, \bibinfo {author} {\bibfnamefont {A.}~\bibnamefont {Dechant}},\ and\ \bibinfo {author} {\bibfnamefont {T.}~\bibnamefont {Sagawa}},\ }\bibfield  {title} {\bibinfo {title} {Estimating entropy production by machine learning of short-time fluctuating currents},\ }\href {https://doi.org/10.1103/PhysRevE.101.062106} {\bibfield  {journal} {\bibinfo  {journal} {Phys. Rev. E}\ }\textbf {\bibinfo {volume} {101}},\ \bibinfo {pages} {062106} (\bibinfo {year} {2020})}\BibitemShut {NoStop}%
\bibitem [{\citenamefont {Proesmans}\ and\ \citenamefont {den Broeck}(2017)}]{Proesmans_EPL_2017}%
  \BibitemOpen
  \bibfield  {author} {\bibinfo {author} {\bibfnamefont {K.}~\bibnamefont {Proesmans}}\ and\ \bibinfo {author} {\bibfnamefont {C.~V.}\ \bibnamefont {den Broeck}},\ }\bibfield  {title} {\bibinfo {title} {Discrete-time thermodynamic uncertainty relation},\ }\href {https://doi.org/10.1209/0295-5075/119/20001} {\bibfield  {journal} {\bibinfo  {journal} {Europhysics Letters}\ }\textbf {\bibinfo {volume} {119}},\ \bibinfo {pages} {20001} (\bibinfo {year} {2017})}\BibitemShut {NoStop}%
\bibitem [{\citenamefont {Barato}\ \emph {et~al.}(2018)\citenamefont {Barato}, \citenamefont {Chetrite}, \citenamefont {Faggionato},\ and\ \citenamefont {Gabrielli}}]{Barato_NJP_2018}%
  \BibitemOpen
  \bibfield  {author} {\bibinfo {author} {\bibfnamefont {A.~C.}\ \bibnamefont {Barato}}, \bibinfo {author} {\bibfnamefont {R.}~\bibnamefont {Chetrite}}, \bibinfo {author} {\bibfnamefont {A.}~\bibnamefont {Faggionato}},\ and\ \bibinfo {author} {\bibfnamefont {D.}~\bibnamefont {Gabrielli}},\ }\bibfield  {title} {\bibinfo {title} {Bounds on current fluctuations in periodically driven systems},\ }\href {https://doi.org/10.1088/1367-2630/aae512} {\bibfield  {journal} {\bibinfo  {journal} {New Journal of Physics}\ }\textbf {\bibinfo {volume} {20}},\ \bibinfo {pages} {103023} (\bibinfo {year} {2018})}\BibitemShut {NoStop}%
\bibitem [{\citenamefont {Holubec}\ and\ \citenamefont {Ryabov}(2018)}]{Holubec_PRL_2018}%
  \BibitemOpen
  \bibfield  {author} {\bibinfo {author} {\bibfnamefont {V.}~\bibnamefont {Holubec}}\ and\ \bibinfo {author} {\bibfnamefont {A.}~\bibnamefont {Ryabov}},\ }\bibfield  {title} {\bibinfo {title} {Cycling tames power fluctuations near optimum efficiency},\ }\href {https://doi.org/10.1103/PhysRevLett.121.120601} {\bibfield  {journal} {\bibinfo  {journal} {Phys. Rev. Lett.}\ }\textbf {\bibinfo {volume} {121}},\ \bibinfo {pages} {120601} (\bibinfo {year} {2018})}\BibitemShut {NoStop}%
\bibitem [{\citenamefont {Macieszczak}\ \emph {et~al.}(2018)\citenamefont {Macieszczak}, \citenamefont {Brandner},\ and\ \citenamefont {Garrahan}}]{Macieszczak_PRL_2018}%
  \BibitemOpen
  \bibfield  {author} {\bibinfo {author} {\bibfnamefont {K.}~\bibnamefont {Macieszczak}}, \bibinfo {author} {\bibfnamefont {K.}~\bibnamefont {Brandner}},\ and\ \bibinfo {author} {\bibfnamefont {J.~P.}\ \bibnamefont {Garrahan}},\ }\bibfield  {title} {\bibinfo {title} {Unified thermodynamic uncertainty relations in linear response},\ }\href {https://doi.org/10.1103/PhysRevLett.121.130601} {\bibfield  {journal} {\bibinfo  {journal} {Phys. Rev. Lett.}\ }\textbf {\bibinfo {volume} {121}},\ \bibinfo {pages} {130601} (\bibinfo {year} {2018})}\BibitemShut {NoStop}%
\bibitem [{\citenamefont {Manikandan}\ \emph {et~al.}(2020)\citenamefont {Manikandan}, \citenamefont {Gupta},\ and\ \citenamefont {Krishnamurthy}}]{Supriya_PRL_2020}%
  \BibitemOpen
  \bibfield  {author} {\bibinfo {author} {\bibfnamefont {S.~K.}\ \bibnamefont {Manikandan}}, \bibinfo {author} {\bibfnamefont {D.}~\bibnamefont {Gupta}},\ and\ \bibinfo {author} {\bibfnamefont {S.}~\bibnamefont {Krishnamurthy}},\ }\bibfield  {title} {\bibinfo {title} {Inferring entropy production from short experiments},\ }\href {https://doi.org/10.1103/PhysRevLett.124.120603} {\bibfield  {journal} {\bibinfo  {journal} {Phys. Rev. Lett.}\ }\textbf {\bibinfo {volume} {124}},\ \bibinfo {pages} {120603} (\bibinfo {year} {2020})}\BibitemShut {NoStop}%
\bibitem [{\citenamefont {Van~Vu}\ \emph {et~al.}(2020)\citenamefont {Van~Vu}, \citenamefont {Vo},\ and\ \citenamefont {Hasegawa}}]{Hasegawa_PRE_2020}%
  \BibitemOpen
  \bibfield  {author} {\bibinfo {author} {\bibfnamefont {T.}~\bibnamefont {Van~Vu}}, \bibinfo {author} {\bibfnamefont {V.~T.}\ \bibnamefont {Vo}},\ and\ \bibinfo {author} {\bibfnamefont {Y.}~\bibnamefont {Hasegawa}},\ }\bibfield  {title} {\bibinfo {title} {Entropy production estimation with optimal current},\ }\href {https://doi.org/10.1103/PhysRevE.101.042138} {\bibfield  {journal} {\bibinfo  {journal} {Phys. Rev. E}\ }\textbf {\bibinfo {volume} {101}},\ \bibinfo {pages} {042138} (\bibinfo {year} {2020})}\BibitemShut {NoStop}%
\bibitem [{\citenamefont {Kim}\ \emph {et~al.}(2020)\citenamefont {Kim}, \citenamefont {Bae}, \citenamefont {Lee},\ and\ \citenamefont {Jeong}}]{Jeong_PRL_2020}%
  \BibitemOpen
  \bibfield  {author} {\bibinfo {author} {\bibfnamefont {D.-K.}\ \bibnamefont {Kim}}, \bibinfo {author} {\bibfnamefont {Y.}~\bibnamefont {Bae}}, \bibinfo {author} {\bibfnamefont {S.}~\bibnamefont {Lee}},\ and\ \bibinfo {author} {\bibfnamefont {H.}~\bibnamefont {Jeong}},\ }\bibfield  {title} {\bibinfo {title} {Learning entropy production via neural networks},\ }\href {https://doi.org/10.1103/PhysRevLett.125.140604} {\bibfield  {journal} {\bibinfo  {journal} {Phys. Rev. Lett.}\ }\textbf {\bibinfo {volume} {125}},\ \bibinfo {pages} {140604} (\bibinfo {year} {2020})}\BibitemShut {NoStop}%
\bibitem [{\citenamefont {Manikandan}\ \emph {et~al.}(2021)\citenamefont {Manikandan}, \citenamefont {Ghosh}, \citenamefont {Kundu}, \citenamefont {Das}, \citenamefont {Agrawal}, \citenamefont {Mitra}, \citenamefont {Banerjee},\ and\ \citenamefont {Krishnamurthy}}]{Supriya_CP_2021}%
  \BibitemOpen
  \bibfield  {author} {\bibinfo {author} {\bibfnamefont {S.~K.}\ \bibnamefont {Manikandan}}, \bibinfo {author} {\bibfnamefont {S.}~\bibnamefont {Ghosh}}, \bibinfo {author} {\bibfnamefont {A.}~\bibnamefont {Kundu}}, \bibinfo {author} {\bibfnamefont {B.}~\bibnamefont {Das}}, \bibinfo {author} {\bibfnamefont {V.}~\bibnamefont {Agrawal}}, \bibinfo {author} {\bibfnamefont {D.}~\bibnamefont {Mitra}}, \bibinfo {author} {\bibfnamefont {A.}~\bibnamefont {Banerjee}},\ and\ \bibinfo {author} {\bibfnamefont {S.}~\bibnamefont {Krishnamurthy}},\ }\bibfield  {title} {\bibinfo {title} {Quantitative analysis of non-equilibrium systems from short-time experimental data},\ }\href {https://doi.org/10.1038/s42005-021-00766-2} {\bibfield  {journal} {\bibinfo  {journal} {Communications Physics}\ }\textbf {\bibinfo {volume} {4}},\ \bibinfo {pages} {258} (\bibinfo {year} {2021})}\BibitemShut {NoStop}%
\bibitem [{\citenamefont {Otsubo}\ \emph {et~al.}(2022)\citenamefont {Otsubo}, \citenamefont {Manikandan}, \citenamefont {Sagawa},\ and\ \citenamefont {Krishnamurthy}}]{Otsubo2022}%
  \BibitemOpen
  \bibfield  {author} {\bibinfo {author} {\bibfnamefont {S.}~\bibnamefont {Otsubo}}, \bibinfo {author} {\bibfnamefont {S.~K.}\ \bibnamefont {Manikandan}}, \bibinfo {author} {\bibfnamefont {T.}~\bibnamefont {Sagawa}},\ and\ \bibinfo {author} {\bibfnamefont {S.}~\bibnamefont {Krishnamurthy}},\ }\bibfield  {title} {\bibinfo {title} {Estimating time-dependent entropy production from non-equilibrium trajectories},\ }\href {https://doi.org/10.1038/s42005-021-00787-x} {\bibfield  {journal} {\bibinfo  {journal} {Communications Physics}\ }\textbf {\bibinfo {volume} {5}},\ \bibinfo {pages} {11} (\bibinfo {year} {2022})}\BibitemShut {NoStop}%
\bibitem [{\citenamefont {Moreira}\ \emph {et~al.}(2025)\citenamefont {Moreira}, \citenamefont {Radaelli}, \citenamefont {Candeloro}, \citenamefont {Binder},\ and\ \citenamefont {Mitchison}}]{moreira2024}%
  \BibitemOpen
  \bibfield  {author} {\bibinfo {author} {\bibfnamefont {S.~V.}\ \bibnamefont {Moreira}}, \bibinfo {author} {\bibfnamefont {M.}~\bibnamefont {Radaelli}}, \bibinfo {author} {\bibfnamefont {A.}~\bibnamefont {Candeloro}}, \bibinfo {author} {\bibfnamefont {F.~C.}\ \bibnamefont {Binder}},\ and\ \bibinfo {author} {\bibfnamefont {M.~T.}\ \bibnamefont {Mitchison}},\ }\bibfield  {title} {\bibinfo {title} {Precision bounds for multiple currents in open quantum systems},\ }\href {https://doi.org/10.1103/PhysRevE.111.064107} {\bibfield  {journal} {\bibinfo  {journal} {Phys. Rev. E}\ }\textbf {\bibinfo {volume} {111}},\ \bibinfo {pages} {064107} (\bibinfo {year} {2025})}\BibitemShut {NoStop}%
\bibitem [{\citenamefont {Barato}\ and\ \citenamefont {Seifert}(2016)}]{Barato2016}%
  \BibitemOpen
  \bibfield  {author} {\bibinfo {author} {\bibfnamefont {A.~C.}\ \bibnamefont {Barato}}\ and\ \bibinfo {author} {\bibfnamefont {U.}~\bibnamefont {Seifert}},\ }\bibfield  {title} {\bibinfo {title} {Cost and precision of brownian clocks},\ }\href {https://doi.org/10.1103/PhysRevX.6.041053} {\bibfield  {journal} {\bibinfo  {journal} {Phys. Rev. X}\ }\textbf {\bibinfo {volume} {6}},\ \bibinfo {pages} {041053} (\bibinfo {year} {2016})}\BibitemShut {NoStop}%
\bibitem [{\citenamefont {Koyuk}\ and\ \citenamefont {Seifert}(2020)}]{Koyuk2020}%
  \BibitemOpen
  \bibfield  {author} {\bibinfo {author} {\bibfnamefont {T.}~\bibnamefont {Koyuk}}\ and\ \bibinfo {author} {\bibfnamefont {U.}~\bibnamefont {Seifert}},\ }\bibfield  {title} {\bibinfo {title} {Thermodynamic uncertainty relation for time-dependent driving},\ }\href {https://doi.org/10.1103/PhysRevLett.125.260604} {\bibfield  {journal} {\bibinfo  {journal} {Phys. Rev. Lett.}\ }\textbf {\bibinfo {volume} {125}},\ \bibinfo {pages} {260604} (\bibinfo {year} {2020})}\BibitemShut {NoStop}%
\bibitem [{\citenamefont {Cangemi}\ \emph {et~al.}(2021)\citenamefont {Cangemi}, \citenamefont {Carrega}, \citenamefont {De~Candia}, \citenamefont {Cataudella}, \citenamefont {De~Filippis}, \citenamefont {Sassetti},\ and\ \citenamefont {Benenti}}]{Cangemi2021}%
  \BibitemOpen
  \bibfield  {author} {\bibinfo {author} {\bibfnamefont {L.~M.}\ \bibnamefont {Cangemi}}, \bibinfo {author} {\bibfnamefont {M.}~\bibnamefont {Carrega}}, \bibinfo {author} {\bibfnamefont {A.}~\bibnamefont {De~Candia}}, \bibinfo {author} {\bibfnamefont {V.}~\bibnamefont {Cataudella}}, \bibinfo {author} {\bibfnamefont {G.}~\bibnamefont {De~Filippis}}, \bibinfo {author} {\bibfnamefont {M.}~\bibnamefont {Sassetti}},\ and\ \bibinfo {author} {\bibfnamefont {G.}~\bibnamefont {Benenti}},\ }\bibfield  {title} {\bibinfo {title} {Optimal energy conversion through antiadiabatic driving breaking time-reversal symmetry},\ }\href {https://doi.org/10.1103/PhysRevResearch.3.013237} {\bibfield  {journal} {\bibinfo  {journal} {Phys. Rev. Res.}\ }\textbf {\bibinfo {volume} {3}},\ \bibinfo {pages} {013237} (\bibinfo {year} {2021})}\BibitemShut {NoStop}%
\bibitem [{\citenamefont {Fischer}\ \emph {et~al.}(2018)\citenamefont {Fischer}, \citenamefont {Pietzonka},\ and\ \citenamefont {Seifert}}]{Fischer2018}%
  \BibitemOpen
  \bibfield  {author} {\bibinfo {author} {\bibfnamefont {L.~P.}\ \bibnamefont {Fischer}}, \bibinfo {author} {\bibfnamefont {P.}~\bibnamefont {Pietzonka}},\ and\ \bibinfo {author} {\bibfnamefont {U.}~\bibnamefont {Seifert}},\ }\bibfield  {title} {\bibinfo {title} {Large deviation function for a driven underdamped particle in a periodic potential},\ }\href {https://doi.org/10.1103/PhysRevE.97.022143} {\bibfield  {journal} {\bibinfo  {journal} {Phys. Rev. E}\ }\textbf {\bibinfo {volume} {97}},\ \bibinfo {pages} {022143} (\bibinfo {year} {2018})}\BibitemShut {NoStop}%
\bibitem [{\citenamefont {Lee}\ \emph {et~al.}(2019)\citenamefont {Lee}, \citenamefont {Park},\ and\ \citenamefont {Park}}]{Lee2019}%
  \BibitemOpen
  \bibfield  {author} {\bibinfo {author} {\bibfnamefont {J.~S.}\ \bibnamefont {Lee}}, \bibinfo {author} {\bibfnamefont {J.-M.}\ \bibnamefont {Park}},\ and\ \bibinfo {author} {\bibfnamefont {H.}~\bibnamefont {Park}},\ }\bibfield  {title} {\bibinfo {title} {Thermodynamic uncertainty relation for underdamped langevin systems driven by a velocity-dependent force},\ }\href {https://doi.org/10.1103/PhysRevE.100.062132} {\bibfield  {journal} {\bibinfo  {journal} {Phys. Rev. E}\ }\textbf {\bibinfo {volume} {100}},\ \bibinfo {pages} {062132} (\bibinfo {year} {2019})}\BibitemShut {NoStop}%
\bibitem [{\citenamefont {Van~Vu}\ and\ \citenamefont {Hasegawa}(2019)}]{Vu2019}%
  \BibitemOpen
  \bibfield  {author} {\bibinfo {author} {\bibfnamefont {T.}~\bibnamefont {Van~Vu}}\ and\ \bibinfo {author} {\bibfnamefont {Y.}~\bibnamefont {Hasegawa}},\ }\bibfield  {title} {\bibinfo {title} {Uncertainty relations for underdamped langevin dynamics},\ }\href {https://doi.org/10.1103/PhysRevE.100.032130} {\bibfield  {journal} {\bibinfo  {journal} {Phys. Rev. E}\ }\textbf {\bibinfo {volume} {100}},\ \bibinfo {pages} {032130} (\bibinfo {year} {2019})}\BibitemShut {NoStop}%
\bibitem [{\citenamefont {Potts}\ and\ \citenamefont {Samuelsson}(2019)}]{Potts2019}%
  \BibitemOpen
  \bibfield  {author} {\bibinfo {author} {\bibfnamefont {P.~P.}\ \bibnamefont {Potts}}\ and\ \bibinfo {author} {\bibfnamefont {P.}~\bibnamefont {Samuelsson}},\ }\bibfield  {title} {\bibinfo {title} {Thermodynamic uncertainty relations including measurement and feedback},\ }\href {https://doi.org/10.1103/PhysRevE.100.052137} {\bibfield  {journal} {\bibinfo  {journal} {Phys. Rev. E}\ }\textbf {\bibinfo {volume} {100}},\ \bibinfo {pages} {052137} (\bibinfo {year} {2019})}\BibitemShut {NoStop}%
\bibitem [{\citenamefont {Timpanaro}\ \emph {et~al.}(2019)\citenamefont {Timpanaro}, \citenamefont {Guarnieri}, \citenamefont {Goold},\ and\ \citenamefont {Landi}}]{Landi_PRL_2019}%
  \BibitemOpen
  \bibfield  {author} {\bibinfo {author} {\bibfnamefont {A.~M.}\ \bibnamefont {Timpanaro}}, \bibinfo {author} {\bibfnamefont {G.}~\bibnamefont {Guarnieri}}, \bibinfo {author} {\bibfnamefont {J.}~\bibnamefont {Goold}},\ and\ \bibinfo {author} {\bibfnamefont {G.~T.}\ \bibnamefont {Landi}},\ }\bibfield  {title} {\bibinfo {title} {Thermodynamic uncertainty relations from exchange fluctuation theorems},\ }\href {https://doi.org/10.1103/PhysRevLett.123.090604} {\bibfield  {journal} {\bibinfo  {journal} {Phys. Rev. Lett.}\ }\textbf {\bibinfo {volume} {123}},\ \bibinfo {pages} {090604} (\bibinfo {year} {2019})}\BibitemShut {NoStop}%
\bibitem [{\citenamefont {Pietzonka}\ \emph {et~al.}(2016{\natexlab{a}})\citenamefont {Pietzonka}, \citenamefont {Barato},\ and\ \citenamefont {Seifert}}]{Pietzonka_2016}%
  \BibitemOpen
  \bibfield  {author} {\bibinfo {author} {\bibfnamefont {P.}~\bibnamefont {Pietzonka}}, \bibinfo {author} {\bibfnamefont {A.~C.}\ \bibnamefont {Barato}},\ and\ \bibinfo {author} {\bibfnamefont {U.}~\bibnamefont {Seifert}},\ }\bibfield  {title} {\bibinfo {title} {Universal bound on the efficiency of molecular motors},\ }\href {https://doi.org/10.1088/1742-5468/2016/12/124004} {\bibfield  {journal} {\bibinfo  {journal} {Journal of Statistical Mechanics: Theory and Experiment}\ }\textbf {\bibinfo {volume} {2016}},\ \bibinfo {pages} {124004} (\bibinfo {year} {2016}{\natexlab{a}})}\BibitemShut {NoStop}%
\bibitem [{\citenamefont {Jack}\ \emph {et~al.}(2020)\citenamefont {Jack}, \citenamefont {L\'opez-Alamilla},\ and\ \citenamefont {Challis}}]{Jack2020}%
  \BibitemOpen
  \bibfield  {author} {\bibinfo {author} {\bibfnamefont {M.~W.}\ \bibnamefont {Jack}}, \bibinfo {author} {\bibfnamefont {N.~J.}\ \bibnamefont {L\'opez-Alamilla}},\ and\ \bibinfo {author} {\bibfnamefont {K.~J.}\ \bibnamefont {Challis}},\ }\bibfield  {title} {\bibinfo {title} {Thermodynamic uncertainty relations and molecular-scale energy conversion},\ }\href {https://doi.org/10.1103/PhysRevE.101.062123} {\bibfield  {journal} {\bibinfo  {journal} {Phys. Rev. E}\ }\textbf {\bibinfo {volume} {101}},\ \bibinfo {pages} {062123} (\bibinfo {year} {2020})}\BibitemShut {NoStop}%
\bibitem [{\citenamefont {Song}\ and\ \citenamefont {Hyeon}(2020)}]{Song2020}%
  \BibitemOpen
  \bibfield  {author} {\bibinfo {author} {\bibfnamefont {Y.}~\bibnamefont {Song}}\ and\ \bibinfo {author} {\bibfnamefont {C.}~\bibnamefont {Hyeon}},\ }\bibfield  {title} {\bibinfo {title} {Thermodynamic cost, speed, fluctuations, and error reduction of biological copy machines},\ }\href {https://doi.org/10.1021/acs.jpclett.0c00545} {\bibfield  {journal} {\bibinfo  {journal} {The Journal of Physical Chemistry Letters}\ }\textbf {\bibinfo {volume} {11}},\ \bibinfo {pages} {3136} (\bibinfo {year} {2020})}\BibitemShut {NoStop}%
\bibitem [{\citenamefont {Campisi}\ and\ \citenamefont {Fazio}(2016)}]{Campisi2016}%
  \BibitemOpen
  \bibfield  {author} {\bibinfo {author} {\bibfnamefont {M.}~\bibnamefont {Campisi}}\ and\ \bibinfo {author} {\bibfnamefont {R.}~\bibnamefont {Fazio}},\ }\bibfield  {title} {\bibinfo {title} {The power of a critical heat engine},\ }\href {https://doi.org/10.1038/ncomms11895} {\bibfield  {journal} {\bibinfo  {journal} {Nature Communications}\ }\textbf {\bibinfo {volume} {7}},\ \bibinfo {pages} {11895} (\bibinfo {year} {2016})}\BibitemShut {NoStop}%
\bibitem [{\citenamefont {Holubec}\ and\ \citenamefont {Ryabov}(2017)}]{Holubec2017}%
  \BibitemOpen
  \bibfield  {author} {\bibinfo {author} {\bibfnamefont {V.}~\bibnamefont {Holubec}}\ and\ \bibinfo {author} {\bibfnamefont {A.}~\bibnamefont {Ryabov}},\ }\bibfield  {title} {\bibinfo {title} {Work and power fluctuations in a critical heat engine},\ }\href {https://doi.org/10.1103/PhysRevE.96.030102} {\bibfield  {journal} {\bibinfo  {journal} {Phys. Rev. E}\ }\textbf {\bibinfo {volume} {96}},\ \bibinfo {pages} {030102} (\bibinfo {year} {2017})}\BibitemShut {NoStop}%
\bibitem [{\citenamefont {Pietzonka}\ and\ \citenamefont {Seifert}(2018)}]{Seifert_PRL_2018}%
  \BibitemOpen
  \bibfield  {author} {\bibinfo {author} {\bibfnamefont {P.}~\bibnamefont {Pietzonka}}\ and\ \bibinfo {author} {\bibfnamefont {U.}~\bibnamefont {Seifert}},\ }\bibfield  {title} {\bibinfo {title} {Universal trade-off between power, efficiency, and constancy in steady-state heat engines},\ }\href {https://doi.org/10.1103/PhysRevLett.120.190602} {\bibfield  {journal} {\bibinfo  {journal} {Phys. Rev. Lett.}\ }\textbf {\bibinfo {volume} {120}},\ \bibinfo {pages} {190602} (\bibinfo {year} {2018})}\BibitemShut {NoStop}%
\bibitem [{\citenamefont {Seifert}(2019)}]{Seifert_2019_AR}%
  \BibitemOpen
  \bibfield  {author} {\bibinfo {author} {\bibfnamefont {U.}~\bibnamefont {Seifert}},\ }\bibfield  {title} {\bibinfo {title} {From stochastic thermodynamics to thermodynamic inference},\ }\href {https://doi.org/https://doi.org/10.1146/annurev-conmatphys-031218-013554} {\bibfield  {journal} {\bibinfo  {journal} {Annual Review of Condensed Matter Physics}\ }\textbf {\bibinfo {volume} {10}},\ \bibinfo {pages} {171} (\bibinfo {year} {2019})}\BibitemShut {NoStop}%
\bibitem [{\citenamefont {Hwang}\ and\ \citenamefont {Hyeon}(2018)}]{Hwang2018}%
  \BibitemOpen
  \bibfield  {author} {\bibinfo {author} {\bibfnamefont {W.}~\bibnamefont {Hwang}}\ and\ \bibinfo {author} {\bibfnamefont {C.}~\bibnamefont {Hyeon}},\ }\bibfield  {title} {\bibinfo {title} {Energetic costs, precision, and transport efficiency of molecular motors},\ }\href {https://doi.org/10.1021/acs.jpclett.7b03197} {\bibfield  {journal} {\bibinfo  {journal} {The Journal of Physical Chemistry Letters}\ }\textbf {\bibinfo {volume} {9}},\ \bibinfo {pages} {513} (\bibinfo {year} {2018})}\BibitemShut {NoStop}%
\bibitem [{\citenamefont {Paneru}\ \emph {et~al.}(2020)\citenamefont {Paneru}, \citenamefont {Dutta}, \citenamefont {Tlusty},\ and\ \citenamefont {Pak}}]{Paneru2020}%
  \BibitemOpen
  \bibfield  {author} {\bibinfo {author} {\bibfnamefont {G.}~\bibnamefont {Paneru}}, \bibinfo {author} {\bibfnamefont {S.}~\bibnamefont {Dutta}}, \bibinfo {author} {\bibfnamefont {T.}~\bibnamefont {Tlusty}},\ and\ \bibinfo {author} {\bibfnamefont {H.~K.}\ \bibnamefont {Pak}},\ }\bibfield  {title} {\bibinfo {title} {Reaching and violating thermodynamic uncertainty bounds in information engines},\ }\href {https://doi.org/10.1103/PhysRevE.102.032126} {\bibfield  {journal} {\bibinfo  {journal} {Phys. Rev. E}\ }\textbf {\bibinfo {volume} {102}},\ \bibinfo {pages} {032126} (\bibinfo {year} {2020})}\BibitemShut {NoStop}%
\bibitem [{\citenamefont {Pal}\ \emph {et~al.}(2020)\citenamefont {Pal}, \citenamefont {Saryal}, \citenamefont {Segal}, \citenamefont {Mahesh},\ and\ \citenamefont {Agarwalla}}]{Pal2020}%
  \BibitemOpen
  \bibfield  {author} {\bibinfo {author} {\bibfnamefont {S.}~\bibnamefont {Pal}}, \bibinfo {author} {\bibfnamefont {S.}~\bibnamefont {Saryal}}, \bibinfo {author} {\bibfnamefont {D.}~\bibnamefont {Segal}}, \bibinfo {author} {\bibfnamefont {T.~S.}\ \bibnamefont {Mahesh}},\ and\ \bibinfo {author} {\bibfnamefont {B.~K.}\ \bibnamefont {Agarwalla}},\ }\bibfield  {title} {\bibinfo {title} {Experimental study of the thermodynamic uncertainty relation},\ }\href {https://doi.org/10.1103/PhysRevResearch.2.022044} {\bibfield  {journal} {\bibinfo  {journal} {Phys. Rev. Res.}\ }\textbf {\bibinfo {volume} {2}},\ \bibinfo {pages} {022044} (\bibinfo {year} {2020})}\BibitemShut {NoStop}%
\bibitem [{\citenamefont {Friedman}\ \emph {et~al.}(2020)\citenamefont {Friedman}, \citenamefont {Agarwalla}, \citenamefont {Shein-Lumbroso}, \citenamefont {Tal},\ and\ \citenamefont {Segal}}]{Friedman2020}%
  \BibitemOpen
  \bibfield  {author} {\bibinfo {author} {\bibfnamefont {H.~M.}\ \bibnamefont {Friedman}}, \bibinfo {author} {\bibfnamefont {B.~K.}\ \bibnamefont {Agarwalla}}, \bibinfo {author} {\bibfnamefont {O.}~\bibnamefont {Shein-Lumbroso}}, \bibinfo {author} {\bibfnamefont {O.}~\bibnamefont {Tal}},\ and\ \bibinfo {author} {\bibfnamefont {D.}~\bibnamefont {Segal}},\ }\bibfield  {title} {\bibinfo {title} {Thermodynamic uncertainty relation in atomic-scale quantum conductors},\ }\href {https://doi.org/10.1103/PhysRevB.101.195423} {\bibfield  {journal} {\bibinfo  {journal} {Phys. Rev. B}\ }\textbf {\bibinfo {volume} {101}},\ \bibinfo {pages} {195423} (\bibinfo {year} {2020})}\BibitemShut {NoStop}%
\bibitem [{\citenamefont {Taddei}\ and\ \citenamefont {Fazio}(2023)}]{Taddei_PRB_2023}%
  \BibitemOpen
  \bibfield  {author} {\bibinfo {author} {\bibfnamefont {F.}~\bibnamefont {Taddei}}\ and\ \bibinfo {author} {\bibfnamefont {R.}~\bibnamefont {Fazio}},\ }\bibfield  {title} {\bibinfo {title} {Thermodynamic uncertainty relations for systems with broken time reversal symmetry: The case of superconducting hybrid systems},\ }\href {https://doi.org/10.1103/PhysRevB.108.115422} {\bibfield  {journal} {\bibinfo  {journal} {Phys. Rev. B}\ }\textbf {\bibinfo {volume} {108}},\ \bibinfo {pages} {115422} (\bibinfo {year} {2023})}\BibitemShut {NoStop}%
\bibitem [{\citenamefont {Brandner}\ \emph {et~al.}(2018)\citenamefont {Brandner}, \citenamefont {Hanazato},\ and\ \citenamefont {Saito}}]{Saito_PRL_2018}%
  \BibitemOpen
  \bibfield  {author} {\bibinfo {author} {\bibfnamefont {K.}~\bibnamefont {Brandner}}, \bibinfo {author} {\bibfnamefont {T.}~\bibnamefont {Hanazato}},\ and\ \bibinfo {author} {\bibfnamefont {K.}~\bibnamefont {Saito}},\ }\bibfield  {title} {\bibinfo {title} {Thermodynamic bounds on precision in ballistic multiterminal transport},\ }\href {https://doi.org/10.1103/PhysRevLett.120.090601} {\bibfield  {journal} {\bibinfo  {journal} {Phys. Rev. Lett.}\ }\textbf {\bibinfo {volume} {120}},\ \bibinfo {pages} {090601} (\bibinfo {year} {2018})}\BibitemShut {NoStop}%
\bibitem [{\citenamefont {Ptaszy\ifmmode~\acute{n}\else \'{n}\fi{}ski}(2018)}]{Krzysztof_PRB_2018}%
  \BibitemOpen
  \bibfield  {author} {\bibinfo {author} {\bibfnamefont {K.}~\bibnamefont {Ptaszy\ifmmode~\acute{n}\else \'{n}\fi{}ski}},\ }\bibfield  {title} {\bibinfo {title} {Coherence-enhanced constancy of a quantum thermoelectric generator},\ }\href {https://doi.org/10.1103/PhysRevB.98.085425} {\bibfield  {journal} {\bibinfo  {journal} {Phys. Rev. B}\ }\textbf {\bibinfo {volume} {98}},\ \bibinfo {pages} {085425} (\bibinfo {year} {2018})}\BibitemShut {NoStop}%
\bibitem [{\citenamefont {Agarwalla}\ and\ \citenamefont {Segal}(2018)}]{Bijay_PRB_2018}%
  \BibitemOpen
  \bibfield  {author} {\bibinfo {author} {\bibfnamefont {B.~K.}\ \bibnamefont {Agarwalla}}\ and\ \bibinfo {author} {\bibfnamefont {D.}~\bibnamefont {Segal}},\ }\bibfield  {title} {\bibinfo {title} {Assessing the validity of the thermodynamic uncertainty relation in quantum systems},\ }\href {https://doi.org/10.1103/PhysRevB.98.155438} {\bibfield  {journal} {\bibinfo  {journal} {Phys. Rev. B}\ }\textbf {\bibinfo {volume} {98}},\ \bibinfo {pages} {155438} (\bibinfo {year} {2018})}\BibitemShut {NoStop}%
\bibitem [{\citenamefont {Cangemi}\ \emph {et~al.}(2020)\citenamefont {Cangemi}, \citenamefont {Cataudella}, \citenamefont {Benenti}, \citenamefont {Sassetti},\ and\ \citenamefont {De~Filippis}}]{Cangemi_PRB_2020}%
  \BibitemOpen
  \bibfield  {author} {\bibinfo {author} {\bibfnamefont {L.~M.}\ \bibnamefont {Cangemi}}, \bibinfo {author} {\bibfnamefont {V.}~\bibnamefont {Cataudella}}, \bibinfo {author} {\bibfnamefont {G.}~\bibnamefont {Benenti}}, \bibinfo {author} {\bibfnamefont {M.}~\bibnamefont {Sassetti}},\ and\ \bibinfo {author} {\bibfnamefont {G.}~\bibnamefont {De~Filippis}},\ }\bibfield  {title} {\bibinfo {title} {Violation of thermodynamics uncertainty relations in a periodically driven work-to-work converter from weak to strong dissipation},\ }\href {https://doi.org/10.1103/PhysRevB.102.165418} {\bibfield  {journal} {\bibinfo  {journal} {Phys. Rev. B}\ }\textbf {\bibinfo {volume} {102}},\ \bibinfo {pages} {165418} (\bibinfo {year} {2020})}\BibitemShut {NoStop}%
\bibitem [{\citenamefont {Rignon-Bret}\ \emph {et~al.}(2021)\citenamefont {Rignon-Bret}, \citenamefont {Guarnieri}, \citenamefont {Goold},\ and\ \citenamefont {Mitchison}}]{Mitchison_PRE_2021}%
  \BibitemOpen
  \bibfield  {author} {\bibinfo {author} {\bibfnamefont {A.}~\bibnamefont {Rignon-Bret}}, \bibinfo {author} {\bibfnamefont {G.}~\bibnamefont {Guarnieri}}, \bibinfo {author} {\bibfnamefont {J.}~\bibnamefont {Goold}},\ and\ \bibinfo {author} {\bibfnamefont {M.~T.}\ \bibnamefont {Mitchison}},\ }\bibfield  {title} {\bibinfo {title} {Thermodynamics of precision in quantum nanomachines},\ }\href {https://doi.org/10.1103/PhysRevE.103.012133} {\bibfield  {journal} {\bibinfo  {journal} {Phys. Rev. E}\ }\textbf {\bibinfo {volume} {103}},\ \bibinfo {pages} {012133} (\bibinfo {year} {2021})}\BibitemShut {NoStop}%
\bibitem [{\citenamefont {Kalaee}\ \emph {et~al.}(2021)\citenamefont {Kalaee}, \citenamefont {Wacker},\ and\ \citenamefont {Potts}}]{Potts_PRE_2021}%
  \BibitemOpen
  \bibfield  {author} {\bibinfo {author} {\bibfnamefont {A.~A.~S.}\ \bibnamefont {Kalaee}}, \bibinfo {author} {\bibfnamefont {A.}~\bibnamefont {Wacker}},\ and\ \bibinfo {author} {\bibfnamefont {P.~P.}\ \bibnamefont {Potts}},\ }\bibfield  {title} {\bibinfo {title} {Violating the thermodynamic uncertainty relation in the three-level maser},\ }\href {https://doi.org/10.1103/PhysRevE.104.L012103} {\bibfield  {journal} {\bibinfo  {journal} {Phys. Rev. E}\ }\textbf {\bibinfo {volume} {104}},\ \bibinfo {pages} {L012103} (\bibinfo {year} {2021})}\BibitemShut {NoStop}%
\bibitem [{\citenamefont {Van~Vu}\ and\ \citenamefont {Saito}(2022)}]{Keiji_PRL_2022}%
  \BibitemOpen
  \bibfield  {author} {\bibinfo {author} {\bibfnamefont {T.}~\bibnamefont {Van~Vu}}\ and\ \bibinfo {author} {\bibfnamefont {K.}~\bibnamefont {Saito}},\ }\bibfield  {title} {\bibinfo {title} {Thermodynamics of precision in markovian open quantum dynamics},\ }\href {https://doi.org/10.1103/PhysRevLett.128.140602} {\bibfield  {journal} {\bibinfo  {journal} {Phys. Rev. Lett.}\ }\textbf {\bibinfo {volume} {128}},\ \bibinfo {pages} {140602} (\bibinfo {year} {2022})}\BibitemShut {NoStop}%
\bibitem [{\citenamefont {Mohan}\ \emph {et~al.}(2025)\citenamefont {Mohan}, \citenamefont {Gangwar}, \citenamefont {Pandit}, \citenamefont {Bera}, \citenamefont {Lewenstein},\ and\ \citenamefont {Bera}}]{mohan2024}%
  \BibitemOpen
  \bibfield  {author} {\bibinfo {author} {\bibfnamefont {B.}~\bibnamefont {Mohan}}, \bibinfo {author} {\bibfnamefont {R.}~\bibnamefont {Gangwar}}, \bibinfo {author} {\bibfnamefont {T.}~\bibnamefont {Pandit}}, \bibinfo {author} {\bibfnamefont {M.~L.}\ \bibnamefont {Bera}}, \bibinfo {author} {\bibfnamefont {M.}~\bibnamefont {Lewenstein}},\ and\ \bibinfo {author} {\bibfnamefont {M.~N.}\ \bibnamefont {Bera}},\ }\bibfield  {title} {\bibinfo {title} {Coherent heat transfer leads to genuine quantum enhancement in the performances of continuous engines},\ }\href {https://doi.org/10.1103/PhysRevApplied.23.044050} {\bibfield  {journal} {\bibinfo  {journal} {Phys. Rev. Appl.}\ }\textbf {\bibinfo {volume} {23}},\ \bibinfo {pages} {044050} (\bibinfo {year} {2025})}\BibitemShut {NoStop}%
\bibitem [{\citenamefont {Saryal}\ \emph {et~al.}(2019)\citenamefont {Saryal}, \citenamefont {Friedman}, \citenamefont {Segal},\ and\ \citenamefont {Agarwalla}}]{Saryal2019}%
  \BibitemOpen
  \bibfield  {author} {\bibinfo {author} {\bibfnamefont {S.}~\bibnamefont {Saryal}}, \bibinfo {author} {\bibfnamefont {H.~M.}\ \bibnamefont {Friedman}}, \bibinfo {author} {\bibfnamefont {D.}~\bibnamefont {Segal}},\ and\ \bibinfo {author} {\bibfnamefont {B.~K.}\ \bibnamefont {Agarwalla}},\ }\bibfield  {title} {\bibinfo {title} {Thermodynamic uncertainty relation in thermal transport},\ }\href {https://doi.org/10.1103/PhysRevE.100.042101} {\bibfield  {journal} {\bibinfo  {journal} {Phys. Rev. E}\ }\textbf {\bibinfo {volume} {100}},\ \bibinfo {pages} {042101} (\bibinfo {year} {2019})}\BibitemShut {NoStop}%
\bibitem [{\citenamefont {Liu}\ and\ \citenamefont {Segal}(2019)}]{Segal_PRE_2019}%
  \BibitemOpen
  \bibfield  {author} {\bibinfo {author} {\bibfnamefont {J.}~\bibnamefont {Liu}}\ and\ \bibinfo {author} {\bibfnamefont {D.}~\bibnamefont {Segal}},\ }\bibfield  {title} {\bibinfo {title} {Thermodynamic uncertainty relation in quantum thermoelectric junctions},\ }\href {https://doi.org/10.1103/PhysRevE.99.062141} {\bibfield  {journal} {\bibinfo  {journal} {Phys. Rev. E}\ }\textbf {\bibinfo {volume} {99}},\ \bibinfo {pages} {062141} (\bibinfo {year} {2019})}\BibitemShut {NoStop}%
\bibitem [{\citenamefont {Landi}\ \emph {et~al.}(2014)\citenamefont {Landi}, \citenamefont {Novais}, \citenamefont {de~Oliveira},\ and\ \citenamefont {Karevski}}]{Landi2014}%
  \BibitemOpen
  \bibfield  {author} {\bibinfo {author} {\bibfnamefont {G.~T.}\ \bibnamefont {Landi}}, \bibinfo {author} {\bibfnamefont {E.}~\bibnamefont {Novais}}, \bibinfo {author} {\bibfnamefont {M.~J.}\ \bibnamefont {de~Oliveira}},\ and\ \bibinfo {author} {\bibfnamefont {D.}~\bibnamefont {Karevski}},\ }\bibfield  {title} {\bibinfo {title} {Flux rectification in the quantum $xxz$ chain},\ }\href {https://doi.org/10.1103/PhysRevE.90.042142} {\bibfield  {journal} {\bibinfo  {journal} {Phys. Rev. E}\ }\textbf {\bibinfo {volume} {90}},\ \bibinfo {pages} {042142} (\bibinfo {year} {2014})}\BibitemShut {NoStop}%
\bibitem [{\citenamefont {Ciccarello}\ \emph {et~al.}(2022)\citenamefont {Ciccarello}, \citenamefont {Lorenzo}, \citenamefont {Giovannetti},\ and\ \citenamefont {Palma}}]{Ciccarello2022}%
  \BibitemOpen
  \bibfield  {author} {\bibinfo {author} {\bibfnamefont {F.}~\bibnamefont {Ciccarello}}, \bibinfo {author} {\bibfnamefont {S.}~\bibnamefont {Lorenzo}}, \bibinfo {author} {\bibfnamefont {V.}~\bibnamefont {Giovannetti}},\ and\ \bibinfo {author} {\bibfnamefont {G.~M.}\ \bibnamefont {Palma}},\ }\bibfield  {title} {\bibinfo {title} {Quantum collision models: Open system dynamics from repeated interactions},\ }\href {https://doi.org/https://doi.org/10.1016/j.physrep.2022.01.001} {\bibfield  {journal} {\bibinfo  {journal} {Physics Reports}\ }\textbf {\bibinfo {volume} {954}},\ \bibinfo {pages} {1} (\bibinfo {year} {2022})}\BibitemShut {NoStop}%
\bibitem [{\citenamefont {Cusumano}(2022)}]{Cusumano2022}%
  \BibitemOpen
  \bibfield  {author} {\bibinfo {author} {\bibfnamefont {S.}~\bibnamefont {Cusumano}},\ }\bibfield  {title} {\bibinfo {title} {Quantum collision models: A beginner guide},\ }\href {https://doi.org/10.3390/e24091258} {\bibfield  {journal} {\bibinfo  {journal} {Entropy}\ }\textbf {\bibinfo {volume} {24}},\ \bibinfo {pages} {1258} (\bibinfo {year} {2022})}\BibitemShut {NoStop}%
\bibitem [{\citenamefont {Ciccarello}\ \emph {et~al.}(2013)\citenamefont {Ciccarello}, \citenamefont {Palma},\ and\ \citenamefont {Giovannetti}}]{Ciccarello2013}%
  \BibitemOpen
  \bibfield  {author} {\bibinfo {author} {\bibfnamefont {F.}~\bibnamefont {Ciccarello}}, \bibinfo {author} {\bibfnamefont {G.~M.}\ \bibnamefont {Palma}},\ and\ \bibinfo {author} {\bibfnamefont {V.}~\bibnamefont {Giovannetti}},\ }\bibfield  {title} {\bibinfo {title} {Collision-model-based approach to non-markovian quantum dynamics},\ }\href {https://doi.org/10.1103/PhysRevA.87.040103} {\bibfield  {journal} {\bibinfo  {journal} {Phys. Rev. A}\ }\textbf {\bibinfo {volume} {87}},\ \bibinfo {pages} {040103} (\bibinfo {year} {2013})}\BibitemShut {NoStop}%
\bibitem [{\citenamefont {Bernardes}\ \emph {et~al.}(2014)\citenamefont {Bernardes}, \citenamefont {Carvalho}, \citenamefont {Monken},\ and\ \citenamefont {Santos}}]{Bernardes2014}%
  \BibitemOpen
  \bibfield  {author} {\bibinfo {author} {\bibfnamefont {N.~K.}\ \bibnamefont {Bernardes}}, \bibinfo {author} {\bibfnamefont {A.~R.~R.}\ \bibnamefont {Carvalho}}, \bibinfo {author} {\bibfnamefont {C.~H.}\ \bibnamefont {Monken}},\ and\ \bibinfo {author} {\bibfnamefont {M.~F.}\ \bibnamefont {Santos}},\ }\bibfield  {title} {\bibinfo {title} {Environmental correlations and markovian to non-markovian transitions in collisional models},\ }\href {https://doi.org/10.1103/PhysRevA.90.032111} {\bibfield  {journal} {\bibinfo  {journal} {Phys. Rev. A}\ }\textbf {\bibinfo {volume} {90}},\ \bibinfo {pages} {032111} (\bibinfo {year} {2014})}\BibitemShut {NoStop}%
\bibitem [{\citenamefont {Kretschmer}\ \emph {et~al.}(2016)\citenamefont {Kretschmer}, \citenamefont {Luoma},\ and\ \citenamefont {Strunz}}]{Kretschmer_2016}%
  \BibitemOpen
  \bibfield  {author} {\bibinfo {author} {\bibfnamefont {S.}~\bibnamefont {Kretschmer}}, \bibinfo {author} {\bibfnamefont {K.}~\bibnamefont {Luoma}},\ and\ \bibinfo {author} {\bibfnamefont {W.~T.}\ \bibnamefont {Strunz}},\ }\bibfield  {title} {\bibinfo {title} {Collision model for non-markovian quantum dynamics},\ }\href {https://doi.org/10.1103/PhysRevA.94.012106} {\bibfield  {journal} {\bibinfo  {journal} {Phys. Rev. A}\ }\textbf {\bibinfo {volume} {94}},\ \bibinfo {pages} {012106} (\bibinfo {year} {2016})}\BibitemShut {NoStop}%
\bibitem [{\citenamefont {Lorenzo}\ \emph {et~al.}(2017)\citenamefont {Lorenzo}, \citenamefont {Ciccarello}, \citenamefont {Palma},\ and\ \citenamefont {Vacchini}}]{Lorenzo2017}%
  \BibitemOpen
  \bibfield  {author} {\bibinfo {author} {\bibfnamefont {S.}~\bibnamefont {Lorenzo}}, \bibinfo {author} {\bibfnamefont {F.}~\bibnamefont {Ciccarello}}, \bibinfo {author} {\bibfnamefont {G.~M.}\ \bibnamefont {Palma}},\ and\ \bibinfo {author} {\bibfnamefont {B.}~\bibnamefont {Vacchini}},\ }\bibfield  {title} {\bibinfo {title} {Quantum non-{M}arkovian piecewise dynamics from collision models},\ }\href {https://doi.org/10.1142/S123016121740011X} {\bibfield  {journal} {\bibinfo  {journal} {Open Syst. Inf. Dyn.}\ }\textbf {\bibinfo {volume} {24}},\ \bibinfo {pages} {1740011} (\bibinfo {year} {2017})}\BibitemShut {NoStop}%
\bibitem [{\citenamefont {Man}\ \emph {et~al.}(2019)\citenamefont {Man}, \citenamefont {Xia},\ and\ \citenamefont {Lo~Franco}}]{Man2019}%
  \BibitemOpen
  \bibfield  {author} {\bibinfo {author} {\bibfnamefont {Z.-X.}\ \bibnamefont {Man}}, \bibinfo {author} {\bibfnamefont {Y.-J.}\ \bibnamefont {Xia}},\ and\ \bibinfo {author} {\bibfnamefont {R.}~\bibnamefont {Lo~Franco}},\ }\bibfield  {title} {\bibinfo {title} {Validity of the landauer principle and quantum memory effects via collisional models},\ }\href {https://doi.org/10.1103/PhysRevA.99.042106} {\bibfield  {journal} {\bibinfo  {journal} {Phys. Rev. A}\ }\textbf {\bibinfo {volume} {99}},\ \bibinfo {pages} {042106} (\bibinfo {year} {2019})}\BibitemShut {NoStop}%
\bibitem [{\citenamefont {\c{S}enya\c{s}a}\ \emph {et~al.}(2022)\citenamefont {\c{S}enya\c{s}a}, \citenamefont {Kesgin}, \citenamefont {Karpat},\ and\ \citenamefont {\c{C}akmak}}]{Baris_Entropy_2022}%
  \BibitemOpen
  \bibfield  {author} {\bibinfo {author} {\bibfnamefont {H.~T.}\ \bibnamefont {\c{S}enya\c{s}a}}, \bibinfo {author} {\bibfnamefont {{\c{S}}.}~\bibnamefont {Kesgin}}, \bibinfo {author} {\bibfnamefont {G.}~\bibnamefont {Karpat}},\ and\ \bibinfo {author} {\bibfnamefont {B.}~\bibnamefont {\c{C}akmak}},\ }\bibfield  {title} {\bibinfo {title} {Entropy production in non-markovian collision models: Information backflow vs. system-environment correlations},\ }\href {https://www.mdpi.com/1099-4300/24/6/824} {\bibfield  {journal} {\bibinfo  {journal} {Entropy}\ }\textbf {\bibinfo {volume} {24}},\ \bibinfo {pages} {824} (\bibinfo {year} {2022})}\BibitemShut {NoStop}%
\bibitem [{\citenamefont {Saha}\ \emph {et~al.}(2024)\citenamefont {Saha}, \citenamefont {Das},\ and\ \citenamefont {Ghosh}}]{Tanmay_NJP_2024}%
  \BibitemOpen
  \bibfield  {author} {\bibinfo {author} {\bibfnamefont {T.}~\bibnamefont {Saha}}, \bibinfo {author} {\bibfnamefont {A.}~\bibnamefont {Das}},\ and\ \bibinfo {author} {\bibfnamefont {S.}~\bibnamefont {Ghosh}},\ }\bibfield  {title} {\bibinfo {title} {Quantum homogenization in non-{M}arkovian collisional model},\ }\href {https://doi.org/10.1088/1367-2630/ad212f} {\bibfield  {journal} {\bibinfo  {journal} {New Journal of Physics}\ }\textbf {\bibinfo {volume} {26}},\ \bibinfo {pages} {023011} (\bibinfo {year} {2024})}\BibitemShut {NoStop}%
\bibitem [{\citenamefont {Levy}\ and\ \citenamefont {Kosloff}(2014)}]{Levy_EPL_2014}%
  \BibitemOpen
  \bibfield  {author} {\bibinfo {author} {\bibfnamefont {A.}~\bibnamefont {Levy}}\ and\ \bibinfo {author} {\bibfnamefont {R.}~\bibnamefont {Kosloff}},\ }\bibfield  {title} {\bibinfo {title} {The local approach to quantum transport may violate the second law of thermodynamics},\ }\href {https://doi.org/10.1209/0295-5075/107/20004} {\bibfield  {journal} {\bibinfo  {journal} {Europhysics Letters}\ }\textbf {\bibinfo {volume} {107}},\ \bibinfo {pages} {20004} (\bibinfo {year} {2014})}\BibitemShut {NoStop}%
\bibitem [{\citenamefont {Chisholm}\ \emph {et~al.}(2021)\citenamefont {Chisholm}, \citenamefont {García-Pérez}, \citenamefont {Rossi}, \citenamefont {Palma},\ and\ \citenamefont {Maniscalco}}]{Chisholm_NJP_2021}%
  \BibitemOpen
  \bibfield  {author} {\bibinfo {author} {\bibfnamefont {D.~A.}\ \bibnamefont {Chisholm}}, \bibinfo {author} {\bibfnamefont {G.}~\bibnamefont {García-Pérez}}, \bibinfo {author} {\bibfnamefont {M.~A.~C.}\ \bibnamefont {Rossi}}, \bibinfo {author} {\bibfnamefont {G.~M.}\ \bibnamefont {Palma}},\ and\ \bibinfo {author} {\bibfnamefont {S.}~\bibnamefont {Maniscalco}},\ }\bibfield  {title} {\bibinfo {title} {Stochastic collision model approach to transport phenomena in quantum networks},\ }\href {https://doi.org/10.1088/1367-2630/abd57d} {\bibfield  {journal} {\bibinfo  {journal} {New Journal of Physics}\ }\textbf {\bibinfo {volume} {23}},\ \bibinfo {pages} {033031} (\bibinfo {year} {2021})}\BibitemShut {NoStop}%
\bibitem [{\citenamefont {Tian}\ \emph {et~al.}(2021)\citenamefont {Tian}, \citenamefont {Zou}, \citenamefont {Li}, \citenamefont {Li},\ and\ \citenamefont {Shao}}]{Tian_Entropy_2021}%
  \BibitemOpen
  \bibfield  {author} {\bibinfo {author} {\bibfnamefont {F.}~\bibnamefont {Tian}}, \bibinfo {author} {\bibfnamefont {J.}~\bibnamefont {Zou}}, \bibinfo {author} {\bibfnamefont {L.}~\bibnamefont {Li}}, \bibinfo {author} {\bibfnamefont {H.}~\bibnamefont {Li}},\ and\ \bibinfo {author} {\bibfnamefont {B.}~\bibnamefont {Shao}},\ }\bibfield  {title} {\bibinfo {title} {Effect of inter-system coupling on heat transport in a microscopic collision model},\ }\href {https://www.mdpi.com/1099-4300/23/4/471} {\bibfield  {journal} {\bibinfo  {journal} {Entropy}\ }\textbf {\bibinfo {volume} {23}},\ \bibinfo {pages} {471} (\bibinfo {year} {2021})}\BibitemShut {NoStop}%
\bibitem [{\citenamefont {Karpat}\ \emph {et~al.}(2019)\citenamefont {Karpat}, \citenamefont {Yal\ifmmode \mbox{\c{c}}\else \c{c}\fi{}\ifmmode \imath \else~\i \fi{}nkaya},\ and\ \citenamefont {\ifmmode~\mbox{\c{C}}\else \c{C}\fi{}akmak}}]{Karpat_PRA_2019}%
  \BibitemOpen
  \bibfield  {author} {\bibinfo {author} {\bibfnamefont {G.}~\bibnamefont {Karpat}}, \bibinfo {author} {\bibfnamefont {i.~d.~I.}\ \bibnamefont {Yal\ifmmode \mbox{\c{c}}\else \c{c}\fi{}\ifmmode \imath \else~\i \fi{}nkaya}},\ and\ \bibinfo {author} {\bibfnamefont {B.}~\bibnamefont {\ifmmode~\mbox{\c{C}}\else \c{C}\fi{}akmak}},\ }\bibfield  {title} {\bibinfo {title} {Quantum synchronization in a collision model},\ }\href {https://doi.org/10.1103/PhysRevA.100.012133} {\bibfield  {journal} {\bibinfo  {journal} {Phys. Rev. A}\ }\textbf {\bibinfo {volume} {100}},\ \bibinfo {pages} {012133} (\bibinfo {year} {2019})}\BibitemShut {NoStop}%
\bibitem [{\citenamefont {{\c{C}}akmak}\ \emph {et~al.}(2019)\citenamefont {{\c{C}}akmak}, \citenamefont {Campbell}, \citenamefont {Vacchini}, \citenamefont {M{\"u}stecapl{\i}o{\u{g}}lu},\ and\ \citenamefont {Paternostro}}]{Paternostro_PRA_2019}%
  \BibitemOpen
  \bibfield  {author} {\bibinfo {author} {\bibfnamefont {B.}~\bibnamefont {{\c{C}}akmak}}, \bibinfo {author} {\bibfnamefont {S.}~\bibnamefont {Campbell}}, \bibinfo {author} {\bibfnamefont {B.}~\bibnamefont {Vacchini}}, \bibinfo {author} {\bibfnamefont {{\"O}.~E.}\ \bibnamefont {M{\"u}stecapl{\i}o{\u{g}}lu}},\ and\ \bibinfo {author} {\bibfnamefont {M.}~\bibnamefont {Paternostro}},\ }\bibfield  {title} {\bibinfo {title} {Robust multipartite entanglement generation via a collision model},\ }\href {https://doi.org/10.1103/PhysRevA.99.012319} {\bibfield  {journal} {\bibinfo  {journal} {Physical Review A}\ }\textbf {\bibinfo {volume} {99}},\ \bibinfo {pages} {012319} (\bibinfo {year} {2019})}\BibitemShut {NoStop}%
\bibitem [{\citenamefont {Molitor}\ and\ \citenamefont {Landi}(2020)}]{Molitor_PRA_2020}%
  \BibitemOpen
  \bibfield  {author} {\bibinfo {author} {\bibfnamefont {O.~A.~D.}\ \bibnamefont {Molitor}}\ and\ \bibinfo {author} {\bibfnamefont {G.~T.}\ \bibnamefont {Landi}},\ }\bibfield  {title} {\bibinfo {title} {Stroboscopic two-stroke quantum heat engines},\ }\href {https://doi.org/10.1103/PhysRevA.102.042217} {\bibfield  {journal} {\bibinfo  {journal} {Phys. Rev. A}\ }\textbf {\bibinfo {volume} {102}},\ \bibinfo {pages} {042217} (\bibinfo {year} {2020})}\BibitemShut {NoStop}%
\bibitem [{\citenamefont {Taranto}\ \emph {et~al.}(2020)\citenamefont {Taranto}, \citenamefont {Bakhshinezhad}, \citenamefont {Sch\"uttelkopf}, \citenamefont {Clivaz},\ and\ \citenamefont {Huber}}]{Taranto_PRAP_2020}%
  \BibitemOpen
  \bibfield  {author} {\bibinfo {author} {\bibfnamefont {P.}~\bibnamefont {Taranto}}, \bibinfo {author} {\bibfnamefont {F.}~\bibnamefont {Bakhshinezhad}}, \bibinfo {author} {\bibfnamefont {P.}~\bibnamefont {Sch\"uttelkopf}}, \bibinfo {author} {\bibfnamefont {F.}~\bibnamefont {Clivaz}},\ and\ \bibinfo {author} {\bibfnamefont {M.}~\bibnamefont {Huber}},\ }\bibfield  {title} {\bibinfo {title} {Exponential improvement for quantum cooling through finite-memory effects},\ }\href {https://doi.org/10.1103/PhysRevApplied.14.054005} {\bibfield  {journal} {\bibinfo  {journal} {Phys. Rev. Appl.}\ }\textbf {\bibinfo {volume} {14}},\ \bibinfo {pages} {054005} (\bibinfo {year} {2020})}\BibitemShut {NoStop}%
\bibitem [{\citenamefont {Abah}\ and\ \citenamefont {Paternostro}(2020)}]{Abah_JPC_2020}%
  \BibitemOpen
  \bibfield  {author} {\bibinfo {author} {\bibfnamefont {O.}~\bibnamefont {Abah}}\ and\ \bibinfo {author} {\bibfnamefont {M.}~\bibnamefont {Paternostro}},\ }\bibfield  {title} {\bibinfo {title} {Implications of non-{M}arkovian dynamics on information-driven engine},\ }\href {https://doi.org/10.1088/2399-6528/abaf99} {\bibfield  {journal} {\bibinfo  {journal} {Journal of Physics Communications}\ }\textbf {\bibinfo {volume} {4}},\ \bibinfo {pages} {085016} (\bibinfo {year} {2020})}\BibitemShut {NoStop}%
\bibitem [{\citenamefont {Seah}\ \emph {et~al.}(2021)\citenamefont {Seah}, \citenamefont {Perarnau-Llobet}, \citenamefont {Haack}, \citenamefont {Brunner},\ and\ \citenamefont {Nimmrichter}}]{Stella_PRL_2021}%
  \BibitemOpen
  \bibfield  {author} {\bibinfo {author} {\bibfnamefont {S.}~\bibnamefont {Seah}}, \bibinfo {author} {\bibfnamefont {M.}~\bibnamefont {Perarnau-Llobet}}, \bibinfo {author} {\bibfnamefont {G.}~\bibnamefont {Haack}}, \bibinfo {author} {\bibfnamefont {N.}~\bibnamefont {Brunner}},\ and\ \bibinfo {author} {\bibfnamefont {S.}~\bibnamefont {Nimmrichter}},\ }\bibfield  {title} {\bibinfo {title} {Quantum speed-up in collisional battery charging},\ }\href {https://doi.org/10.1103/PhysRevLett.127.100601} {\bibfield  {journal} {\bibinfo  {journal} {Phys. Rev. Lett.}\ }\textbf {\bibinfo {volume} {127}},\ \bibinfo {pages} {100601} (\bibinfo {year} {2021})}\BibitemShut {NoStop}%
\bibitem [{\citenamefont {Scarani}\ \emph {et~al.}(2002)\citenamefont {Scarani}, \citenamefont {Ziman}, \citenamefont {\ifmmode \check{S}\else \v{S}\fi{}telmachovi\ifmmode~\check{c}\else \v{c}\fi{}}, \citenamefont {Gisin},\ and\ \citenamefont {Bu\ifmmode~\check{z}\else \v{z}\fi{}ek}}]{Scarani_PRL_2002}%
  \BibitemOpen
  \bibfield  {author} {\bibinfo {author} {\bibfnamefont {V.}~\bibnamefont {Scarani}}, \bibinfo {author} {\bibfnamefont {M.}~\bibnamefont {Ziman}}, \bibinfo {author} {\bibfnamefont {P.}~\bibnamefont {\ifmmode \check{S}\else \v{S}\fi{}telmachovi\ifmmode~\check{c}\else \v{c}\fi{}}}, \bibinfo {author} {\bibfnamefont {N.}~\bibnamefont {Gisin}},\ and\ \bibinfo {author} {\bibfnamefont {V.}~\bibnamefont {Bu\ifmmode~\check{z}\else \v{z}\fi{}ek}},\ }\bibfield  {title} {\bibinfo {title} {Thermalizing quantum machines: Dissipation and entanglement},\ }\href {https://doi.org/10.1103/PhysRevLett.88.097905} {\bibfield  {journal} {\bibinfo  {journal} {Phys. Rev. Lett.}\ }\textbf {\bibinfo {volume} {88}},\ \bibinfo {pages} {097905} (\bibinfo {year} {2002})}\BibitemShut {NoStop}%
\bibitem [{\citenamefont {Bruneau}\ \emph {et~al.}(2014)\citenamefont {Bruneau}, \citenamefont {Joye},\ and\ \citenamefont {Merkli}}]{Bruneau_JMP_2014}%
  \BibitemOpen
  \bibfield  {author} {\bibinfo {author} {\bibfnamefont {L.}~\bibnamefont {Bruneau}}, \bibinfo {author} {\bibfnamefont {A.}~\bibnamefont {Joye}},\ and\ \bibinfo {author} {\bibfnamefont {M.}~\bibnamefont {Merkli}},\ }\bibfield  {title} {\bibinfo {title} {{Repeated interactions in open quantum systems}},\ }\href {https://doi.org/10.1063/1.4879240} {\bibfield  {journal} {\bibinfo  {journal} {Journal of Mathematical Physics}\ }\textbf {\bibinfo {volume} {55}},\ \bibinfo {pages} {075204} (\bibinfo {year} {2014})}\BibitemShut {NoStop}%
\bibitem [{\citenamefont {Grimmer}\ \emph {et~al.}(2016)\citenamefont {Grimmer}, \citenamefont {Layden}, \citenamefont {Mann},\ and\ \citenamefont {Mart\'{\i}n-Mart\'{\i}nez}}]{Grimmer_PRA_2016}%
  \BibitemOpen
  \bibfield  {author} {\bibinfo {author} {\bibfnamefont {D.}~\bibnamefont {Grimmer}}, \bibinfo {author} {\bibfnamefont {D.}~\bibnamefont {Layden}}, \bibinfo {author} {\bibfnamefont {R.~B.}\ \bibnamefont {Mann}},\ and\ \bibinfo {author} {\bibfnamefont {E.}~\bibnamefont {Mart\'{\i}n-Mart\'{\i}nez}},\ }\bibfield  {title} {\bibinfo {title} {Open dynamics under rapid repeated interaction},\ }\href {https://doi.org/10.1103/PhysRevA.94.032126} {\bibfield  {journal} {\bibinfo  {journal} {Phys. Rev. A}\ }\textbf {\bibinfo {volume} {94}},\ \bibinfo {pages} {032126} (\bibinfo {year} {2016})}\BibitemShut {NoStop}%
\bibitem [{\citenamefont {Seah}\ \emph {et~al.}(2019)\citenamefont {Seah}, \citenamefont {Nimmrichter},\ and\ \citenamefont {Scarani}}]{Seah_PRE_2019}%
  \BibitemOpen
  \bibfield  {author} {\bibinfo {author} {\bibfnamefont {S.}~\bibnamefont {Seah}}, \bibinfo {author} {\bibfnamefont {S.}~\bibnamefont {Nimmrichter}},\ and\ \bibinfo {author} {\bibfnamefont {V.}~\bibnamefont {Scarani}},\ }\bibfield  {title} {\bibinfo {title} {Nonequilibrium dynamics with finite-time repeated interactions},\ }\href {https://doi.org/10.1103/PhysRevE.99.042103} {\bibfield  {journal} {\bibinfo  {journal} {Phys. Rev. E}\ }\textbf {\bibinfo {volume} {99}},\ \bibinfo {pages} {042103} (\bibinfo {year} {2019})}\BibitemShut {NoStop}%
\bibitem [{\citenamefont {Cattaneo}\ \emph {et~al.}(2021)\citenamefont {Cattaneo}, \citenamefont {De~Chiara}, \citenamefont {Maniscalco}, \citenamefont {Zambrini},\ and\ \citenamefont {Giorgi}}]{Cattaneo_PRL_2021}%
  \BibitemOpen
  \bibfield  {author} {\bibinfo {author} {\bibfnamefont {M.}~\bibnamefont {Cattaneo}}, \bibinfo {author} {\bibfnamefont {G.}~\bibnamefont {De~Chiara}}, \bibinfo {author} {\bibfnamefont {S.}~\bibnamefont {Maniscalco}}, \bibinfo {author} {\bibfnamefont {R.}~\bibnamefont {Zambrini}},\ and\ \bibinfo {author} {\bibfnamefont {G.~L.}\ \bibnamefont {Giorgi}},\ }\bibfield  {title} {\bibinfo {title} {Collision models can efficiently simulate any multipartite {M}arkovian quantum dynamics},\ }\href {https://doi.org/10.1103/PhysRevLett.126.130403} {\bibfield  {journal} {\bibinfo  {journal} {Phys. Rev. Lett.}\ }\textbf {\bibinfo {volume} {126}},\ \bibinfo {pages} {130403} (\bibinfo {year} {2021})}\BibitemShut {NoStop}%
\bibitem [{\citenamefont {Strasberg}(2019)}]{Strasberg_PRL_2019}%
  \BibitemOpen
  \bibfield  {author} {\bibinfo {author} {\bibfnamefont {P.}~\bibnamefont {Strasberg}},\ }\bibfield  {title} {\bibinfo {title} {Repeated interactions and quantum stochastic thermodynamics at strong coupling},\ }\href {https://doi.org/10.1103/PhysRevLett.123.180604} {\bibfield  {journal} {\bibinfo  {journal} {Phys. Rev. Lett.}\ }\textbf {\bibinfo {volume} {123}},\ \bibinfo {pages} {180604} (\bibinfo {year} {2019})}\BibitemShut {NoStop}%
\bibitem [{\citenamefont {Bernardes}\ \emph {et~al.}(2016)\citenamefont {Bernardes}, \citenamefont {Peterson}, \citenamefont {Sarthour}, \citenamefont {Souza}, \citenamefont {Monken}, \citenamefont {Roditi}, \citenamefont {Oliveira},\ and\ \citenamefont {Santos}}]{Bernardes_SR_2016}%
  \BibitemOpen
  \bibfield  {author} {\bibinfo {author} {\bibfnamefont {N.~K.}\ \bibnamefont {Bernardes}}, \bibinfo {author} {\bibfnamefont {J.~P.~S.}\ \bibnamefont {Peterson}}, \bibinfo {author} {\bibfnamefont {R.~S.}\ \bibnamefont {Sarthour}}, \bibinfo {author} {\bibfnamefont {A.~M.}\ \bibnamefont {Souza}}, \bibinfo {author} {\bibfnamefont {C.~H.}\ \bibnamefont {Monken}}, \bibinfo {author} {\bibfnamefont {I.}~\bibnamefont {Roditi}}, \bibinfo {author} {\bibfnamefont {I.~S.}\ \bibnamefont {Oliveira}},\ and\ \bibinfo {author} {\bibfnamefont {M.~F.}\ \bibnamefont {Santos}},\ }\bibfield  {title} {\bibinfo {title} {High resolution non-{M}arkovianity in {NMR}},\ }\href {https://doi.org/10.1038/srep33945} {\bibfield  {journal} {\bibinfo  {journal} {Scientific Reports}\ }\textbf {\bibinfo {volume} {6}},\ \bibinfo {pages} {33945} (\bibinfo {year} {2016})}\BibitemShut {NoStop}%
\bibitem [{\citenamefont {Barreiro}\ \emph {et~al.}(2011)\citenamefont {Barreiro}, \citenamefont {M{\"u}ller}, \citenamefont {Schindler}, \citenamefont {Nigg}, \citenamefont {Monz}, \citenamefont {Chwalla}, \citenamefont {Hennrich}, \citenamefont {Roos}, \citenamefont {Zoller},\ and\ \citenamefont {Blatt}}]{Barreiro_Nature_2011}%
  \BibitemOpen
  \bibfield  {author} {\bibinfo {author} {\bibfnamefont {J.~T.}\ \bibnamefont {Barreiro}}, \bibinfo {author} {\bibfnamefont {M.}~\bibnamefont {M{\"u}ller}}, \bibinfo {author} {\bibfnamefont {P.}~\bibnamefont {Schindler}}, \bibinfo {author} {\bibfnamefont {D.}~\bibnamefont {Nigg}}, \bibinfo {author} {\bibfnamefont {T.}~\bibnamefont {Monz}}, \bibinfo {author} {\bibfnamefont {M.}~\bibnamefont {Chwalla}}, \bibinfo {author} {\bibfnamefont {M.}~\bibnamefont {Hennrich}}, \bibinfo {author} {\bibfnamefont {C.~F.}\ \bibnamefont {Roos}}, \bibinfo {author} {\bibfnamefont {P.}~\bibnamefont {Zoller}},\ and\ \bibinfo {author} {\bibfnamefont {R.}~\bibnamefont {Blatt}},\ }\bibfield  {title} {\bibinfo {title} {An open-system quantum simulator with trapped ions},\ }\href {https://doi.org/10.1038/nature09801} {\bibfield  {journal} {\bibinfo  {journal} {Nature}\ }\textbf {\bibinfo {volume} {470}},\ \bibinfo {pages} {486} (\bibinfo {year} {2011})}\BibitemShut {NoStop}%
\bibitem [{\citenamefont {Cuevas}\ \emph {et~al.}(2019)\citenamefont {Cuevas}, \citenamefont {Geraldi}, \citenamefont {Liorni}, \citenamefont {Bonavena}, \citenamefont {De~Pasquale}, \citenamefont {Sciarrino}, \citenamefont {Giovannetti},\ and\ \citenamefont {Mataloni}}]{Cuevas_SR_2019}%
  \BibitemOpen
  \bibfield  {author} {\bibinfo {author} {\bibfnamefont {{\'A}.}~\bibnamefont {Cuevas}}, \bibinfo {author} {\bibfnamefont {A.}~\bibnamefont {Geraldi}}, \bibinfo {author} {\bibfnamefont {C.}~\bibnamefont {Liorni}}, \bibinfo {author} {\bibfnamefont {L.~D.}\ \bibnamefont {Bonavena}}, \bibinfo {author} {\bibfnamefont {A.}~\bibnamefont {De~Pasquale}}, \bibinfo {author} {\bibfnamefont {F.}~\bibnamefont {Sciarrino}}, \bibinfo {author} {\bibfnamefont {V.}~\bibnamefont {Giovannetti}},\ and\ \bibinfo {author} {\bibfnamefont {P.}~\bibnamefont {Mataloni}},\ }\bibfield  {title} {\bibinfo {title} {All-optical implementation of collision-based evolutions of open quantum systems},\ }\href {https://doi.org/10.1038/s41598-019-39832-9} {\bibfield  {journal} {\bibinfo  {journal} {Scientific Reports}\ }\textbf {\bibinfo {volume} {9}},\ \bibinfo {pages} {3205} (\bibinfo {year} {2019})}\BibitemShut {NoStop}%
\bibitem [{\citenamefont {Garc{\'i}a-P{\'e}rez}\ \emph {et~al.}(2020)\citenamefont {Garc{\'i}a-P{\'e}rez}, \citenamefont {Rossi},\ and\ \citenamefont {Maniscalco}}]{IBM_NPJ_2020}%
  \BibitemOpen
  \bibfield  {author} {\bibinfo {author} {\bibfnamefont {G.}~\bibnamefont {Garc{\'i}a-P{\'e}rez}}, \bibinfo {author} {\bibfnamefont {M.~A.~C.}\ \bibnamefont {Rossi}},\ and\ \bibinfo {author} {\bibfnamefont {S.}~\bibnamefont {Maniscalco}},\ }\bibfield  {title} {\bibinfo {title} {{IBM} {Q} experience as a versatile experimental testbed for simulating open quantum systems},\ }\href {https://doi.org/10.1038/s41534-019-0235-y} {\bibfield  {journal} {\bibinfo  {journal} {npj Quantum Information}\ }\textbf {\bibinfo {volume} {6}},\ \bibinfo {pages} {1} (\bibinfo {year} {2020})}\BibitemShut {NoStop}%
\bibitem [{\citenamefont {Burger}\ \emph {et~al.}(2022)\citenamefont {Burger}, \citenamefont {Kwek},\ and\ \citenamefont {Poletti}}]{Burger_Entropy_2022}%
  \BibitemOpen
  \bibfield  {author} {\bibinfo {author} {\bibfnamefont {A.}~\bibnamefont {Burger}}, \bibinfo {author} {\bibfnamefont {L.~C.}\ \bibnamefont {Kwek}},\ and\ \bibinfo {author} {\bibfnamefont {D.}~\bibnamefont {Poletti}},\ }\bibfield  {title} {\bibinfo {title} {Digital quantum simulation of the spin-boson model under {M}arkovian open-system dynamics},\ }\href {https://doi.org/10.3390/e24121766} {\bibfield  {journal} {\bibinfo  {journal} {Entropy}\ }\textbf {\bibinfo {volume} {24}},\ \bibinfo {pages} {1766} (\bibinfo {year} {2022})}\BibitemShut {NoStop}%
\bibitem [{\citenamefont {Cattaneo}\ \emph {et~al.}(2023)\citenamefont {Cattaneo}, \citenamefont {Rossi}, \citenamefont {Garc\'{\i}a-P\'erez}, \citenamefont {Zambrini},\ and\ \citenamefont {Maniscalco}}]{Sabrina_PRXQ_2023}%
  \BibitemOpen
  \bibfield  {author} {\bibinfo {author} {\bibfnamefont {M.}~\bibnamefont {Cattaneo}}, \bibinfo {author} {\bibfnamefont {M.~A.~C.}\ \bibnamefont {Rossi}}, \bibinfo {author} {\bibfnamefont {G.}~\bibnamefont {Garc\'{\i}a-P\'erez}}, \bibinfo {author} {\bibfnamefont {R.}~\bibnamefont {Zambrini}},\ and\ \bibinfo {author} {\bibfnamefont {S.}~\bibnamefont {Maniscalco}},\ }\bibfield  {title} {\bibinfo {title} {Quantum simulation of dissipative collective effects on noisy quantum computers},\ }\href {https://doi.org/10.1103/PRXQuantum.4.010324} {\bibfield  {journal} {\bibinfo  {journal} {PRX Quantum}\ }\textbf {\bibinfo {volume} {4}},\ \bibinfo {pages} {010324} (\bibinfo {year} {2023})}\BibitemShut {NoStop}%
\bibitem [{\citenamefont {Erbanni}\ \emph {et~al.}(2023)\citenamefont {Erbanni}, \citenamefont {Xu}, \citenamefont {Demarie},\ and\ \citenamefont {Poletti}}]{Erbanni_PRA_2023}%
  \BibitemOpen
  \bibfield  {author} {\bibinfo {author} {\bibfnamefont {R.}~\bibnamefont {Erbanni}}, \bibinfo {author} {\bibfnamefont {X.}~\bibnamefont {Xu}}, \bibinfo {author} {\bibfnamefont {T.~F.}\ \bibnamefont {Demarie}},\ and\ \bibinfo {author} {\bibfnamefont {D.}~\bibnamefont {Poletti}},\ }\bibfield  {title} {\bibinfo {title} {Simulating quantum transport via collisional models on a digital quantum computer},\ }\href {https://doi.org/10.1103/PhysRevA.108.032619} {\bibfield  {journal} {\bibinfo  {journal} {Phys. Rev. A}\ }\textbf {\bibinfo {volume} {108}},\ \bibinfo {pages} {032619} (\bibinfo {year} {2023})}\BibitemShut {NoStop}%
\bibitem [{\citenamefont {Mi}\ \emph {et~al.}(2024)\citenamefont {Mi}, \citenamefont {Michailidis}, \citenamefont {Shabani}, \citenamefont {Miao}, \citenamefont {Klimov}, \citenamefont {Lloyd}, \citenamefont {Rosenberg}, \citenamefont {Acharya}, \citenamefont {Aleiner}, \citenamefont {\text{Andersen}} \emph {et~al.}}]{X_Science_2024}%
  \BibitemOpen
  \bibfield  {author} {\bibinfo {author} {\bibfnamefont {X.}~\bibnamefont {Mi}}, \bibinfo {author} {\bibfnamefont {A.~A.}\ \bibnamefont {Michailidis}}, \bibinfo {author} {\bibfnamefont {S.}~\bibnamefont {Shabani}}, \bibinfo {author} {\bibfnamefont {K.~C.}\ \bibnamefont {Miao}}, \bibinfo {author} {\bibfnamefont {P.~V.}\ \bibnamefont {Klimov}}, \bibinfo {author} {\bibfnamefont {J.}~\bibnamefont {Lloyd}}, \bibinfo {author} {\bibfnamefont {E.}~\bibnamefont {Rosenberg}}, \bibinfo {author} {\bibfnamefont {R.}~\bibnamefont {Acharya}}, \bibinfo {author} {\bibfnamefont {I.}~\bibnamefont {Aleiner}}, \bibinfo {author} {\bibfnamefont {T.~I.}\ \bibnamefont {\text{Andersen}}}, \emph {et~al.},\ }\bibfield  {title} {\bibinfo {title} {Stable quantum-correlated many-body states through engineered dissipation},\ }\href {https://doi.org/10.1126/science.adh9932} {\bibfield  {journal} {\bibinfo  {journal} {Science}\ }\textbf {\bibinfo {volume} {383}},\ \bibinfo {pages} {1332} (\bibinfo {year} {2024})}\BibitemShut {NoStop}%
\bibitem [{\citenamefont {Karevski}\ and\ \citenamefont {Platini}(2009)}]{Dragi_PRL_2009}%
  \BibitemOpen
  \bibfield  {author} {\bibinfo {author} {\bibfnamefont {D.}~\bibnamefont {Karevski}}\ and\ \bibinfo {author} {\bibfnamefont {T.}~\bibnamefont {Platini}},\ }\bibfield  {title} {\bibinfo {title} {Quantum nonequilibrium steady states induced by repeated interactions},\ }\href {https://doi.org/10.1103/PhysRevLett.102.207207} {\bibfield  {journal} {\bibinfo  {journal} {Phys. Rev. Lett.}\ }\textbf {\bibinfo {volume} {102}},\ \bibinfo {pages} {207207} (\bibinfo {year} {2009})}\BibitemShut {NoStop}%
\bibitem [{\citenamefont {Guarnieri}\ \emph {et~al.}(2020)\citenamefont {Guarnieri}, \citenamefont {Morrone}, \citenamefont {Çakmak}, \citenamefont {Plastina},\ and\ \citenamefont {Campbell}}]{Campbell_PLA_2020}%
  \BibitemOpen
  \bibfield  {author} {\bibinfo {author} {\bibfnamefont {G.}~\bibnamefont {Guarnieri}}, \bibinfo {author} {\bibfnamefont {D.}~\bibnamefont {Morrone}}, \bibinfo {author} {\bibfnamefont {B.}~\bibnamefont {Çakmak}}, \bibinfo {author} {\bibfnamefont {F.}~\bibnamefont {Plastina}},\ and\ \bibinfo {author} {\bibfnamefont {S.}~\bibnamefont {Campbell}},\ }\bibfield  {title} {\bibinfo {title} {Non-equilibrium steady-states of memoryless quantum collision models},\ }\href {https://doi.org/https://doi.org/10.1016/j.physleta.2020.126576} {\bibfield  {journal} {\bibinfo  {journal} {Physics Letters A}\ }\textbf {\bibinfo {volume} {384}},\ \bibinfo {pages} {126576} (\bibinfo {year} {2020})}\BibitemShut {NoStop}%
\bibitem [{\citenamefont {Breuer}\ \emph {et~al.}(2009)\citenamefont {Breuer}, \citenamefont {Laine},\ and\ \citenamefont {Piilo}}]{BLP_PRL_2009}%
  \BibitemOpen
  \bibfield  {author} {\bibinfo {author} {\bibfnamefont {H.-P.}\ \bibnamefont {Breuer}}, \bibinfo {author} {\bibfnamefont {E.-M.}\ \bibnamefont {Laine}},\ and\ \bibinfo {author} {\bibfnamefont {J.}~\bibnamefont {Piilo}},\ }\bibfield  {title} {\bibinfo {title} {Measure for the degree of non-{M}arkovian behavior of quantum processes in open systems},\ }\href {https://doi.org/10.1103/PhysRevLett.103.210401} {\bibfield  {journal} {\bibinfo  {journal} {Phys. Rev. Lett.}\ }\textbf {\bibinfo {volume} {103}},\ \bibinfo {pages} {210401} (\bibinfo {year} {2009})}\BibitemShut {NoStop}%
\bibitem [{\citenamefont {Laine}\ \emph {et~al.}(2010)\citenamefont {Laine}, \citenamefont {Piilo},\ and\ \citenamefont {Breuer}}]{BLP_PRA_2010}%
  \BibitemOpen
  \bibfield  {author} {\bibinfo {author} {\bibfnamefont {E.-M.}\ \bibnamefont {Laine}}, \bibinfo {author} {\bibfnamefont {J.}~\bibnamefont {Piilo}},\ and\ \bibinfo {author} {\bibfnamefont {H.-P.}\ \bibnamefont {Breuer}},\ }\bibfield  {title} {\bibinfo {title} {Measure for the non-{M}arkovianity of quantum processes},\ }\href {https://doi.org/10.1103/PhysRevA.81.062115} {\bibfield  {journal} {\bibinfo  {journal} {Phys. Rev. A}\ }\textbf {\bibinfo {volume} {81}},\ \bibinfo {pages} {062115} (\bibinfo {year} {2010})}\BibitemShut {NoStop}%
\bibitem [{\citenamefont {Rodrigues}\ \emph {et~al.}(2019)\citenamefont {Rodrigues}, \citenamefont {De~Chiara}, \citenamefont {Paternostro},\ and\ \citenamefont {Landi}}]{Rodrigues_PRL_2019}%
  \BibitemOpen
  \bibfield  {author} {\bibinfo {author} {\bibfnamefont {F.~L.~S.}\ \bibnamefont {Rodrigues}}, \bibinfo {author} {\bibfnamefont {G.}~\bibnamefont {De~Chiara}}, \bibinfo {author} {\bibfnamefont {M.}~\bibnamefont {Paternostro}},\ and\ \bibinfo {author} {\bibfnamefont {G.~T.}\ \bibnamefont {Landi}},\ }\bibfield  {title} {\bibinfo {title} {Thermodynamics of weakly coherent collisional models},\ }\href {https://doi.org/10.1103/PhysRevLett.123.140601} {\bibfield  {journal} {\bibinfo  {journal} {Phys. Rev. Lett.}\ }\textbf {\bibinfo {volume} {123}},\ \bibinfo {pages} {140601} (\bibinfo {year} {2019})}\BibitemShut {NoStop}%
\bibitem [{\citenamefont {Gorini}\ \emph {et~al.}(1976)\citenamefont {Gorini}, \citenamefont {Kossakowski},\ and\ \citenamefont {Sudarshan}}]{Sudarshan_JMP_1976}%
  \BibitemOpen
  \bibfield  {author} {\bibinfo {author} {\bibfnamefont {V.}~\bibnamefont {Gorini}}, \bibinfo {author} {\bibfnamefont {A.}~\bibnamefont {Kossakowski}},\ and\ \bibinfo {author} {\bibfnamefont {E.~C.~G.}\ \bibnamefont {Sudarshan}},\ }\bibfield  {title} {\bibinfo {title} {{Completely positive dynamical semigroups of $N$‐level systems}},\ }\href {https://doi.org/10.1063/1.522979} {\bibfield  {journal} {\bibinfo  {journal} {Journal of Mathematical Physics}\ }\textbf {\bibinfo {volume} {17}},\ \bibinfo {pages} {821} (\bibinfo {year} {1976})}\BibitemShut {NoStop}%
\bibitem [{\citenamefont {{Lindblad}}(1976)}]{Lindblad_CMP_1976}%
  \BibitemOpen
  \bibfield  {author} {\bibinfo {author} {\bibfnamefont {G.}~\bibnamefont {{Lindblad}}},\ }\bibfield  {title} {\bibinfo {title} {{On the generators of quantum dynamical semigroups}},\ }\href {https://doi.org/10.1007/BF01608499} {\bibfield  {journal} {\bibinfo  {journal} {Communications in Mathematical Physics}\ }\textbf {\bibinfo {volume} {48}},\ \bibinfo {pages} {119} (\bibinfo {year} {1976})}\BibitemShut {NoStop}%
\bibitem [{\citenamefont {Breuer}\ and\ \citenamefont {Petruccione}(2002)}]{Petruccione_book}%
  \BibitemOpen
  \bibfield  {author} {\bibinfo {author} {\bibfnamefont {H.~P.}\ \bibnamefont {Breuer}}\ and\ \bibinfo {author} {\bibfnamefont {F.}~\bibnamefont {Petruccione}},\ }\href@noop {} {\emph {\bibinfo {title} {The {T}heory of {O}pen {Q}uantum {S}ystems}}}\ (\bibinfo  {publisher} {Oxford University Press},\ \bibinfo {address} {New York},\ \bibinfo {year} {2002})\BibitemShut {NoStop}%
\bibitem [{\citenamefont {Alicki}\ and\ \citenamefont {Lendi}(2007)}]{Alicki_2007}%
  \BibitemOpen
  \bibfield  {author} {\bibinfo {author} {\bibfnamefont {R.}~\bibnamefont {Alicki}}\ and\ \bibinfo {author} {\bibfnamefont {K.}~\bibnamefont {Lendi}},\ }\href@noop {} {\emph {\bibinfo {title} {Quantum Dynamical Semigroups and Applications}}}\ (\bibinfo  {publisher} {Springer, Berlin},\ \bibinfo {year} {2007})\BibitemShut {NoStop}%
\bibitem [{\citenamefont {Rivas}\ and\ \citenamefont {Huelga}(2011)}]{Rivas_Huelga_book}%
  \BibitemOpen
  \bibfield  {author} {\bibinfo {author} {\bibfnamefont {A.}~\bibnamefont {Rivas}}\ and\ \bibinfo {author} {\bibfnamefont {S.~F.}\ \bibnamefont {Huelga}},\ }\href {https://doi.org/https://doi.org/10.1007/978-3-642-23354-8} {\emph {\bibinfo {title} {Open Quantum Systems}}}\ (\bibinfo  {publisher} {Springer, Berlin},\ \bibinfo {year} {2011})\BibitemShut {NoStop}%
\bibitem [{\citenamefont {Lidar}()}]{Lidar_2020_lecture}%
  \BibitemOpen
  \bibfield  {author} {\bibinfo {author} {\bibfnamefont {D.~A.}\ \bibnamefont {Lidar}},\ }\href@noop {} {\bibinfo {title} {Lecture notes on the theory of open quantum systems}},\ \Eprint {https://arxiv.org/abs/1902.00967} {arXiv:1902.00967 [quant-ph]} \BibitemShut {NoStop}%
\bibitem [{\citenamefont {Spohn}(1978)}]{Spohn_JMP_1978}%
  \BibitemOpen
  \bibfield  {author} {\bibinfo {author} {\bibfnamefont {H.}~\bibnamefont {Spohn}},\ }\bibfield  {title} {\bibinfo {title} {Entropy production for quantum dynamical semigroups},\ }\href {https://doi.org/10.1063/1.523789} {\bibfield  {journal} {\bibinfo  {journal} {Journal of Mathematical Physics}\ }\textbf {\bibinfo {volume} {19}},\ \bibinfo {pages} {1227} (\bibinfo {year} {1978})}\BibitemShut {NoStop}%
\bibitem [{\citenamefont {Spohn}\ and\ \citenamefont {Lebowitz}(1978)}]{Lebowitz}%
  \BibitemOpen
  \bibfield  {author} {\bibinfo {author} {\bibfnamefont {H.}~\bibnamefont {Spohn}}\ and\ \bibinfo {author} {\bibfnamefont {J.~L.}\ \bibnamefont {Lebowitz}},\ }\bibfield  {title} {\bibinfo {title} {Irreversible thermodynamics for quantum systems weakly coupled to thermal reservoirs},\ }\href {https://doi.org/https://doi.org/10.1002/9780470142578.ch2} {\bibfield  {journal} {\bibinfo  {journal} {Adv. Chem. Phys.}\ }\textbf {\bibinfo {volume} {38}},\ \bibinfo {pages} {109} (\bibinfo {year} {1978})}\BibitemShut {NoStop}%
\bibitem [{\citenamefont {de~Groot}\ and\ \citenamefont {Mazur}(2013)}]{Groot}%
  \BibitemOpen
  \bibfield  {author} {\bibinfo {author} {\bibfnamefont {S.~R.}\ \bibnamefont {de~Groot}}\ and\ \bibinfo {author} {\bibfnamefont {P.}~\bibnamefont {Mazur}},\ }\href@noop {} {\emph {\bibinfo {title} {Non-Equilibrium Thermodynamics}}}\ (\bibinfo  {publisher} {Dover, New York},\ \bibinfo {year} {2013})\BibitemShut {NoStop}%
\bibitem [{\citenamefont {Pietzonka}\ \emph {et~al.}(2016{\natexlab{b}})\citenamefont {Pietzonka}, \citenamefont {Barato},\ and\ \citenamefont {Seifert}}]{Seifert_PRE_2016}%
  \BibitemOpen
  \bibfield  {author} {\bibinfo {author} {\bibfnamefont {P.}~\bibnamefont {Pietzonka}}, \bibinfo {author} {\bibfnamefont {A.~C.}\ \bibnamefont {Barato}},\ and\ \bibinfo {author} {\bibfnamefont {U.}~\bibnamefont {Seifert}},\ }\bibfield  {title} {\bibinfo {title} {Universal bounds on current fluctuations},\ }\href {https://doi.org/10.1103/PhysRevE.93.052145} {\bibfield  {journal} {\bibinfo  {journal} {Phys. Rev. E}\ }\textbf {\bibinfo {volume} {93}},\ \bibinfo {pages} {052145} (\bibinfo {year} {2016}{\natexlab{b}})}\BibitemShut {NoStop}%
\bibitem [{\citenamefont {Liu}\ and\ \citenamefont {Segal}(2021)}]{Segal_PRE_2021}%
  \BibitemOpen
  \bibfield  {author} {\bibinfo {author} {\bibfnamefont {J.}~\bibnamefont {Liu}}\ and\ \bibinfo {author} {\bibfnamefont {D.}~\bibnamefont {Segal}},\ }\bibfield  {title} {\bibinfo {title} {Coherences and the thermodynamic uncertainty relation: Insights from quantum absorption refrigerators},\ }\href {https://doi.org/10.1103/PhysRevE.103.032138} {\bibfield  {journal} {\bibinfo  {journal} {Phys. Rev. E}\ }\textbf {\bibinfo {volume} {103}},\ \bibinfo {pages} {032138} (\bibinfo {year} {2021})}\BibitemShut {NoStop}%
\bibitem [{\citenamefont {Esposito}\ \emph {et~al.}(2009)\citenamefont {Esposito}, \citenamefont {Harbola},\ and\ \citenamefont {Mukamel}}]{Esposito_RMP_2009}%
  \BibitemOpen
  \bibfield  {author} {\bibinfo {author} {\bibfnamefont {M.}~\bibnamefont {Esposito}}, \bibinfo {author} {\bibfnamefont {U.}~\bibnamefont {Harbola}},\ and\ \bibinfo {author} {\bibfnamefont {S.}~\bibnamefont {Mukamel}},\ }\bibfield  {title} {\bibinfo {title} {Nonequilibrium fluctuations, fluctuation theorems, and counting statistics in quantum systems},\ }\href {https://doi.org/10.1103/RevModPhys.81.1665} {\bibfield  {journal} {\bibinfo  {journal} {Rev. Mod. Phys.}\ }\textbf {\bibinfo {volume} {81}},\ \bibinfo {pages} {1665} (\bibinfo {year} {2009})}\BibitemShut {NoStop}%
\bibitem [{\citenamefont {Bruderer}\ \emph {et~al.}(2014)\citenamefont {Bruderer}, \citenamefont {Contreras-Pulido}, \citenamefont {Thaller}, \citenamefont {Sironi}, \citenamefont {Obreschkow},\ and\ \citenamefont {Plenio}}]{Bruderer_2014_NJP}%
  \BibitemOpen
  \bibfield  {author} {\bibinfo {author} {\bibfnamefont {M.}~\bibnamefont {Bruderer}}, \bibinfo {author} {\bibfnamefont {L.~D.}\ \bibnamefont {Contreras-Pulido}}, \bibinfo {author} {\bibfnamefont {M.}~\bibnamefont {Thaller}}, \bibinfo {author} {\bibfnamefont {L.}~\bibnamefont {Sironi}}, \bibinfo {author} {\bibfnamefont {D.}~\bibnamefont {Obreschkow}},\ and\ \bibinfo {author} {\bibfnamefont {M.~B.}\ \bibnamefont {Plenio}},\ }\bibfield  {title} {\bibinfo {title} {Inverse counting statistics for stochastic and open quantum systems: the characteristic polynomial approach},\ }\href {https://doi.org/10.1088/1367-2630/16/3/033030} {\bibfield  {journal} {\bibinfo  {journal} {New Journal of Physics}\ }\textbf {\bibinfo {volume} {16}},\ \bibinfo {pages} {033030} (\bibinfo {year} {2014})}\BibitemShut {NoStop}%
\bibitem [{\citenamefont {Prech}\ \emph {et~al.}(2023)\citenamefont {Prech}, \citenamefont {Johansson}, \citenamefont {Nyholm}, \citenamefont {Landi}, \citenamefont {Verdozzi}, \citenamefont {Samuelsson},\ and\ \citenamefont {Potts}}]{Kacper_2023_PRR}%
  \BibitemOpen
  \bibfield  {author} {\bibinfo {author} {\bibfnamefont {K.}~\bibnamefont {Prech}}, \bibinfo {author} {\bibfnamefont {P.}~\bibnamefont {Johansson}}, \bibinfo {author} {\bibfnamefont {E.}~\bibnamefont {Nyholm}}, \bibinfo {author} {\bibfnamefont {G.~T.}\ \bibnamefont {Landi}}, \bibinfo {author} {\bibfnamefont {C.}~\bibnamefont {Verdozzi}}, \bibinfo {author} {\bibfnamefont {P.}~\bibnamefont {Samuelsson}},\ and\ \bibinfo {author} {\bibfnamefont {P.~P.}\ \bibnamefont {Potts}},\ }\bibfield  {title} {\bibinfo {title} {Entanglement and thermokinetic uncertainty relations in coherent mesoscopic transport},\ }\href {https://doi.org/10.1103/PhysRevResearch.5.023155} {\bibfield  {journal} {\bibinfo  {journal} {Phys. Rev. Res.}\ }\textbf {\bibinfo {volume} {5}},\ \bibinfo {pages} {023155} (\bibinfo {year} {2023})}\BibitemShut {NoStop}%
\bibitem [{\citenamefont {Landi}\ \emph {et~al.}(2024)\citenamefont {Landi}, \citenamefont {Kewming}, \citenamefont {Mitchison},\ and\ \citenamefont {Potts}}]{Landi_PRX_Quantum_2024}%
  \BibitemOpen
  \bibfield  {author} {\bibinfo {author} {\bibfnamefont {G.~T.}\ \bibnamefont {Landi}}, \bibinfo {author} {\bibfnamefont {M.~J.}\ \bibnamefont {Kewming}}, \bibinfo {author} {\bibfnamefont {M.~T.}\ \bibnamefont {Mitchison}},\ and\ \bibinfo {author} {\bibfnamefont {P.~P.}\ \bibnamefont {Potts}},\ }\bibfield  {title} {\bibinfo {title} {Current fluctuations in open quantum systems: Bridging the gap between quantum continuous measurements and full counting statistics},\ }\href {https://doi.org/10.1103/PRXQuantum.5.020201} {\bibfield  {journal} {\bibinfo  {journal} {PRX Quantum}\ }\textbf {\bibinfo {volume} {5}},\ \bibinfo {pages} {020201} (\bibinfo {year} {2024})}\BibitemShut {NoStop}%
\bibitem [{\citenamefont {Kie\ss{}lich}\ \emph {et~al.}(2006)\citenamefont {Kie\ss{}lich}, \citenamefont {Samuelsson}, \citenamefont {Wacker},\ and\ \citenamefont {Sch\"oll}}]{G_PRB_2006}%
  \BibitemOpen
  \bibfield  {author} {\bibinfo {author} {\bibfnamefont {G.}~\bibnamefont {Kie\ss{}lich}}, \bibinfo {author} {\bibfnamefont {P.}~\bibnamefont {Samuelsson}}, \bibinfo {author} {\bibfnamefont {A.}~\bibnamefont {Wacker}},\ and\ \bibinfo {author} {\bibfnamefont {E.}~\bibnamefont {Sch\"oll}},\ }\bibfield  {title} {\bibinfo {title} {Counting statistics and decoherence in coupled quantum dots},\ }\href {https://doi.org/10.1103/PhysRevB.73.033312} {\bibfield  {journal} {\bibinfo  {journal} {Phys. Rev. B}\ }\textbf {\bibinfo {volume} {73}},\ \bibinfo {pages} {033312} (\bibinfo {year} {2006})}\BibitemShut {NoStop}%
\bibitem [{\citenamefont {Latune}\ \emph {et~al.}(2020)\citenamefont {Latune}, \citenamefont {Sinayskiy},\ and\ \citenamefont {Petruccione}}]{Latune_PRA_2020}%
  \BibitemOpen
  \bibfield  {author} {\bibinfo {author} {\bibfnamefont {C.~L.}\ \bibnamefont {Latune}}, \bibinfo {author} {\bibfnamefont {I.}~\bibnamefont {Sinayskiy}},\ and\ \bibinfo {author} {\bibfnamefont {F.}~\bibnamefont {Petruccione}},\ }\bibfield  {title} {\bibinfo {title} {Negative contributions to entropy production induced by quantum coherences},\ }\href {https://doi.org/10.1103/PhysRevA.102.042220} {\bibfield  {journal} {\bibinfo  {journal} {Phys. Rev. A}\ }\textbf {\bibinfo {volume} {102}},\ \bibinfo {pages} {042220} (\bibinfo {year} {2020})}\BibitemShut {NoStop}%
\bibitem [{\citenamefont {Scandi}\ \emph {et~al.}(2020)\citenamefont {Scandi}, \citenamefont {Miller}, \citenamefont {Anders},\ and\ \citenamefont {Perarnau-Llobet}}]{Scandi_PRR_2020}%
  \BibitemOpen
  \bibfield  {author} {\bibinfo {author} {\bibfnamefont {M.}~\bibnamefont {Scandi}}, \bibinfo {author} {\bibfnamefont {H.~J.~D.}\ \bibnamefont {Miller}}, \bibinfo {author} {\bibfnamefont {J.}~\bibnamefont {Anders}},\ and\ \bibinfo {author} {\bibfnamefont {M.}~\bibnamefont {Perarnau-Llobet}},\ }\bibfield  {title} {\bibinfo {title} {Quantum work statistics close to equilibrium},\ }\href {https://doi.org/10.1103/PhysRevResearch.2.023377} {\bibfield  {journal} {\bibinfo  {journal} {Phys. Rev. Res.}\ }\textbf {\bibinfo {volume} {2}},\ \bibinfo {pages} {023377} (\bibinfo {year} {2020})}\BibitemShut {NoStop}%
\bibitem [{\citenamefont {Brask}\ and\ \citenamefont {Brunner}(2015)}]{Brask_PRE_2015}%
  \BibitemOpen
  \bibfield  {author} {\bibinfo {author} {\bibfnamefont {J.~B.}\ \bibnamefont {Brask}}\ and\ \bibinfo {author} {\bibfnamefont {N.}~\bibnamefont {Brunner}},\ }\bibfield  {title} {\bibinfo {title} {Small quantum absorption refrigerator in the transient regime: Time scales, enhanced cooling, and entanglement},\ }\href {https://doi.org/10.1103/PhysRevE.92.062101} {\bibfield  {journal} {\bibinfo  {journal} {Phys. Rev. E}\ }\textbf {\bibinfo {volume} {92}},\ \bibinfo {pages} {062101} (\bibinfo {year} {2015})}\BibitemShut {NoStop}%
\bibitem [{\citenamefont {Mitchison}\ \emph {et~al.}(2015)\citenamefont {Mitchison}, \citenamefont {Woods}, \citenamefont {Prior},\ and\ \citenamefont {Huber}}]{Mitchison_NJP_2015}%
  \BibitemOpen
  \bibfield  {author} {\bibinfo {author} {\bibfnamefont {M.~T.}\ \bibnamefont {Mitchison}}, \bibinfo {author} {\bibfnamefont {M.~P.}\ \bibnamefont {Woods}}, \bibinfo {author} {\bibfnamefont {J.}~\bibnamefont {Prior}},\ and\ \bibinfo {author} {\bibfnamefont {M.}~\bibnamefont {Huber}},\ }\bibfield  {title} {\bibinfo {title} {Coherence-assisted single-shot cooling by quantum absorption refrigerators},\ }\href {https://doi.org/10.1088/1367-2630/17/11/115013} {\bibfield  {journal} {\bibinfo  {journal} {New Journal of Physics}\ }\textbf {\bibinfo {volume} {17}},\ \bibinfo {pages} {115013} (\bibinfo {year} {2015})}\BibitemShut {NoStop}%
\bibitem [{\citenamefont {Das}\ \emph {et~al.}(2019)\citenamefont {Das}, \citenamefont {Misra}, \citenamefont {Pal}, \citenamefont {Sen(De)},\ and\ \citenamefont {Sen}}]{Sreetama_EPL_2019}%
  \BibitemOpen
  \bibfield  {author} {\bibinfo {author} {\bibfnamefont {S.}~\bibnamefont {Das}}, \bibinfo {author} {\bibfnamefont {A.}~\bibnamefont {Misra}}, \bibinfo {author} {\bibfnamefont {A.~K.}\ \bibnamefont {Pal}}, \bibinfo {author} {\bibfnamefont {A.}~\bibnamefont {Sen(De)}},\ and\ \bibinfo {author} {\bibfnamefont {U.}~\bibnamefont {Sen}},\ }\bibfield  {title} {\bibinfo {title} {Necessarily transient quantum refrigerator},\ }\href {https://doi.org/10.1209/0295-5075/125/20007} {\bibfield  {journal} {\bibinfo  {journal} {Europhysics Letters}\ }\textbf {\bibinfo {volume} {125}},\ \bibinfo {pages} {20007} (\bibinfo {year} {2019})}\BibitemShut {NoStop}%
\bibitem [{\citenamefont {Ghosh}\ \emph {et~al.}(2021)\citenamefont {Ghosh}, \citenamefont {Ghoshal},\ and\ \citenamefont {Sen}}]{Ghosh_PRA_2021}%
  \BibitemOpen
  \bibfield  {author} {\bibinfo {author} {\bibfnamefont {R.}~\bibnamefont {Ghosh}}, \bibinfo {author} {\bibfnamefont {A.}~\bibnamefont {Ghoshal}},\ and\ \bibinfo {author} {\bibfnamefont {U.}~\bibnamefont {Sen}},\ }\bibfield  {title} {\bibinfo {title} {Quantum thermal transistors: Operation characteristics in steady state versus transient regimes},\ }\href {https://doi.org/10.1103/PhysRevA.103.052613} {\bibfield  {journal} {\bibinfo  {journal} {Phys. Rev. A}\ }\textbf {\bibinfo {volume} {103}},\ \bibinfo {pages} {052613} (\bibinfo {year} {2021})}\BibitemShut {NoStop}%
\bibitem [{\citenamefont {Maity}\ and\ \citenamefont {Ghoshal}(2024)}]{Maity2024}%
  \BibitemOpen
  \bibfield  {author} {\bibinfo {author} {\bibfnamefont {A.}~\bibnamefont {Maity}}\ and\ \bibinfo {author} {\bibfnamefont {A.}~\bibnamefont {Ghoshal}},\ }\bibfield  {title} {\bibinfo {title} {Using non-{M}arkovian dynamics in effective-negative-temperature-based transient quantum {O}tto engines},\ }\href {https://doi.org/10.1103/PhysRevA.109.022207} {\bibfield  {journal} {\bibinfo  {journal} {Phys. Rev. A}\ }\textbf {\bibinfo {volume} {109}},\ \bibinfo {pages} {022207} (\bibinfo {year} {2024})}\BibitemShut {NoStop}%
\bibitem [{\citenamefont {Razzoli}\ \emph {et~al.}(2024)\citenamefont {Razzoli}, \citenamefont {Carrega}, \citenamefont {Cavaliere}, \citenamefont {Benenti},\ and\ \citenamefont {Sassetti}}]{Razzoli_QST_2024}%
  \BibitemOpen
  \bibfield  {author} {\bibinfo {author} {\bibfnamefont {L.}~\bibnamefont {Razzoli}}, \bibinfo {author} {\bibfnamefont {M.}~\bibnamefont {Carrega}}, \bibinfo {author} {\bibfnamefont {F.}~\bibnamefont {Cavaliere}}, \bibinfo {author} {\bibfnamefont {G.}~\bibnamefont {Benenti}},\ and\ \bibinfo {author} {\bibfnamefont {M.}~\bibnamefont {Sassetti}},\ }\bibfield  {title} {\bibinfo {title} {Synchronization-induced violation of thermodynamic uncertainty relations},\ }\href {https://doi.org/10.1088/2058-9565/ad6fc9} {\bibfield  {journal} {\bibinfo  {journal} {Quantum Science and Technology}\ }\textbf {\bibinfo {volume} {9}},\ \bibinfo {pages} {045032} (\bibinfo {year} {2024})}\BibitemShut {NoStop}%
\bibitem [{\citenamefont {Luoma}\ and\ \citenamefont {Piilo}(2016)}]{Luoma_2016}%
  \BibitemOpen
  \bibfield  {author} {\bibinfo {author} {\bibfnamefont {K.}~\bibnamefont {Luoma}}\ and\ \bibinfo {author} {\bibfnamefont {J.}~\bibnamefont {Piilo}},\ }\bibfield  {title} {\bibinfo {title} {Discrete dynamics and non-markovianity},\ }\href {https://doi.org/10.1088/0953-4075/49/12/125501} {\bibfield  {journal} {\bibinfo  {journal} {Journal of Physics B: Atomic, Molecular and Optical Physics}\ }\textbf {\bibinfo {volume} {49}},\ \bibinfo {pages} {125501} (\bibinfo {year} {2016})}\BibitemShut {NoStop}%
\bibitem [{\citenamefont {Wi\ss{}mann}\ \emph {et~al.}(2012)\citenamefont {Wi\ss{}mann}, \citenamefont {Karlsson}, \citenamefont {Laine}, \citenamefont {Piilo},\ and\ \citenamefont {Breuer}}]{Piilo_2012}%
  \BibitemOpen
  \bibfield  {author} {\bibinfo {author} {\bibfnamefont {S.}~\bibnamefont {Wi\ss{}mann}}, \bibinfo {author} {\bibfnamefont {A.}~\bibnamefont {Karlsson}}, \bibinfo {author} {\bibfnamefont {E.-M.}\ \bibnamefont {Laine}}, \bibinfo {author} {\bibfnamefont {J.}~\bibnamefont {Piilo}},\ and\ \bibinfo {author} {\bibfnamefont {H.-P.}\ \bibnamefont {Breuer}},\ }\bibfield  {title} {\bibinfo {title} {Optimal state pairs for non-{M}arkovian quantum dynamics},\ }\href {https://doi.org/10.1103/PhysRevA.86.062108} {\bibfield  {journal} {\bibinfo  {journal} {Phys. Rev. A}\ }\textbf {\bibinfo {volume} {86}},\ \bibinfo {pages} {062108} (\bibinfo {year} {2012})}\BibitemShut {NoStop}%
\bibitem [{\citenamefont {Popovic}\ \emph {et~al.}(2018)\citenamefont {Popovic}, \citenamefont {Vacchini},\ and\ \citenamefont {Campbell}}]{Campbell_PRA_2018}%
  \BibitemOpen
  \bibfield  {author} {\bibinfo {author} {\bibfnamefont {M.}~\bibnamefont {Popovic}}, \bibinfo {author} {\bibfnamefont {B.}~\bibnamefont {Vacchini}},\ and\ \bibinfo {author} {\bibfnamefont {S.}~\bibnamefont {Campbell}},\ }\bibfield  {title} {\bibinfo {title} {Entropy production and correlations in a controlled non-{M}arkovian setting},\ }\href {https://doi.org/10.1103/PhysRevA.98.012130} {\bibfield  {journal} {\bibinfo  {journal} {Phys. Rev. A}\ }\textbf {\bibinfo {volume} {98}},\ \bibinfo {pages} {012130} (\bibinfo {year} {2018})}\BibitemShut {NoStop}%
\bibitem [{\citenamefont {Bhattacharya}\ \emph {et~al.}(2017)\citenamefont {Bhattacharya}, \citenamefont {Misra}, \citenamefont {Mukhopadhyay},\ and\ \citenamefont {Pati}}]{Pati_PRA_2017}%
  \BibitemOpen
  \bibfield  {author} {\bibinfo {author} {\bibfnamefont {S.}~\bibnamefont {Bhattacharya}}, \bibinfo {author} {\bibfnamefont {A.}~\bibnamefont {Misra}}, \bibinfo {author} {\bibfnamefont {C.}~\bibnamefont {Mukhopadhyay}},\ and\ \bibinfo {author} {\bibfnamefont {A.~K.}\ \bibnamefont {Pati}},\ }\bibfield  {title} {\bibinfo {title} {Exact master equation for a spin interacting with a spin bath: Non-markovianity and negative entropy production rate},\ }\href {https://doi.org/10.1103/PhysRevA.95.012122} {\bibfield  {journal} {\bibinfo  {journal} {Phys. Rev. A}\ }\textbf {\bibinfo {volume} {95}},\ \bibinfo {pages} {012122} (\bibinfo {year} {2017})}\BibitemShut {NoStop}%
\bibitem [{\citenamefont {Marcantoni}\ \emph {et~al.}(2017)\citenamefont {Marcantoni}, \citenamefont {Alipour}, \citenamefont {Benatti}, \citenamefont {Floreanini},\ and\ \citenamefont {Rezakhani}}]{Marcantoni_SR_2017}%
  \BibitemOpen
  \bibfield  {author} {\bibinfo {author} {\bibfnamefont {S.}~\bibnamefont {Marcantoni}}, \bibinfo {author} {\bibfnamefont {S.}~\bibnamefont {Alipour}}, \bibinfo {author} {\bibfnamefont {F.}~\bibnamefont {Benatti}}, \bibinfo {author} {\bibfnamefont {R.}~\bibnamefont {Floreanini}},\ and\ \bibinfo {author} {\bibfnamefont {A.~T.}\ \bibnamefont {Rezakhani}},\ }\bibfield  {title} {\bibinfo {title} {Entropy production and non-{M}arkovian dynamical maps},\ }\href {https://doi.org/10.1038/s41598-017-12595-x} {\bibfield  {journal} {\bibinfo  {journal} {Scientific Reports}\ }\textbf {\bibinfo {volume} {7}},\ \bibinfo {pages} {12447} (\bibinfo {year} {2017})}\BibitemShut {NoStop}%
\bibitem [{\citenamefont {Argentieri}\ \emph {et~al.}(2014)\citenamefont {Argentieri}, \citenamefont {Benatti}, \citenamefont {Floreanini},\ and\ \citenamefont {Pezzutto}}]{Argentieri_EPL_2014}%
  \BibitemOpen
  \bibfield  {author} {\bibinfo {author} {\bibfnamefont {G.}~\bibnamefont {Argentieri}}, \bibinfo {author} {\bibfnamefont {F.}~\bibnamefont {Benatti}}, \bibinfo {author} {\bibfnamefont {R.}~\bibnamefont {Floreanini}},\ and\ \bibinfo {author} {\bibfnamefont {M.}~\bibnamefont {Pezzutto}},\ }\bibfield  {title} {\bibinfo {title} {Violations of the second law of thermodynamics by a non-completely positive dynamics},\ }\href {https://doi.org/10.1209/0295-5075/107/50007} {\bibfield  {journal} {\bibinfo  {journal} {Europhysics Letters}\ }\textbf {\bibinfo {volume} {107}},\ \bibinfo {pages} {50007} (\bibinfo {year} {2014})}\BibitemShut {NoStop}%
\bibitem [{\citenamefont {Xu}\ \emph {et~al.}(2018)\citenamefont {Xu}, \citenamefont {Liu},\ and\ \citenamefont {Feng}}]{Xu_PRE_2018}%
  \BibitemOpen
  \bibfield  {author} {\bibinfo {author} {\bibfnamefont {Y.~Y.}\ \bibnamefont {Xu}}, \bibinfo {author} {\bibfnamefont {J.}~\bibnamefont {Liu}},\ and\ \bibinfo {author} {\bibfnamefont {M.}~\bibnamefont {Feng}},\ }\bibfield  {title} {\bibinfo {title} {Positive entropy production rate induced by non-markovianity},\ }\href {https://doi.org/10.1103/PhysRevE.98.032102} {\bibfield  {journal} {\bibinfo  {journal} {Phys. Rev. E}\ }\textbf {\bibinfo {volume} {98}},\ \bibinfo {pages} {032102} (\bibinfo {year} {2018})}\BibitemShut {NoStop}%
\bibitem [{\citenamefont {Strasberg}\ and\ \citenamefont {Esposito}(2019)}]{Strasberg_PRE_2019}%
  \BibitemOpen
  \bibfield  {author} {\bibinfo {author} {\bibfnamefont {P.}~\bibnamefont {Strasberg}}\ and\ \bibinfo {author} {\bibfnamefont {M.}~\bibnamefont {Esposito}},\ }\bibfield  {title} {\bibinfo {title} {Non-{M}arkovianity and negative entropy production rates},\ }\href {https://doi.org/10.1103/PhysRevE.99.012120} {\bibfield  {journal} {\bibinfo  {journal} {Phys. Rev. E}\ }\textbf {\bibinfo {volume} {99}},\ \bibinfo {pages} {012120} (\bibinfo {year} {2019})}\BibitemShut {NoStop}%
\bibitem [{\citenamefont {Rivas}\ \emph {et~al.}(2014)\citenamefont {Rivas}, \citenamefont {Huelga},\ and\ \citenamefont {Plenio}}]{Rivas_RPP_2014}%
  \BibitemOpen
  \bibfield  {author} {\bibinfo {author} {\bibfnamefont {{\'A}.}~\bibnamefont {Rivas}}, \bibinfo {author} {\bibfnamefont {S.~F.}\ \bibnamefont {Huelga}},\ and\ \bibinfo {author} {\bibfnamefont {M.~B.}\ \bibnamefont {Plenio}},\ }\bibfield  {title} {\bibinfo {title} {Quantum non-{M}arkovianity: {C}haracterization, quantification and detection},\ }\href {https://doi.org/10.1088/0034-4885/77/9/094001} {\bibfield  {journal} {\bibinfo  {journal} {Reports on Progress in Physics}\ }\textbf {\bibinfo {volume} {77}},\ \bibinfo {pages} {094001} (\bibinfo {year} {2014})}\BibitemShut {NoStop}%
\bibitem [{\citenamefont {Breuer}\ \emph {et~al.}(2016)\citenamefont {Breuer}, \citenamefont {Laine}, \citenamefont {Piilo},\ and\ \citenamefont {Vacchini}}]{Breuer_RMP_2016}%
  \BibitemOpen
  \bibfield  {author} {\bibinfo {author} {\bibfnamefont {H.-P.}\ \bibnamefont {Breuer}}, \bibinfo {author} {\bibfnamefont {E.-M.}\ \bibnamefont {Laine}}, \bibinfo {author} {\bibfnamefont {J.}~\bibnamefont {Piilo}},\ and\ \bibinfo {author} {\bibfnamefont {B.}~\bibnamefont {Vacchini}},\ }\bibfield  {title} {\bibinfo {title} {\textit{Colloquium}: Non-{M}arkovian dynamics in open quantum systems},\ }\href {https://doi.org/10.1103/RevModPhys.88.021002} {\bibfield  {journal} {\bibinfo  {journal} {Rev. Mod. Phys.}\ }\textbf {\bibinfo {volume} {88}},\ \bibinfo {pages} {021002} (\bibinfo {year} {2016})}\BibitemShut {NoStop}%
\bibitem [{\citenamefont {Esposito}\ \emph {et~al.}(2010)\citenamefont {Esposito}, \citenamefont {Lindenberg},\ and\ \citenamefont {den Broeck}}]{Esposito_NJP_2010}%
  \BibitemOpen
  \bibfield  {author} {\bibinfo {author} {\bibfnamefont {M.}~\bibnamefont {Esposito}}, \bibinfo {author} {\bibfnamefont {K.}~\bibnamefont {Lindenberg}},\ and\ \bibinfo {author} {\bibfnamefont {C.~V.}\ \bibnamefont {den Broeck}},\ }\bibfield  {title} {\bibinfo {title} {Entropy production as correlation between system and reservoir},\ }\href {https://doi.org/10.1088/1367-2630/12/1/013013} {\bibfield  {journal} {\bibinfo  {journal} {New Journal of Physics}\ }\textbf {\bibinfo {volume} {12}},\ \bibinfo {pages} {013013} (\bibinfo {year} {2010})}\BibitemShut {NoStop}%
\bibitem [{\citenamefont {Landi}\ and\ \citenamefont {Paternostro}(2021)}]{Landi_RMP_2021}%
  \BibitemOpen
  \bibfield  {author} {\bibinfo {author} {\bibfnamefont {G.~T.}\ \bibnamefont {Landi}}\ and\ \bibinfo {author} {\bibfnamefont {M.}~\bibnamefont {Paternostro}},\ }\bibfield  {title} {\bibinfo {title} {Irreversible entropy production: From classical to quantum},\ }\href {https://doi.org/10.1103/RevModPhys.93.035008} {\bibfield  {journal} {\bibinfo  {journal} {Rev. Mod. Phys.}\ }\textbf {\bibinfo {volume} {93}},\ \bibinfo {pages} {035008} (\bibinfo {year} {2021})}\BibitemShut {NoStop}%
\bibitem [{\citenamefont {Sanderson}\ and\ \citenamefont {Curtin}(2016)}]{Sanderson}%
  \BibitemOpen
  \bibfield  {author} {\bibinfo {author} {\bibfnamefont {C.}~\bibnamefont {Sanderson}}\ and\ \bibinfo {author} {\bibfnamefont {R.}~\bibnamefont {Curtin}},\ }\bibfield  {title} {\bibinfo {title} {Armadillo: A template-based {C}++ library for linear algebra},\ }\href {https://doi.org/10.21105/joss.00026} {\bibfield  {journal} {\bibinfo  {journal} {J. Open Source Softw}\ }\textbf {\bibinfo {volume} {1}},\ \bibinfo {pages} {26} (\bibinfo {year} {2016})}\BibitemShut {NoStop}%
\bibitem [{\citenamefont {Sanderson}\ and\ \citenamefont {Curtin}(2018)}]{Sanderson1}%
  \BibitemOpen
  \bibfield  {author} {\bibinfo {author} {\bibfnamefont {C.}~\bibnamefont {Sanderson}}\ and\ \bibinfo {author} {\bibfnamefont {R.}~\bibnamefont {Curtin}},\ }\bibfield  {title} {\bibinfo {title} {A user-friendly hybrid sparse matrix class in {C}++, {L}ecture notes in computer science}\ }(\bibinfo  {publisher} {Springer International Publishing},\ \bibinfo {address} {Cham},\ \bibinfo {year} {2018})\BibitemShut {NoStop}%
\bibitem [{\citenamefont {Johnson}()}]{NLOPT}%
  \BibitemOpen
  \bibfield  {author} {\bibinfo {author} {\bibfnamefont {S.~G.}\ \bibnamefont {Johnson}},\ }\bibfield  {title} {\bibinfo {title} {The nlopt nonlinear-optimization package},\ }\href {http://github.com/stevengj/nlopt} {\ }\Eprint {https://arxiv.org/abs/http://github.com/stevengj/nlopt} {http://github.com/stevengj/nlopt} \BibitemShut {NoStop}%
\bibitem [{\citenamefont {Chanda}()}]{QIClib}%
  \BibitemOpen
  \bibfield  {author} {\bibinfo {author} {\bibfnamefont {T.}~\bibnamefont {Chanda}},\ }\bibfield  {title} {\bibinfo {title} {Qiclib},\ }\href {https://titaschanda.github.io/QIClib} {\ }\Eprint {https://arxiv.org/abs/https://titaschanda.github.io/QIClib} {https://titaschanda.github.io/QIClib} \BibitemShut {NoStop}%
\end{thebibliography}%


\end{document}